\documentclass{aa}
\usepackage[varg]{txfonts}
\usepackage{subfigure}
\usepackage{rotating}
\usepackage{xcolor}
\usepackage{multirow}
\usepackage{float}
\usepackage{makecell}
\usepackage{color}
\definecolor{mycolor}{RGB}{0,75,151}
\usepackage[colorlinks=true,urlcolor=purple, citecolor=mycolor, linkcolor=mycolor]{hyperref}
\usepackage{multicol} 
\usepackage{float}

\begin{document}

\title{The interstellar medium distribution, gas kinematics, and system dynamics of the far-infrared luminous quasar SDSS J2310+1855 at $z=6.0$}

\author{Yali Shao\inst{1, 2}\thanks{yshao@mpifr-bonn.mpg.de}
\and Ran Wang\inst{2}
\and Axel Weiss\inst{1}
\and Jeff Wagg\inst{3}
\and Chris L. Carilli\inst{4}
\and Michael A. Strauss\inst{5}
\and Fabian Walter\inst{6, 4}
\and Pierre Cox\inst{7}
\and Xiaohui Fan\inst{8}
\and Karl M. Menten\inst{1}
\and Desika Narayanan\inst{9}
\and Dominik Riechers\inst{10}
\and Frank Bertoldi\inst{11}
\and Alain Omont\inst{7}
\and Linhua Jiang\inst{2}}

\institute{Max-Planck-Institut f\"ur Radioastronomie, Auf dem H\"{u}gel 69, 53121 Bonn, Germany
\and Kavli Institute for Astronomy and Astrophysics, Peking University, Beijing 100871, China
\and SKA Observatory, Lower Withington Macclesfield, Cheshire SK11 9FT, UK
\and National Radio Astronomy Observatory, Socorro, NM 87801-0387, USA
\and Department of Astrophysical Sciences, Princeton University, Princeton, NJ 08544, USA
\and Max-Planck-Institut f\"ur Astronomie, K\"onigstuhl 17, D-69117 Heidelberg, Germany
\and Sorbonne Universit\'{e}, CNRS UMR 7095, Institut d'Astrophysique de Paris, 98bis bvd Arago, 75014 Paris, France
\and Steward Observatory, University of Arizona, 933 North Cherry Avenue, Tucson, AZ 85721, USA
\and Department of Astronomy, University of Florida, 211 Bryant Space Science Center, Gainesville, FL 32611, USA
\and I. Physikalisches Institut, Universit\"at zu K\"oln, Z\"ulpicher Str. 77, D-50937 K\"oln, Germany
\and Argelander-Institut f\"{u}r Astronomie, University at Bonn, Auf dem H\"{u}gel 71, D-53121 Bonn, Germany
}

\abstract{
We present Atacama Large Millimeter/submillimeter Array (ALMA)  sub-kiloparsec- to kiloparsec-scale resolution observations of the [\ion{C}{II}], CO\,(9--8), and OH$^{+}$\,($1_{1}$--$0_{1}$)  lines along with their dust continuum emission toward the far-infrared (FIR) luminous quasar SDSS J231038.88+185519.7 at $z = 6.0031$, to study the interstellar medium distribution, the gas kinematics, and the quasar-host system dynamics. We decompose the intensity maps of the [\ion{C}{II}] and CO\,(9--8) lines and  the dust continuum with two-dimensional elliptical S{\'e}rsic models. The [\ion{C}{II}] brightness follows a flat distribution with a S{\'e}rsic index of 0.59. The CO\,(9--8) line and the dust continuum can be  fit with an unresolved nuclear component and an extended S{\'e}rsic component with a S{\'e}rsic index of $\sim$1, which may correspond to the emission from an active galactic nucleus dusty molecular torus and a quasar host galaxy, respectively. The different [\ion{C}{II}] spatial distribution may be due to the effect of the high dust opacity, which increases the FIR background radiation on the [\ion{C}{II}] line, especially in the galaxy center, significantly suppressing the [\ion{C}{II}] emission profile. The dust temperature drops with distance from the center. The effective radius of the dust continuum is smaller than that of the line emission and the dust mass  surface density, but is consistent with that of the star formation rate surface density. This may indicate that the dust emission is a less robust tracer of the dust and gas distribution but is a decent tracer of the obscured star formation activity. The OH$^{+}$\,($1_{1}$--$0_{1}$) line shows a P-Cygni profile with an absorption at $\sim$--400 km/s, which may indicate an outflow with a neutral gas mass of $(6.2\pm1.2)\times10^{8} M_{\odot}$ along the line of sight. We employed a three-dimensional tilted ring model to fit the [\ion{C}{II}] and CO\,(9--8)  data cubes. The two lines   are both rotation dominated and trace identical disk geometries and gas motions. This suggest that the [\ion{C}{II}] and CO\,(9--8) gas are coplanar and corotating in this quasar host galaxy. The consistent circular velocities measured with  [\ion{C}{II}] and CO\,(9--8)  lines indicate that these two lines trace a similar gravitational potential. We decompose the circular rotation curve measured from the kinematic model fit to the [\ion{C}{II}]  line into four matter components (black hole, stars, gas, and dark matter). The quasar-starburst system is dominated by  baryonic matter inside the central few kiloparsecs. We constrain the black hole mass to be $2.97^{+0.51}_{-0.77}\times 10^{9}\,M_{\odot}$; this is the first time that the dynamical mass of a black hole has been measured at $z\sim6$. This mass is consistent with that determined using the scaling relations from quasar emission lines. A massive stellar component (on the order of $10^{9}\,M_{\odot}$) may have already existed when the Universe was only $\sim$0.93 Gyr old. The relations between the black hole mass and  the baryonic mass of this quasar indicate that the central supermassive black hole may have formed before its host galaxy.
}


\titlerunning{SDSS J2310+1855 at $z=6.0$}
\maketitle

\section{Introduction}
\label{sec_intro}

Almost 400 quasars at $z\ge5.7$ have been discovered in the past $\sim$20 years, mostly from optical wide-field multiband surveys  (e.g., \citealt{Fan2000}; \citealt{Matsuoka2022}). These quasars provide a unique opportunity to study a number of key issues, for example the formation of  young luminous quasars, the evolving impact of the central black holes  on the host galaxies, and the typical interstellar medium (ISM) conditions  in the quasar host galaxies, during the epoch at which the intergalactic medium was being reionized by the first luminous sources. 

Toward some of these $z\sim6$ quasars, bright millimeter dust continuum emission has been detected  at millijansky or sub-millijansky levels (e.g., \citealt{Bertoldi2003}; \citealt{Wang2007, Wang2008thermal, Wang2011fir}; \citealt{Omont2013}; \citealt{Venemans2018}; \citealt{LiQ2020}) using the IRAM facilities, the \textit{James Clerk Maxwell} Telescope (JCMT), and the Atacama Large Millimeter/submillimeter Array  (ALMA), indicating far-infrared (FIR) luminosities of $10^{11}$ to $10^{13}\,L_{\odot}$ and star formation at rates of hundreds to thousands $M_{\odot}$/yr in these young quasar host galaxies. 
A few of these $z\sim6$ FIR luminous quasars  have been detected in multi-$J$ CO transitions from $J=$ 2$-$1 to 17$-$16 (e.g., \citealt{Bertoldi2003hco}; \citealt{Walter2003, Walter2004}; \citealt{Carilli2007};  \citealt{Riechers2009}; \citealt{Wang2010, Wang2011fir, Wang2011}; \citealt{Gallerani2014}; \citealt{Shao2019}; \citealt{LiJ2020}; \citealt{Pensabene2021}) using the IRAM facilities and ALMA, indicating abundant gas reservoirs on the order of $10^{10} M_{\odot}$ and highly excited molecular gas in the quasar host galaxies. 
Most of these high-$z$ FIR-luminous quasars have been detected in [\ion{C}{II}] 158 $\mu$m fine-structure line emission down to arcsecond- and subarcsecond-scale resolution using the IRAM facilities and  ALMA (e.g., \citealt{Maiolino2005}; \citealt{Walter2009}; \citealt{Wang2013}; \citealt{Shao2017}; \citealt{Decarli2018}; \citealt{Izumi2018}; \citealt{Venemans2020}; \citealt{Yue2021}; \citealt{Walter2022}; \citealt{Meyer2022}), suggesting that high-luminosity quasar host galaxies have lower dynamical masses than local galaxies with similar black hole masses, although the ratios of the black hole mass to the host galaxy dynamical mass of most of the low-luminosity quasars  are consistent with the local value. In addition, high-resolution CO, [\ion{C}{II}], and ultraviolet (UV) continuum observations reveal close companions for some $z\sim6$ quasars (e.g., \citealt{Wang2011, Wang2019}; \citealt{McGreer2014}; \citealt{Decarli2017}; \citealt{Miller2020}; \citealt{Izumi2021}; \citealt{Pensabene2021}).  These studies suggest an overall scenario of the  coevolution of quasars and their host galaxies: the central supermassive black holes (SMBHs) may grow rapidly through major galaxy  mergers and the accretion of large amounts of gas, which may trigger high rates of star formation in the quasar host galaxies and thus influence the relationship between the black hole mass and the dynamical mass of these quasar-host systems.

However, only a few of these studies are based on high-resolution (i.e., sub-kiloparsec scale) and high sensitivity observations in the millimeter (e.g., \citealt{Yue2021}; \citealt{Walter2022}). The quasar host galaxies at $z\sim6$ are at most a few beam sizes across. Thus,  the inferred ISM properties and the dynamical information of the quasar-host systems are mostly based on  unresolved or slightly resolved spatially integrated flux stacking along the velocity or frequency axes (i.e., the intensity maps) and on the distribution of spatially integrated flux for each velocity interval as a function of apparent radial Doppler velocity (i.e., the integrated spectra). High-resolution (i.e., a few hundred parsecs) ALMA observations of the ISM in these quasar host galaxies at $z\sim6$ can provide insight into these issues, for example the spatial distribution of different ISM tracers in addition to the star formation activity, the gas kinematics, and the overall dynamics probed by different ISM tracers in the SMBH-host system at the earliest epochs.

Measuring the sizes and  morphologies of galaxies  in the early Universe is critical for understanding  the initial stage of galaxy formation and evolution. The long-wavelength FIR dust continuum traces regions of young and compact star formation that are severely obscured by dust at optical wavelengths. In both observed galaxies  (e.g., \citealt{Riechers2013, Riechers2014}; \citealt{Ikarashi2015}; \citealt{Hodge2016}; \citealt{Wang2019}; \citealt{Tadaki2020}) and simulated galaxies (e.g., \citealt{Cochrane2019}; \citealt{Popping2022}), the distribution of the dust continuum emission generally is more compact than both the cold gas and the dust mass, but is more extended (more compact) than the stellar component at $z\le3$ ($z\ge4$). The [\ion{C}{II}] fine-structure transition at 157.74 $\mu$m is one of the primary coolants of the star-forming ISM in galaxies (e.g., \citealt{Stacey1991}), and it traces both the neutral atomic and ionized gas. The different rotational transitions of CO trace  cold and warm gas that will form stars in the future. Their sizes and morphologies are strong probes of ISM properties. Thus, comparing the size and morphology between the dust emission and the [\ion{C}{II}] and CO lines will provide insights into the evolutionary process of  the ISM in which star formation takes place.

The kinematics of the ISM can yield important  information on the dynamical structure of the quasar host, as well as the distribution and fraction of baryonic and dark matter in the system (e.g., \citealt{vandeHulst1957}; \citealt{Rubin1983}; \citealt{deBlok2008}). 
The  rotation velocities of different phases of the ISM frequently show a complex picture and differ from source to source. The rotation velocities measured from the molecular components (i.e., CO) and atomic components (i.e., \ion{H}{I}) are generally in agreement (e.g., \citealt{Wong2004}; \citealt{Frank2016}). \citet{Ubler2018} find that the kinematics of CO and \ion{H}{$\alpha$}  are in good agreement in a $z = 1.4$ star-forming galaxy. However, \citet{Simon2005} find that the \ion{H}{$\alpha$} line in NGC 4605 shows systematically slower rotation than the CO. \citet{deBlok2016} find that the velocity of the [\ion{C}{II}] line is systematically larger than that of the CO or \ion{H}{I} lines in a few nearby galaxies, which they attribute to systematics in the data reduction and the low velocity resolution of the [\ion{C}{II}] data. It is still not clear why discrepancies between different kinematic tracers are observed in some galaxies.  

In the nearby Universe, the stellar bulge can be observed directly through optical and near-infrared (NIR) imaging, and its dynamics can be probed via  investigations of the rotation curve (e.g., \citealt{Begeman1991}; \citealt{deBlok2002}; \citealt{Sofue2009}; \citealt{Gao2018}). At high redshifts, the stellar bulge may already exist long before the peak of cosmic star formation, as demonstrated by the [\ion{C}{II}] gas kinematics of $4\le z \le5$ dusty star-forming galaxies and quasars (e.g., \citealt{Rizzo2020, Rizzo2021}; \citealt{Tsukui2021}). However, at higher redshifts, into the reionization epoch,  it remains unknown whether the stellar bulge is already present.

In this paper we report on ALMA sub-kiloparsec- to kiloparsec-resolution observations of the [\ion{C}{II}], CO\,(9--8), and OH$^{+}$\,($1_{1}$--$0_{1}$)  lines and their underlying  dust continuum toward the FIR-luminous quasar SDSS J231038.88+185519.7 (hereafter J2310$+$1855) at $z = 6.0031$, to study the ISM distribution, the gas kinematics, and the quasar-host system dynamics. 
This quasar is first reported by \citet{Wang2013} with 250 GHz dust continuum and CO\,(6--5) line observations using the IRAM facilities and [\ion{C}{II}] line observations using the ALMA.
\citet{Jiang2016} are credited with its discovery; they discovered this optically bright quasar with $m_{1450\AA}$ = 19.30 mag in the Sloan Digital Sky Survey (SDSS) imaging data. The black hole mass has been measured to be $(4.17\pm1.02)\times10^{9} M_{\odot}$ and $(3.92\pm0.48)\times10^{9} M_{\odot}$ from the GEMINI/GNIRS spectra of  \ion{Mg}{II} and \ion{C}{IV} lines, respectively \citep{Jiang2016}. A smaller black hole mass of $(1.8\pm0.5)\times10^{9} M_{\odot}$  based on the \ion{C}{IV} emission line detected in the X-SHOOTER/VLT spectrum is reported by \citet{Feruglio2018}. All of these estimates of the black hole mass are based on the local scaling relations (e.g., \citealt{Shen2013}). However, it is still under debate if the local relationship is suitable at high redshift.

J2310$+$1855 has a well-measured dust content and a detailed rest-frame NIR-to-FIR spectral energy distribution (SED) observed by SDSS, WISE, \textit{Herschel}, IRAM, and ALMA. \citep{Shao2019}. These observations reveal a dust temperature of $\sim$40 K and a star formation rate (SFR) of $\sim$2000 $M_{\odot}$/yr under the optically thin approximation. However, the dust and star formation distribution within the host are unknown, and the origins of different dust components have not yet been identified.
The multi-$J$ CO transitions (from $J=$ 2--1 to 13--12) have been detected using the Karl G. Jansky Very Large Array (VLA), the IRAM facilities, and   ALMA (\citealt{Wang2013}; \citealt{Feruglio2018}; \citealt{Shao2019}; \citealt{Carniani2019}; \citealt{LiJ2020}; Riechers  et al. in preparation). A detailed CO spectral line energy distribution analysis reveals that, in addition to the far-UV radiation from young and massive stars, another gas heating mechanism (e.g., X-ray radiation and/or  shocks) may be needed to explain the observed CO luminosities (\citealt{Carniani2019}; \citealt{LiJ2020}). 
The low excitation water para-H$_{2}$O\,(2$_{02}$-1$_{11}$) and para-H$_{2}$O\,(3$_{22}$-3$_{13}$),  OH$^{+}$\,($1_{1}$--$0_{1}$) lines, and lines from ionized gas such as [\ion{O}{I}] 146 $\mu$m, [\ion{O}{III}] 88 $\mu$m,  and [\ion{N}{II}] 122 $\mu$m have been detected with ALMA toward J2310+1855, and a solar-level metallicity is proposed based on the [\ion{O}{III}]/[\ion{N}{II}] ratio (\citealt{Hashimoto2019}; \citealt{LiJ22020}; \citealt{Tripodi2022}).  We should note that [\ion{N}{II}] 122 $\mu$m is an upper [\ion{N}{II}] line (i.e., the transition of $\rm^{3}P_{2}$-$\rm^{3}P_{1}$), so there are potential excitation effects. In addition,  [\ion{O}{III}] 88 $\mu$m and [\ion{N}{II}] 122 $\mu$m lines have significantly different ionization potentials ($\sim$14 versus $\sim$35 eV), so they do not necessarily trace the same parts of the \ion{H}{II} regions.
The bright [{\ion{C}{II}] line has been detected with  ALMA at low angular resolution ($>0\farcs7$; \citealt{Wang2013}; \citealt{Feruglio2018}). A velocity gradient is obvious, which may indicate that the disk is dominated by  rotating gas. The $\sim$4 kpc resolution [\ion{C}{II}]  data reveal a dynamical mass of 9.6 $\times$ 10$^{10}$ $M_{\odot}$ with an approximate estimate of the inclination angle (46$\degr$, determined from the ratio between the minor and major axis), suggesting a $M_{\rm BH}/M_{\rm bulge}$ value that is higher than the local value \citep{Wang2013}. \citet{Tripodi2022} used $\sim$1 kpc resolution [\ion{C}{II}]  data  to find a dynamical mass of 5.2 $\times$ 10$^{10}$ $M_{\odot}$ within a 1.7 kpc region, based on a kinematic modeling on the data cube with a rotating disk inclination angle of 25$\degr$. However, the limited spatial resolution introduces large uncertainties in the determination of the gas kinematics, making it difficult to perform a dynamical decomposition of the quasar-host system. 
In this paper we use ALMA high-resolution observations of the [\ion{C}{II}] and CO\,(9--8)  lines ($\sim$0.6 and $\sim$1 kpc, respectively) and their underlying dust continuum in this quasar to explore its dynamics in detail.

The outline of this paper is as follows.
In Sect. \ref{sec_obs} we describe our ALMA observations and the data reduction. 
In Sect. \ref{sec_res} we present the measurements of the [\ion{C}{II}], CO\,(9--8),  and OH$^{+}$\,($1_{1}$--$0_{1}$)  lines and the underlying dust continuum.
In Sect. \ref{sec_ana} we fit the line and continuum intensity maps to 2D S{\'e}rsic functions, constrain the dust properties, apply a 3D tilted ring model to the [\ion{C}{II}] and CO\,(9--8)  data cubes, and decompose the circular rotation curve measured from the high-resolution [\ion{C}{II}] line into multiple components.
In Sect. \ref{sec_disc}  we discuss the spatial distribution and extent of the ISM,  the surface density of the gas and the star formation, the ionized and molecular gas kinematics, the gas outflow, and the dynamics of the quasar--host system.
In Sect. \ref{sec_sum} we summarize our results. 
Finally, in Appendices \ref{sec_channel}--\ref{sec_til}, we present the channel maps of [\ion{C}{II}] and CO\,(9--8) lines and describe the 2D S{\'e}rsic function and the tilted ring model.
Throughout the paper we adopt a  $\Lambda$CDM cosmology with $H_{0}$ = 67.8 km/s/Mpc, $\Omega_{\rm M}$ = 0.3089, and $\Omega_{\Lambda}$ = 0.6911 \citep{PlanckCollaboration2016}. Under this cosmological assumption and at $z=6.0031$, 1$^{\arcsec}$ on the sky corresponds to a physical size of  5.84 kpc, the luminosity distance is $D_{\rm L}=59.0$ Gpc, and the age  was 0.9297 Gyr since the Big Bang.

\section{ALMA observations and data reduction} 
\label{sec_obs}

We conducted ALMA band-6 observations of the [\ion{C}{II}]  line ($\nu_{\rm rest}$ = 1900.5369 GHz; redshifted to $\nu_{\rm obs}$ = 271.3851 GHz), along with band-4 observations of the CO\,(9--8)  line ($\nu_{\rm rest}$ = 1036.9124 GHz; redshifted to $\nu_{\rm obs}$ = 148.0648 GHz) and the OH$^{+}$\,($1_{1}$--$0_{1}$) line ($\nu_{\rm rest}$ = 1033.0582 GHz; redshifted to $\nu_{\rm obs}$ = 147.5144 GHz; hereafter OH$^{+}$),  toward J2310+1855 at $z=6.0031$ from 2019 August 08 to 20 (PI: Yali Shao; Project code: 2018.1.00597.S).
We used 40--47 12-m antennas in the C43-7 configuration with a maximum projected baseline of 3.6 km for observations  with both bands. We centered one of the 2 GHz spectral windows (four in total) on the redshifted [\ion{C}{II}]/CO\,(9--8)   line frequency, and used the rest of the  spectral windows to observe the dust continuum. 
The total observing times are 1.60 and 1.44 hours, resulting in on-source integration times of 0.66 and 0.80 hours for the [\ion{C}{II}] and CO\,(9--8) /OH$^{+}$ observations, respectively. 
We established the flux density scale using scans of the standard ALMA calibrator J2253+1608.
The flux calibration uncertainties are $\sim$5--10$\%$ for ALMA Cycle 6 bands-4 and -6 observations \citep{Warmels2018}. We checked the phase and water vapor by observing nearby calibrators of J2316+1618 and J2307+1450. 
The data were calibrated using the ALMA standard  pipeline using the Common Astronomy Software Application ({\footnotesize{CASA}}\footnote{\url{https://casa.nrao.edu/}}). The original channel width is 15.625 MHz, corresponding to $\sim$17 and 32 km/s for the band-6 and -4 observations, respectively, which can sample the intrinsic line widths well. 
The underlying dust continuum emission was subtracted in the uv-plane for both data sets. 

In order to improve the uv coverage and thus the final image  and spectrum sensitivity, we also included archival data from projects -- 2011.0.00206.S (PI: Ran Wang), 2015.1.00997.S (PI: Roberto Maiolino), and 2015.1.01265.S (PI: Ran Wang), which observed the [\ion{C}{II}] line in the 12-m array and 12-m + 7-m arrays, and the CO\,(9--8) /OH$^{+}$ ($1_{1}$--$0_{1}$) line in the 12-m array, toward J2310+1855, respectively. We only consider the frequency range overlapping with our science goals.
As we used ALMA Cycle 0 observations for which the data weights are not correct, before the final combination of multi-epoch data, we re-weighted the calibrated target data with the {\footnotesize{STATWT}}  task in {\footnotesize{CASA}}, which attempts to assess the sensitivity per visibility and adjust the weights accordingly with line-free data.
Finally, we made the line  data cube from the combined calibrated data using the {\footnotesize{TCLEAN}} task in {\footnotesize{CASA}} with robust weighting (robust = 0.5) for the [\ion{C}{II}] and CO\,(9--8)  line in order to optimize the sensitivity per frequency bin and the resolution of the final maps, and natural weighting (robust = 2.0) for the OH$^{+}$  line in order to improve the sensitivity. For the continuum images, we used robust weighting (robust = 0.5), which allows us to compare the emission of the continuum with that of the covered lines ([\ion{C}{II}] and CO\,(9--8))  at similar angular resolution. As the OH$^{+}$  line is very weak, we {\footnotesize{TCLEAN}} the data cube deeply with a threshold of 1$\sigma$  using a $1\arcsec\times1\arcsec$ square centered on the quasar position. For the rest of the lines and continuum, we {\footnotesize{TCLEAN}} to a level of 3$\sigma$. In addition, during {\footnotesize{TCLEAN}} for the 12-m and 7-m combined data, we used mosaic gridder mode, which can correctly image data with different antenna sizes.
The synthesized beam size of the final [\ion{C}{II}] and CO\,(9--8)  images are $0\farcs111$ $\times$ $0\farcs092$ and $0\farcs187$ $\times$ $0\farcs153$,  corresponding to 0.65 kpc $\times$ 0.54 kpc and 1.09 kpc $\times$ 0.89 kpc, respectively, at the quasar redshift. The noise levels in a 15.625 MHz channel are 0.17 and 0.10 mJy/beam for the [\ion{C}{II}] and CO\,(9--8) lines, respectively.   The root mean square (rms) noise in the continuum maps at 262 and 147 GHz are 0.02 and 0.01 mJy/beam, respectively.

\begin{table*}
\renewcommand\arraystretch{1.6}
\caption{Measurements from ALMA observations.}
\label{obs}
\centering
\begin{tabular}{llccccccccc}
\hline\hline
&&[\ion{C}{II}]&CO\,(9--8) &OH$^{+}$\\
\hline
\multicolumn{2}{l}{\underline{Line Intensity Map}}\\
\multicolumn{2}{l}{Weighting}&0.5&0.5&2.0\\
Size$_{\rm beam}$ &($\arcsec$)&$0.111\times0.092$&$0.187\times0.153$&$0.243\times0.198$\\
\hline
\multicolumn{2}{l}{\underline{Line Spectrum}}\\
$z$&&$6.0032\pm0.0001^{a}$&$6.0028\pm0.0002^{a}$&-\\
FWHM &(km/s)&$384\pm5^{a}$&$388\pm16^{a}$&$225\pm117^{c}$; $240\pm74^{d}$; $330\pm198^{e}$\\
$\nu S_{\nu\_\rm spectrum}$ &(Jy km/s)&$6.50\pm0.17^{a}$; $6.31\pm0.18^{b}$&$1.44\pm0.11^{a}$; $1.39\pm0.11^{b}$&$0.072\pm0.049^{c}$; $0.072\pm0.023^{d}$; $-0.065\pm0.044^{e}$\\
$L_{\rm spectrum}$ &($10^{9}L_{\odot}$)&$6.39\pm0.17^{a}$; $6.21\pm0.18^{b}$&$0.77\pm0.06^{a}$; $0.74\pm0.06^{b}$&$0.038\pm0.026^{c}$; $0.038\pm0.013^{d}$; $-0.035\pm0.023^{e}$\\
\hline
\multicolumn{2}{l}{\underline{Continuum Map}}\\
Size$_{\rm beam\_con}$ &($\arcsec$)&$0.113\times0.092$&$0.192\times0.156$&-\\
$\nu_{\rm con}$ &(GHz)&262&147&-\\
\hline
\end{tabular}
\tablefoot{The parameters for the line spectra are measured using a single and double Gaussian fit for [\ion{C}{ii}] and CO\,(9--8) lines, and a triple-Gaussian fit for the OH$^{+}$ line. The adopted [\ion{C}{ii}],  CO\,(9--8) and OH$^{+}$ line spectra are shown in the right panels of Figs. \ref{ciimap}, \ref{co98map} and \ref{ohmap}. For the  [\ion{C}{ii}] and CO\,(9--8) lines: $^{a}$single Gaussian fit results; $^{b}$double-Gaussian fit results. For the OH$^{+}$ line: $^{c}$blue part emission component; $^{d}$red part emission component; $^{e}$absorption component.}
\end{table*}

\section{Results} 
\label{sec_res}

The [\ion{C}{II}] and CO\,(9--8)  lines and their underlying dust continuum are all spatially resolved. The intensity peak of OH$^{+}$ ($1_{1}$--$0_{1}$) is detected at $>$5$\sigma$ significance.  We list the observational results in Table \ref{obs}.

\subsection{The [\ion{C}{II}] line}
\label{sec_rescii}

The velocity-integrated intensity map, the intensity-weighted velocity map and the spectrum of the [\ion{C}{II}]  line  are presented in Fig. \ref{ciimap}. A clear velocity gradient can be seen in the velocity map. In addition, the [\ion{C}{II}] emission peak moves in a circular, counterclockwise path from 219 to $-$281 km/s shown in the [\ion{C}{II}]  line channel maps (Fig. \ref{ciichannel}). These are the main characteristics of a disk with rotating gas. Similar rotating disks have been widely detected in the local and high-$z$ Universe in both quasar hosts and galaxies without an active galactic nucleus (AGN;\citealt{Wang2013}; \citealt{Lucero2015}; \citealt{Shao2017}; \citealt{Jones2017, Jones2021}; \citealt{Banerji2021}; \citealt{Alonso-Herrero2018}; \citealt{BewketuBelete2021}).  
We measured the line spectrum by integrating the intensity of each channel from the [\ion{C}{II}] line  data cube, including pixels determined in the line-emitting region above 2$\sigma$ in the [\ion{C}{II}] intensity map. The line spectrum with original spectral resolution (15.625 MHz) is shown in the right panel of  Fig. \ref{ciimap}, which reveals that this line has an asymmetric profile with an enhancement to positive velocities.

The single Gaussian fit to the [\ion{C}{II}] spectrum shows that the line width and the source redshift are consistent with our previous  Cycle 0 observations ($393\pm21$ km/s and $6.0031\pm0.0002,$ respectively; \citealt{Wang2013}). The [\ion{C}{II}]  line flux calculated from the Gaussian fit to the line spectrum agrees with our previous ALMA observations at $0\farcs7$ resolution ($8.83\pm0.44$ Jy km/s; \citealt{Wang2013}) within the $\sim$ 15$\%$ calibration uncertainty. Our double-Gaussian fit results are consistent with that of the single Gaussian fit. We also measured a consistent value by modeling the observed intensity map with the 2D elliptical S{\'e}rsic model as described in Sect. \ref{sec_sb} and shown in Table \ref{par_ima}.  

\begin{figure*}
\centering
\subfigure{\includegraphics[width=0.31\textwidth]{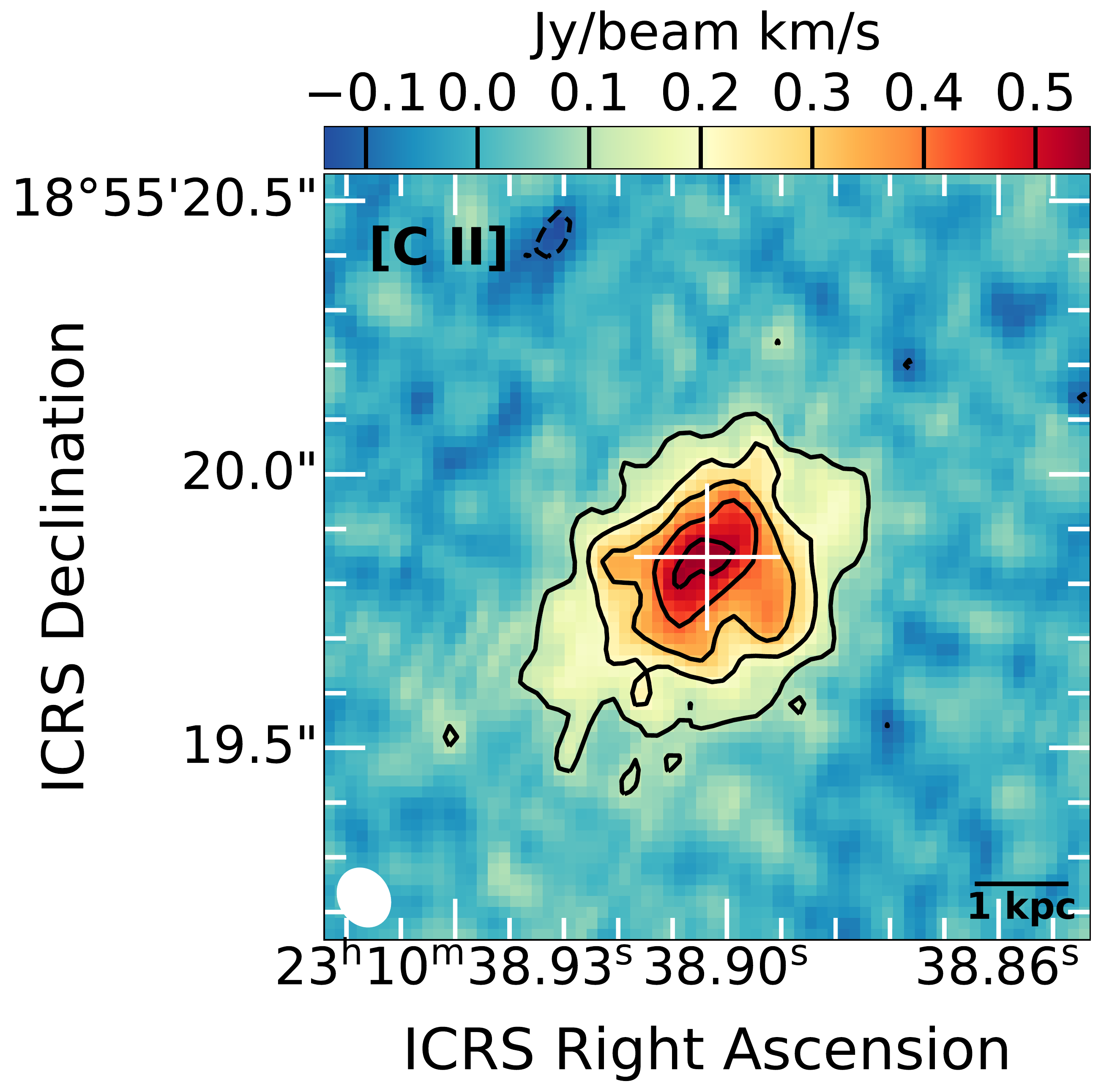}} 
\subfigure{\includegraphics[width=0.31\textwidth]{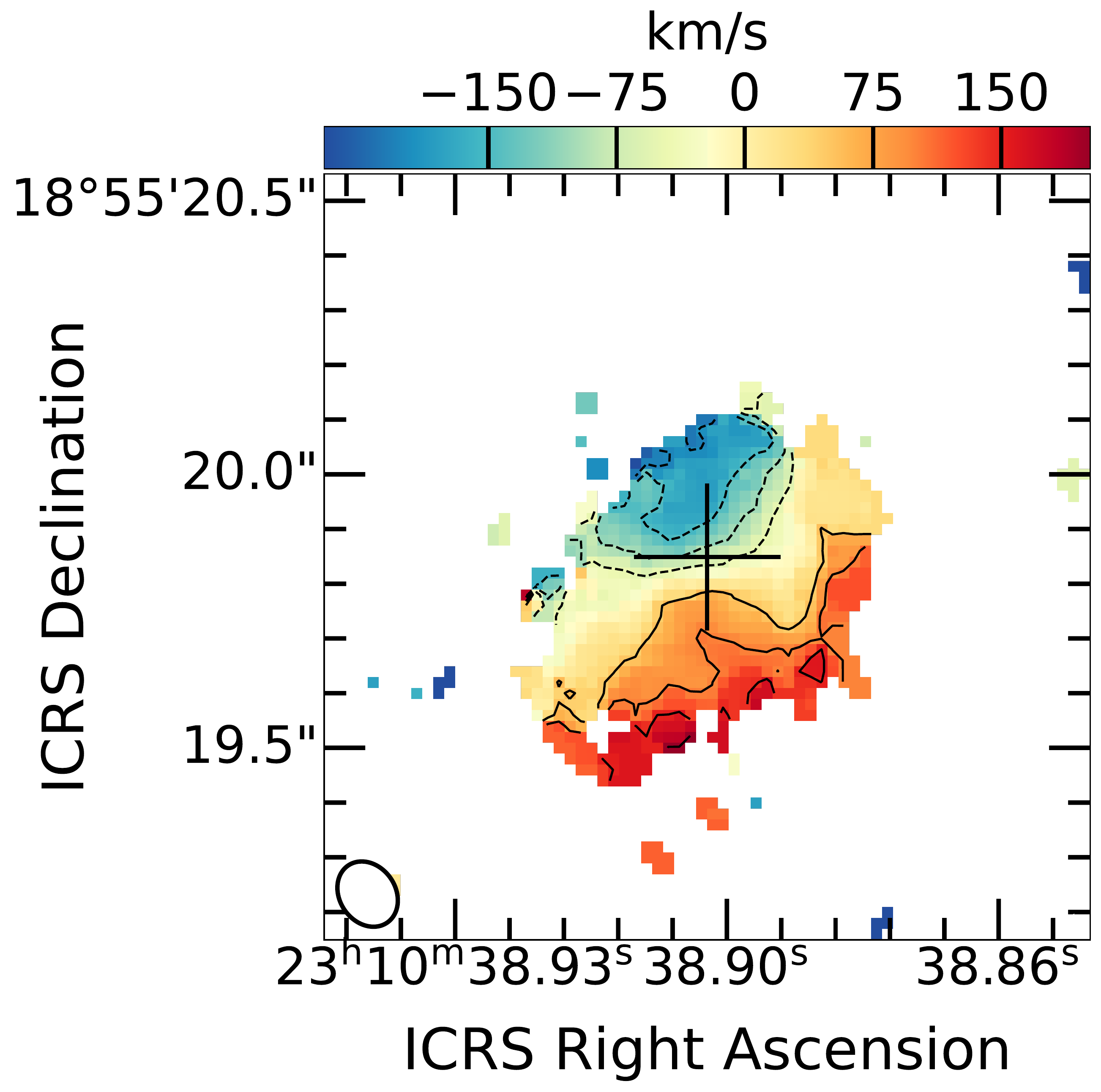}}
\subfigure{\includegraphics[width=0.36\textwidth]{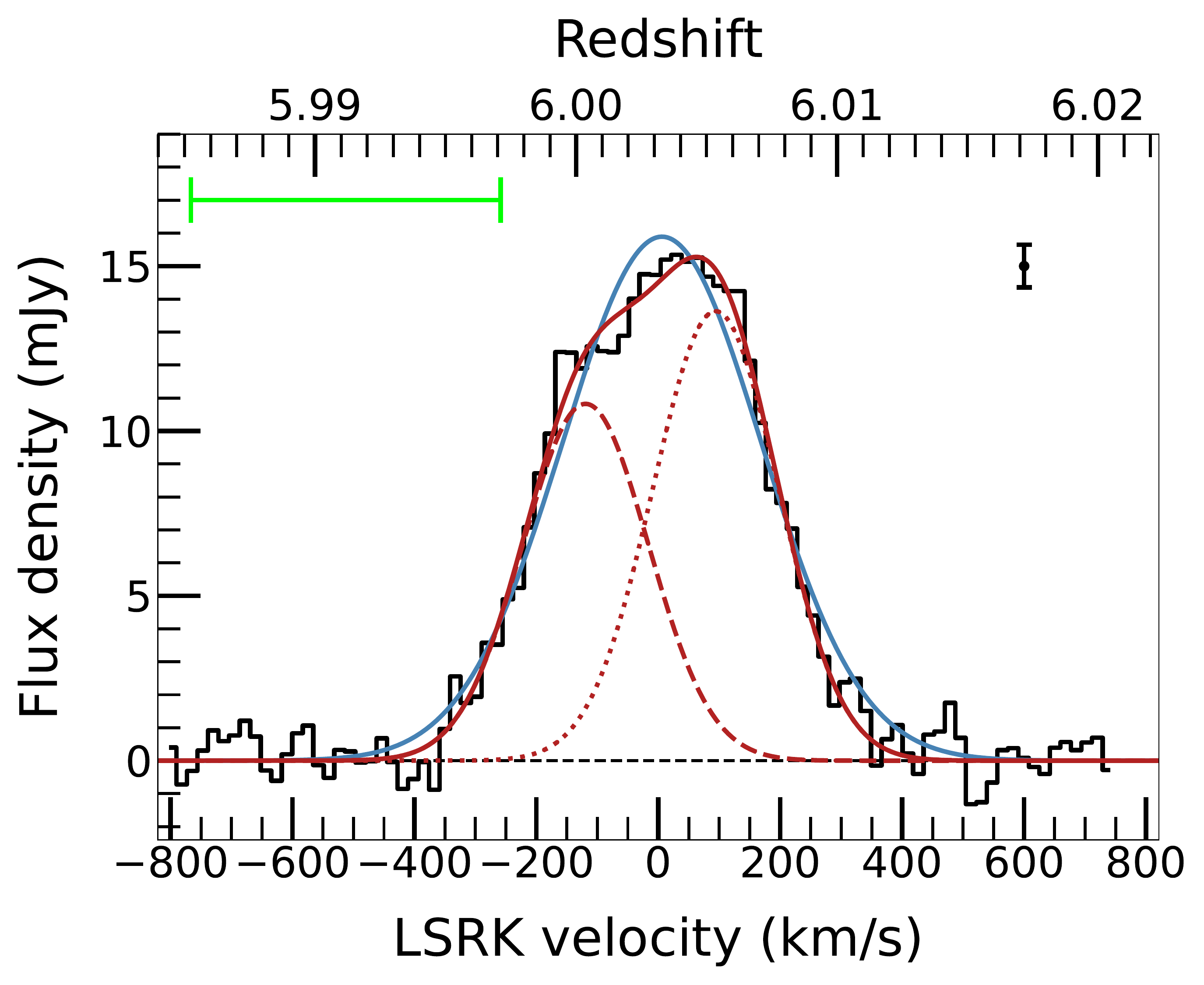}}
\caption{
ALMA observed [\ion{C}{II}] line. Left panel: [\ion{C}{II}] velocity-integrated map. The white plus sign is the \textit{Hubble} Space Telescope (HST) quasar position (RA of 23:10:38.90; Dec of 18:55:19.85). The shape of the restoring beam with a FWHM size of $0\farcs111$ $\times$ $0\farcs092$ is plotted  in the bottom-left corner. The contour levels are  [$-$3, 3, 6, 9, 12, 15] $\times$ rms$_{\rm[\ion{C}{II}]}$, where rms$_{\rm[\ion{C}{II}]}=0.035$ Jy/beam km/s. 
Middle panel: [\ion{C}{II}] velocity map produced using pixels above 4$\sigma$ in the channel maps. The black plus sign is the HST quasar position. A clear velocity gradient can be seen.
Right panel: [\ion{C}{II}] line spectrum (black histogram) overplotted with the best-fit Gaussian profiles. The spectrum is extracted from the 2$\sigma$ region of the source emitting area in the [\ion{C}{II}] intensity map. The solid blue line presents the best-fit single Gaussian model. The solid red line represents the best-fit double-Gaussian model, and the dashed and dotted red lines are for each of the double Gaussians. The  kinematic local standard of rest (LSRK) velocity scale is relative to the [\ion{C}{II}] redshift from our ALMA Cycle 0 observations \citep{Wang2013}. The spectral resolution is 15.625 MHz, corresponding to 17 km/s. The rms of the spectrum measured using the line-free spectrum is 0.65 mJy and is shown as a black bar. The  [\ion{C}{II}] spectral profile is asymmetric, with enhancement on the red side. The green bar indicates the velocity range of the OH$^{+}$ absorption.
}
\label{ciimap}
\end{figure*}

\subsection{The CO\,(9--8), OH$^{+}$ lines and the possible detection of para-H$_{2}$O ground-state emission line}
\label{sec_resco98}

The velocity-integrated intensity map, the intensity-weighted velocity map and the spectrum of the CO\,(9--8)  line  are presented in Fig. \ref{co98map}.  A similar asymmetric component at $\sim$--50 km/s is seen as in the  [\ion{C}{II}] line. 

The Gaussian fit of the CO\,(9--8)  line reveals a line width and source redshift that are consistent with the previous ALMA Cycle 3 observations ($376\pm18$ km/s and $6.0031\pm0.0002,$ respectively; \citealt{LiJ2020}). The CO\,(9--8) line flux calculated from the Gaussian fit on the line spectrum agrees with the previous ALMA observations at $\sim0\farcs7$ resolution ($ 1.31\pm0.06$ Jy km/s; \citealt{LiJ2020}). We also obtained a consistent value by modeling the observed intensity map with  the 2D elliptical  S{\'e}rsic model shown in Table \ref{par_ima} (Sect. \ref{sec_sb}).  

\begin{figure*}
\centering
\subfigure{\includegraphics[width=0.31\textwidth]{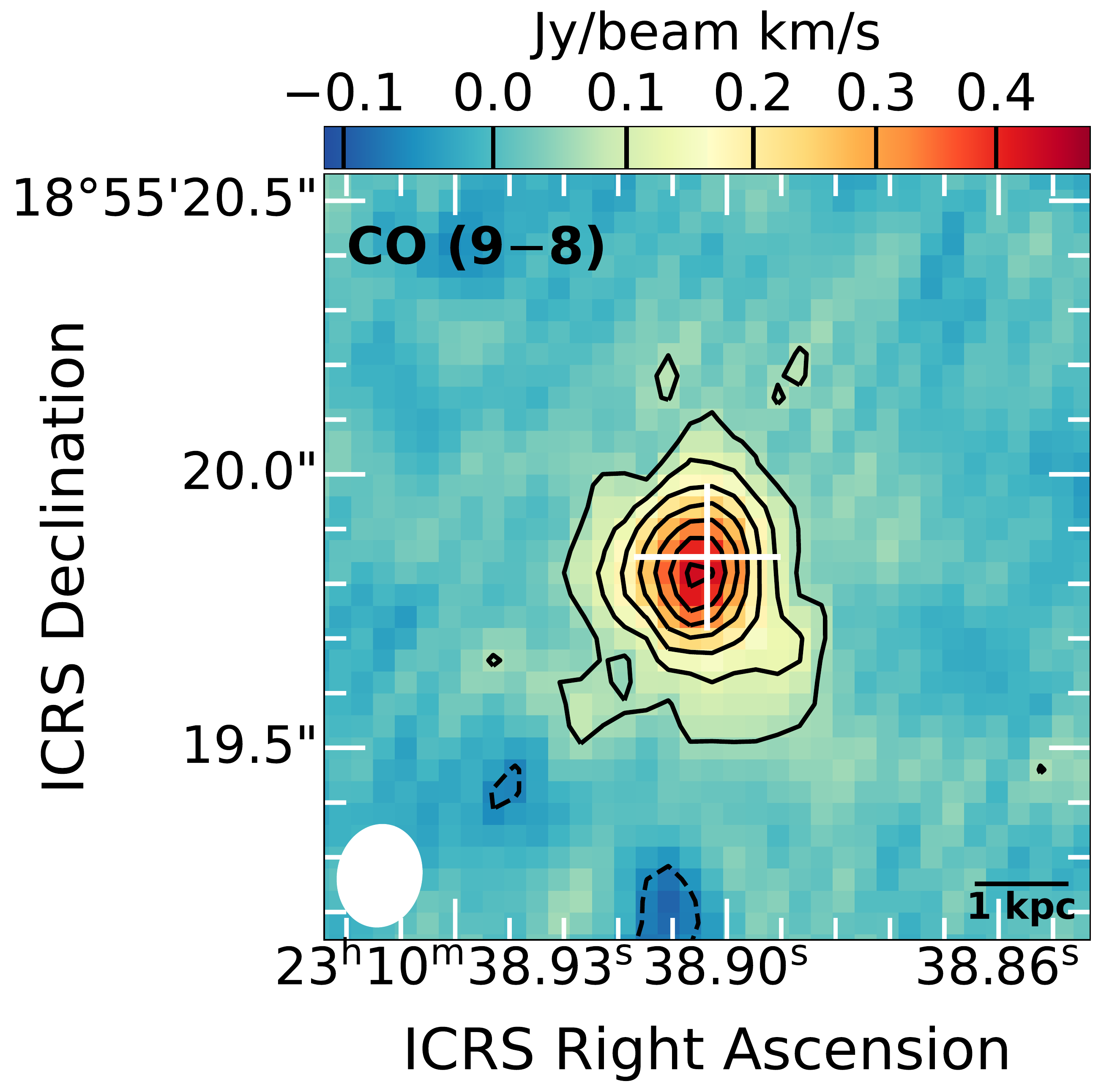}} 
\subfigure{\includegraphics[width=0.31\textwidth]{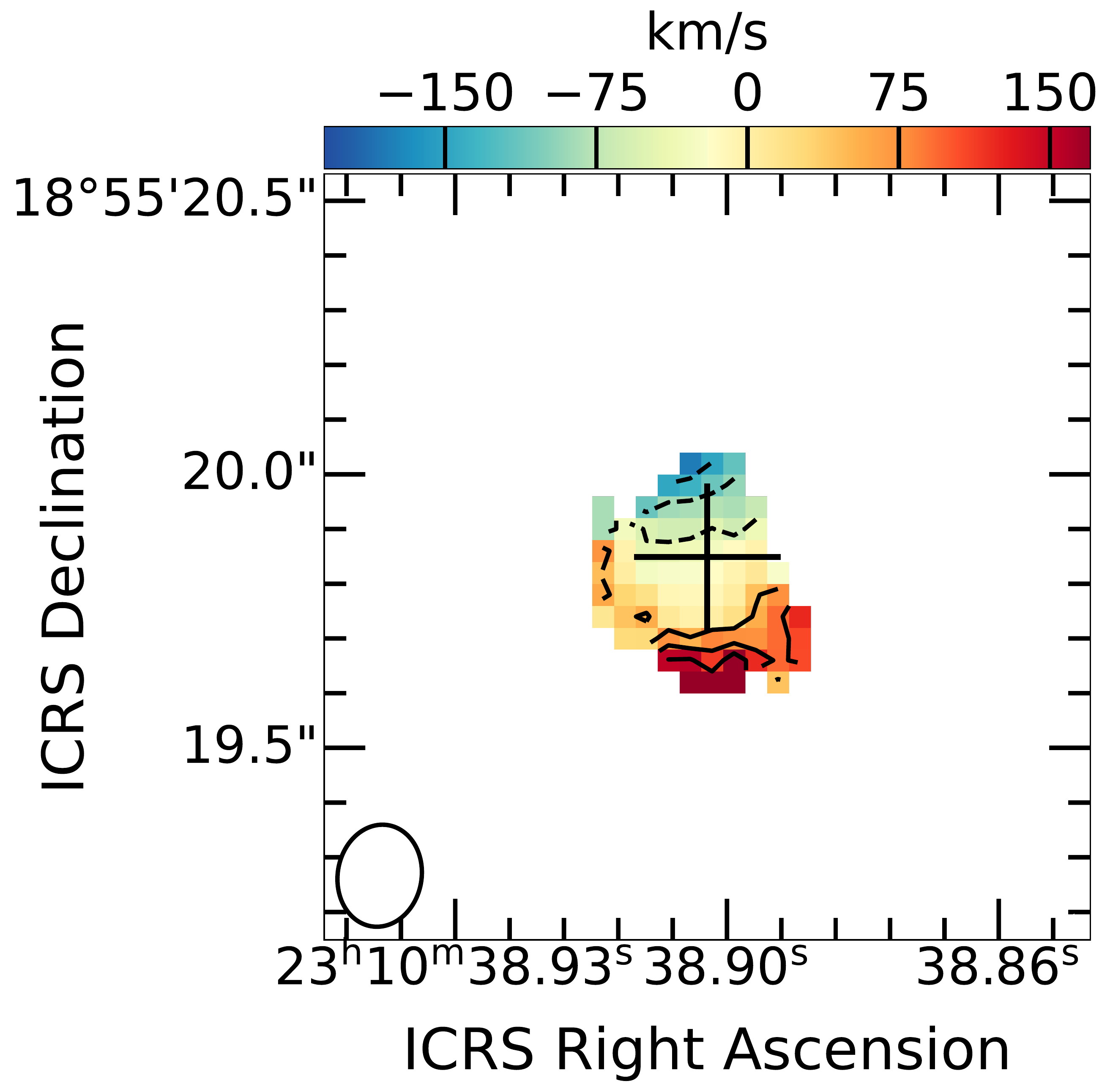}}
\subfigure{\includegraphics[width=0.36\textwidth]{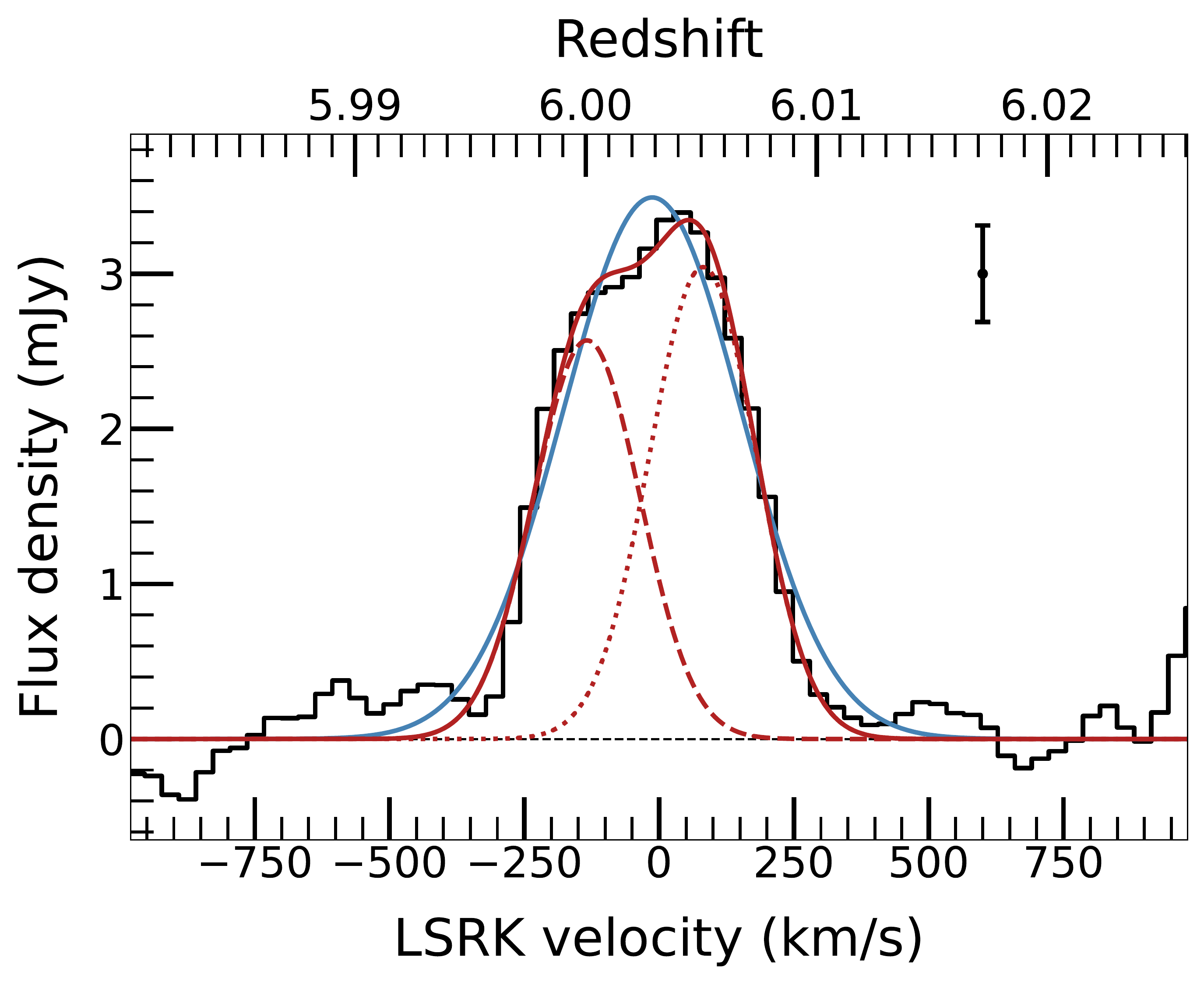}} 
\caption{
Similar to Fig. \ref{ciimap}, but for the CO\,(9--8) line.
The FWHM size of the restoring beam is $0\farcs187$ $\times$ $0\farcs153$. The contour levels are  [$-$3, 3, 6, 9, 12, 15, 18, 21] $\times$ rms$_{\rm co}$, where rms$_{\rm co}=0.020$ Jy/beam km/s. 
The spectral resolution is 15.625 MHz, corresponding to 32 km/s. The rms of the spectrum is 0.31 mJy. The  CO\,(9--8) spectral profile is also asymmetric, with enhancement on the red side.
}
\label{co98map}
\end{figure*}

The OH$^{+}$ velocity-integrated  intensity map and spectrum are shown in Fig. \ref{ohmap}.  The peak flux density is $0.066\pm0.012$ Jy/beam km/s ($>$5$\sigma$) in the intensity map averaged only over the emission component.  There appears a double-horn emission profile and an absorption at $\sim$--400 km/s. We used a triple-Gaussian model (red line in Fig. \ref{ohmap}) to fit the spectrum, and we present the flux and line parameters in Table \ref{obs}. The P-Cygni profile (with an absorption at $-384\pm128$ km/s from the Gaussian fit) of the OH$^{+}$ spectrum may indicate an outflow.

\begin{figure*}
\centering
\subfigure{\includegraphics[width=0.45\textwidth]{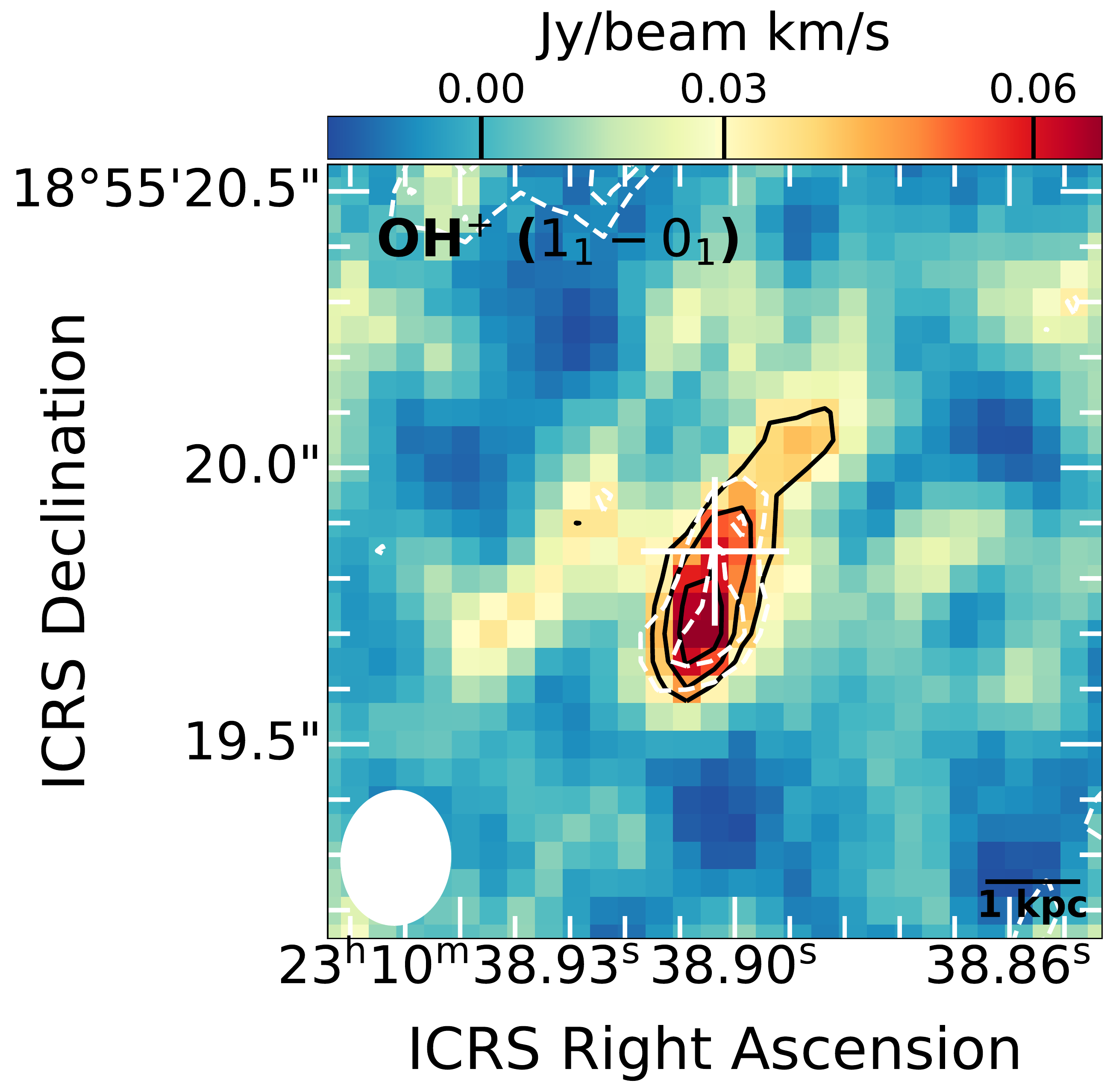}} 
\subfigure{\includegraphics[width=0.54\textwidth]{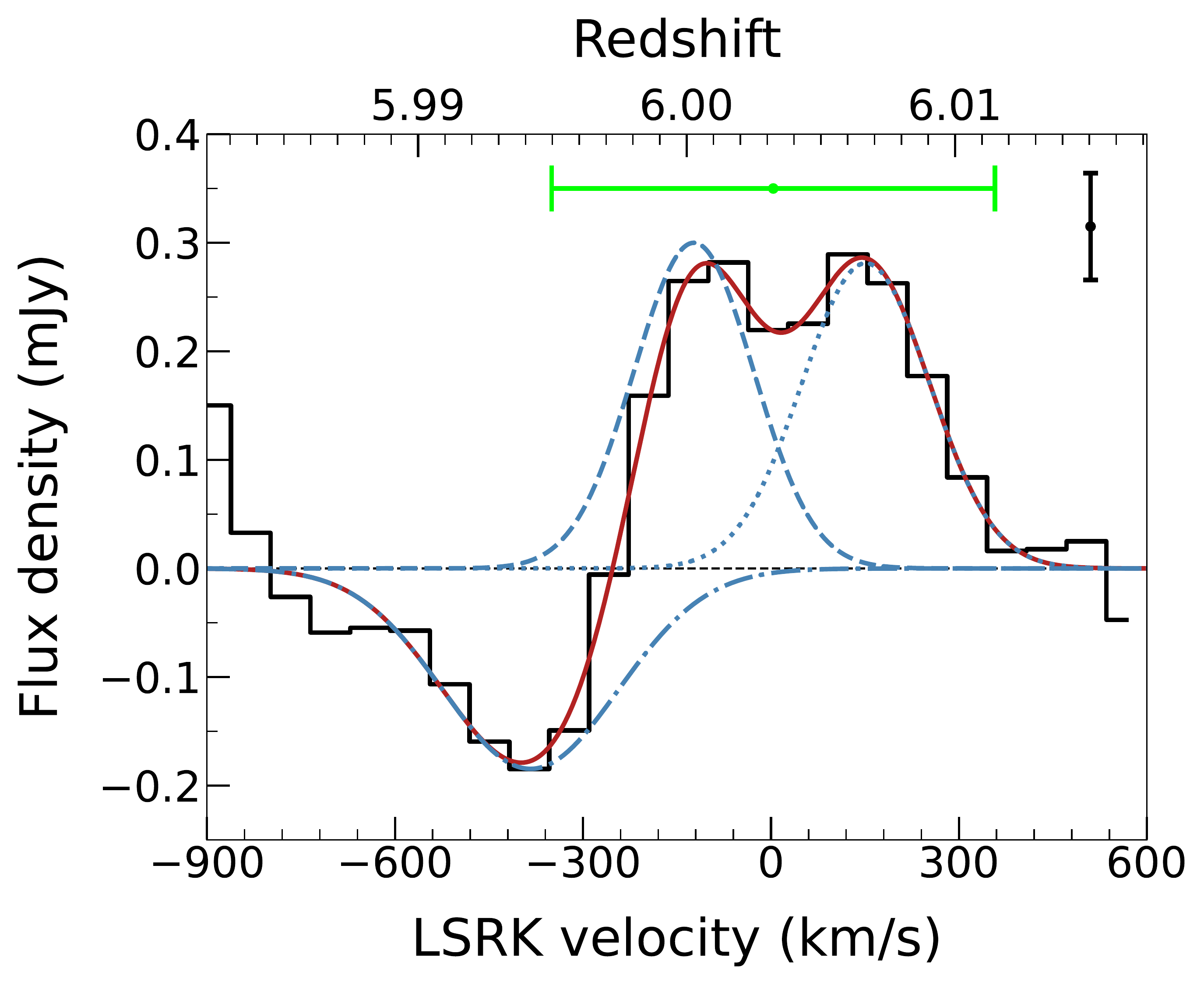}}   
\caption{
ALMA observed OH$^{+}$ line. Left panel: OH$^{+}$ emission line velocity-integrated map. The white plus sign is the HST quasar position. The shape of the restoring beam with a FWHM size of $0\farcs243$ $\times$ $0\farcs198$ is plotted  in the bottom-left corner. The contour levels in black are  [3, 4, 5] $\times$ rms$_{\rm OH^{+}\_em}$, where  rms$_{\rm OH^{+}\_em}=0.012$ Jy/beam km/s. The peak flux density is $0.066\pm0.012$ Jy/beam km/s. The overplotted  dashed white contours are from the OH$^{+}$ absorption line, with contour levels of [$-$3, $-$2] $\times$ rms$_{\rm OH^{+}\_ab}$, where  rms$_{\rm OH^{+}\_ab}=0.008$ Jy/beam km/s.
Right panel: OH$^{+}$   line spectrum (black histogram) overplotted with the best-fit triple-Gaussian profile (red line). The dashed and dotted blue lines present the blueward and redward emission components, respectively, and the dash-dotted blue line represent the absorption component. The spectrum is extracted from the 2$\sigma$ region of the source emitting area in the OH$^{+}$ intensity map.  The LSRK velocity scale is relative to the [\ion{C}{II}] redshift from our ALMA Cycle 0 observations \citep{Wang2013}. The spectral resolution is 31.25 MHz, corresponding to 64 km/s. The rms of the spectrum is 0.05 mJy and is shown as a black bar. The green bar indicates the full width at zero intensity of the [\ion{C}{II}] line, which is roughly consistent with that of the best-fit model for the OH$^{+}$ emission line.}
\label{ohmap}
\end{figure*}

Our new ALMA band-4 observations serendipitously partially cover the para-H$_{2}$O\,(1$_{11}$-0$_{00}$) line at $\nu_{\rm rest}=1113.3430$ GHz (redshifted to 158.9786 GHz), which is detected in emission in J2310+1855. 
The  spectrum is shown in Fig. \ref{watermap}, where we overlaid a spectrum of the  para-H$_{2}$O\,(2$_{02}$-1$_{11}$) line from the ALMA archive.
The para-H$_{2}$O\,(1$_{11}$-0$_{00}$) line is usually detected as an absorption line in most galaxies (e.g., \citealt{Weiss2010}; \citealt{Yang2013}); however it shows up  as a conspicuous but weak emission feature in a few galaxies (e.g., \citealt{Gonzalez-Alfonso2010}; \citealt{Spinoglio2012}; \citealt{Liu2017}). The cold ($T_{\rm dust}\sim$20--30 K) and widespread diffuse ISM component gives rise to the ground-state para-H$_{2}$O line (\citealt{Weiss2010}; \citealt{Liu2017}). \citet{Liu2017} find the ratio of para-H$_{2}$O\,(1$_{11}$-0$_{00}$) and para-H$_{2}$O\,(2$_{02}$-1$_{11}$) to be $<1$ in galaxies in which both lines are detected as emission lines. As the ground-state para-H$_{2}$O line is very close to the edge of the spectral window in our observations, we only detect it partially and may not cover the peak of its spectrum. Thus, we only simply report the detection of this line and will not discuss it further in what follows.

\begin{figure}
\centering
\subfigure{\includegraphics[width=0.48\textwidth]{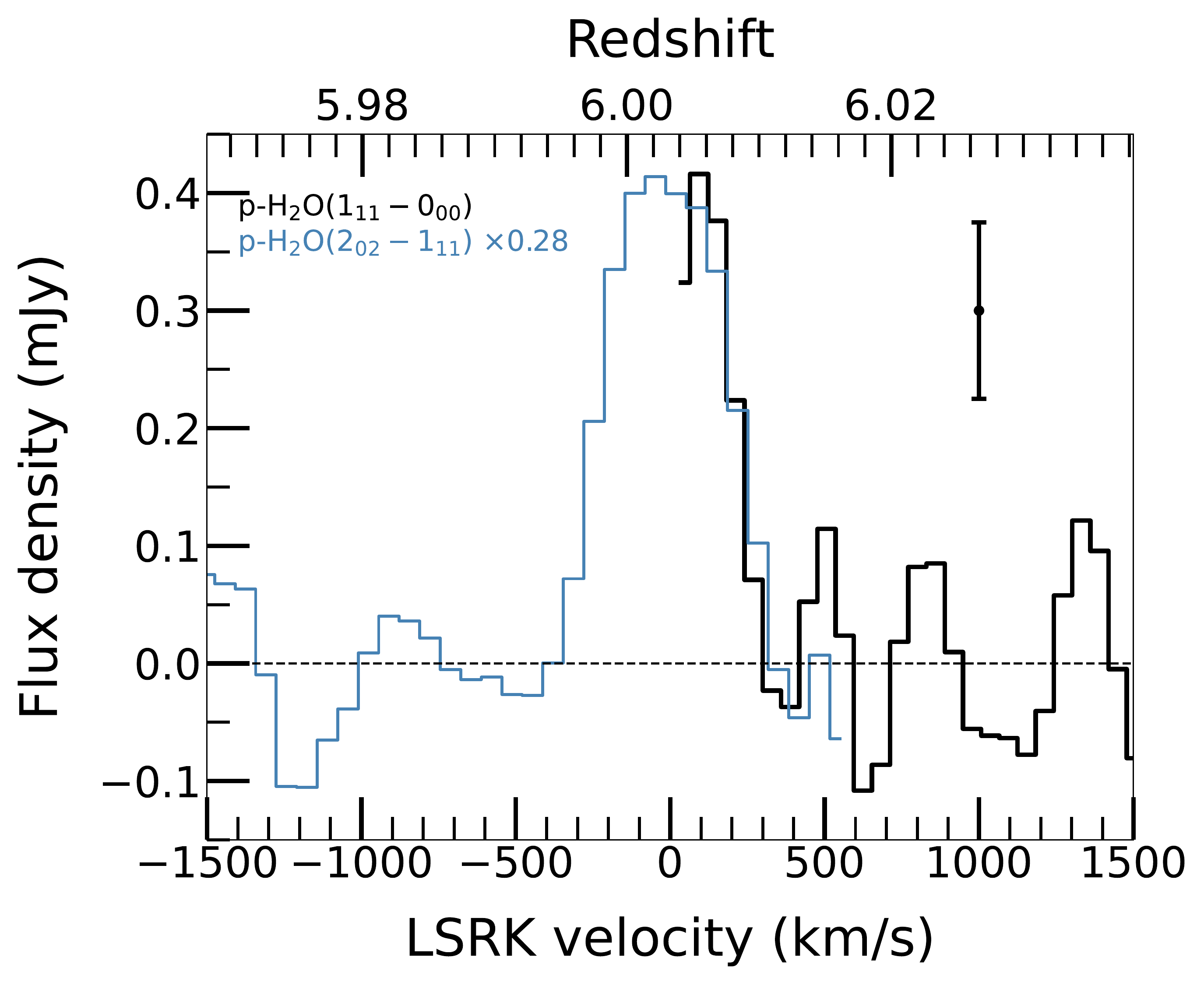}} 
\caption{
Spectra of the para-H$_{2}$O\,(1$_{11}$-0$_{00}$) line (black histogram) and the scaled para-H$_{2}$O\,(2$_{02}$-1$_{11}$) line (blue histogram). The LSRK velocity scale is relative to the [\ion{C}{ii}] redshift from our ALMA Cycle 0 observations \citep{Wang2013}. The spectral resolution for both lines is 15.625 MHz, corresponding to $\sim$32 km/s. The observations may only cover the red tail of the para-H$_{2}$O\,(1$_{11}$-0$_{00}$) emission line.}
\label{watermap}
\end{figure}

\subsection{The dust continuum emission}
\label{sec_resdust}

The dust continuum distribution, overplotted with that of the [\ion{C}{II}] and CO\,(9--8) lines, is shown in the left and middle panels of Fig. \ref{conmap}. We study the brightness distribution and the effective radius (half-light radius) for both lines and their underlying dust continuum emission in Sect. \ref{sec_sb}. The measured intensities are shown in Table \ref{par_ima}. The $z\sim6$ source is spatially resolved  with our data, so we investigate the brightness distributions with 2D elliptical Gaussian functions. However, for the low-resolution data, the observed brightness distribution is dominated by the large beam, so the {\footnotesize{CASA}} {\footnotesize{IMFIT}} tool, which fits a 2D elliptical Gaussian, can be used effectively. The flux densities (listed in Table \ref{par_ima}) of the [\ion{C}{II}] and CO\,(9--8) underlying dust continua measured using the fits in Sect. \ref{sec_sb} are consistent with the ALMA Cycle 0 data ($8.91\pm0.08$ mJy; \citealt{Wang2013}) and ALMA Cycle 3 data ($1.59\pm0.04$ mJy; \citealt{LiJ2020}),  respectively, measured using the  {\footnotesize{CASA}} {\footnotesize{IMFIT}} tool including additional calibration uncertainties. 
As shown in Table \ref{par_ima}, the effective radii of the [\ion{C}{II}] and CO\,(9--8)  lines are larger than that of their underlying dust continuum emission,  as we discuss in Sect. \ref{sec_dis}.

\begin{figure*}
\centering
\subfigure{\includegraphics[width=0.32\textwidth]{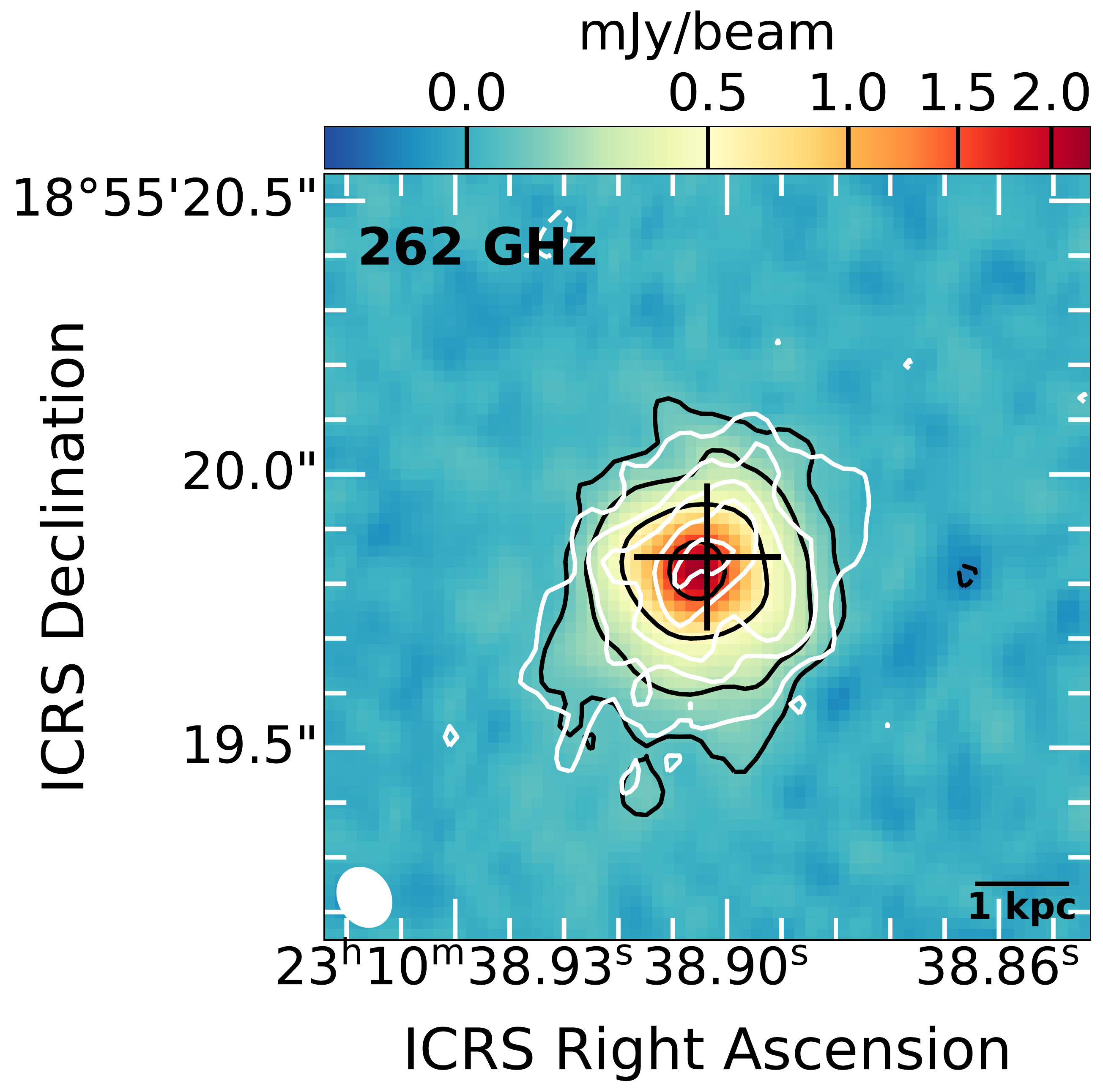}} 
\subfigure{\includegraphics[width=0.32\textwidth]{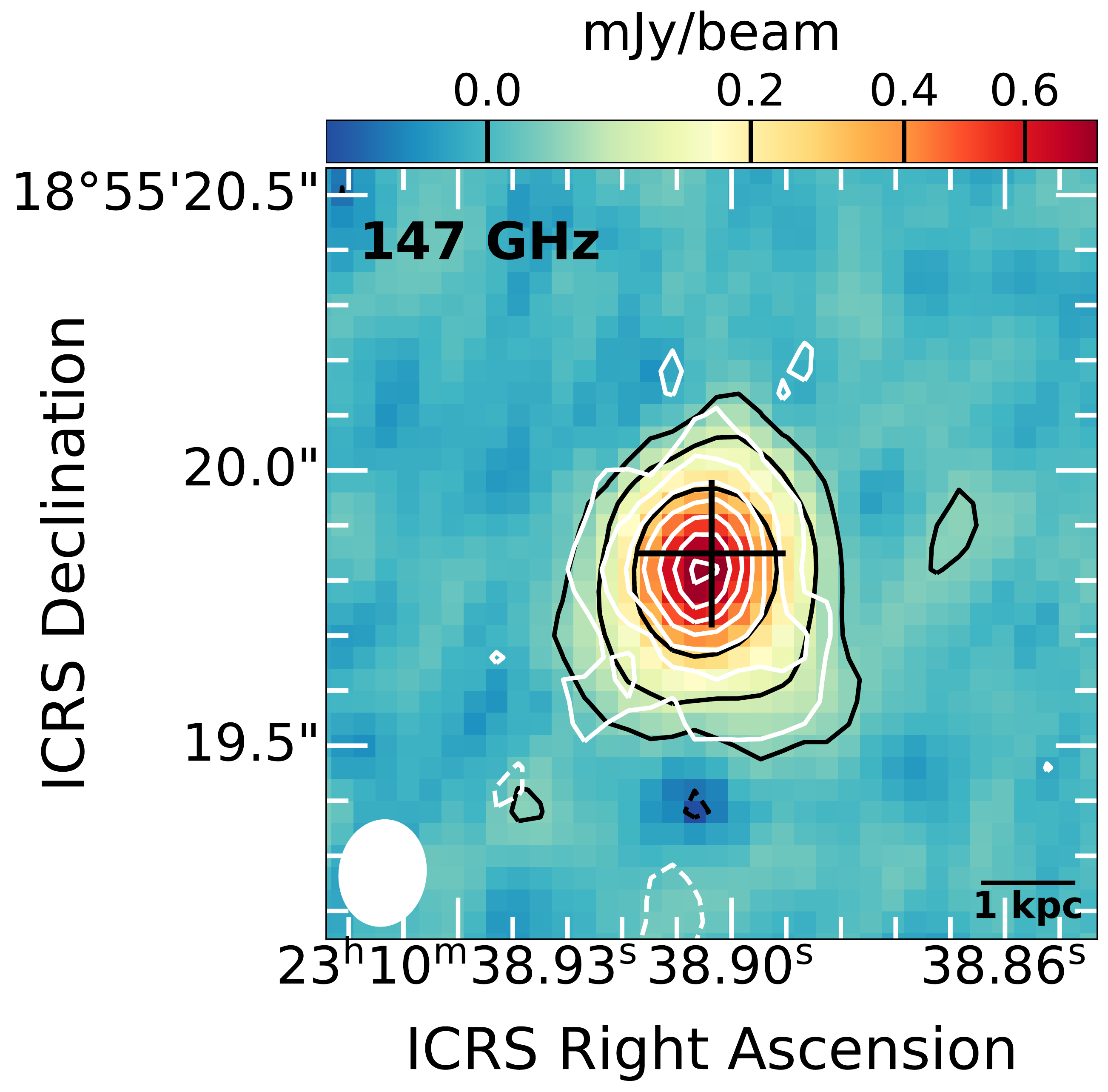}} 
\subfigure{\includegraphics[width=0.34\textwidth]{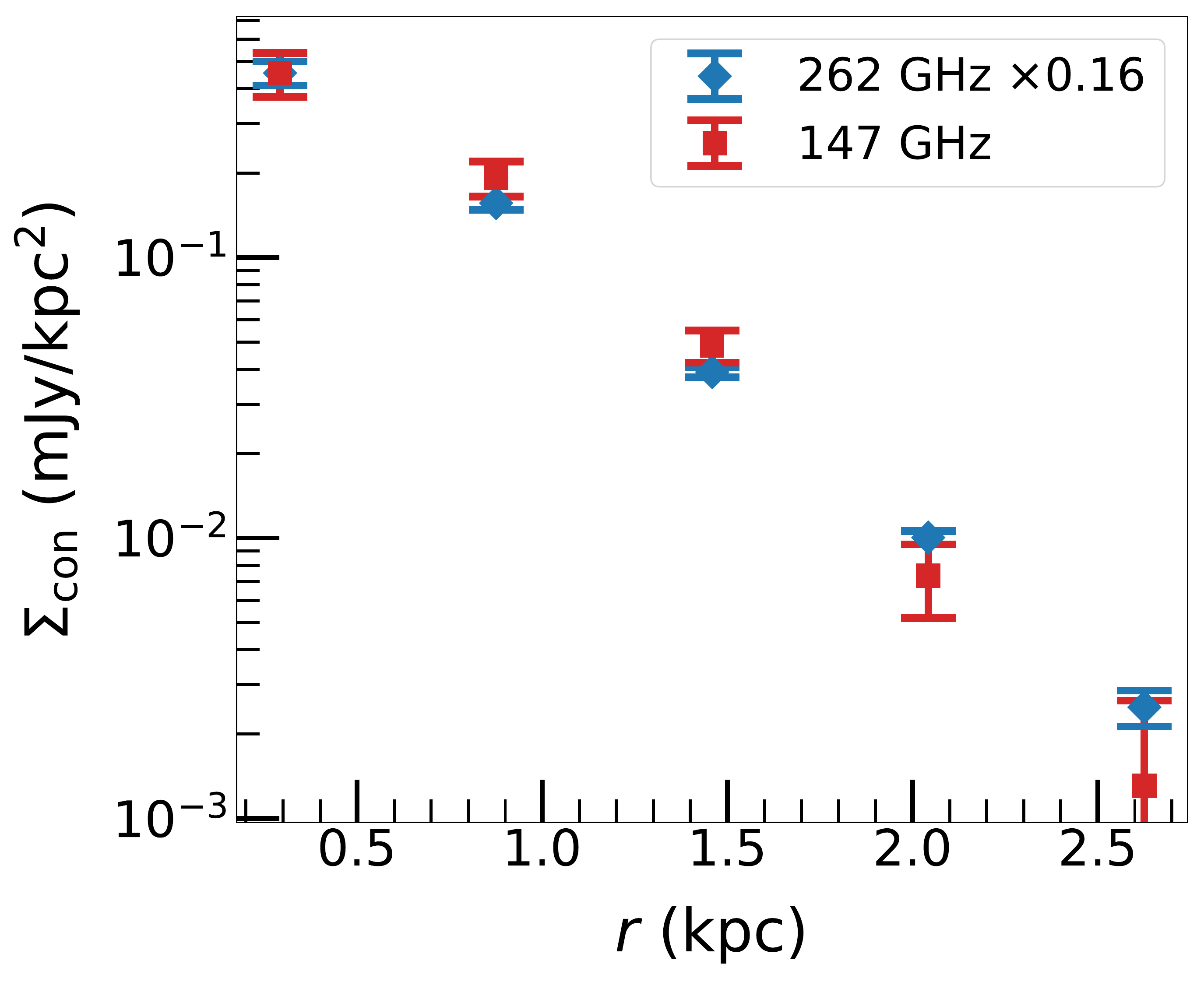}} 
\caption{Dust continuum images at different frequencies and their radial profile comparison. Left and right panels: [\ion{C}{II}] and CO\,(9--8)  underlying dust continuum maps at 262 and 147 GHz, respectively. The black plus sign  is the quasar position from HST snapshot observations. The bottom-left ellipse in each panel shows the restoring beam with FWHM sizes of $0\farcs113$ $\times$ $0\farcs092$ and $0\farcs192$ $\times$ $0\farcs156$, respectively. The black contours are [$-$3, 3, 9, 27, 81] $\times$ rms$_{\rm con}$ for both maps. The rms$_{\rm con}$ at 262 and 147 GHz are 0.02 and 0.01 mJy/beam, respectively. The overplotted white contours are from [\ion{C}{II}] and CO\,(9--8) emission lines, and the contour levels are the same as those of the  black contours in the left panels of Figs. \ref{ciimap} and \ref{co98map}. Right panel: Comparison of the surface brightness of the continuum at two different frequencies. The blue and red symbols with error bars are for the scaled 262 and 147 GHz dust continuum, respectively. The error bar represents the deviation of the values of all pixels in each ring used to do the aperture photometry. } 
\label{conmap}
\end{figure*}

\section{Analysis}
\label{sec_ana}

\subsection{Image decomposition}
\label{sec_sb}

In order to investigate the ISM distribution, we used a 2D elliptical S{\'e}rsic function (Eq. \ref{2dsersicfunc}) to reproduce the observed intensity maps of the [\ion{C}{II}] and CO\,(9--8)  lines and their underlying dust continuum. The 2D elliptical models represent the inclined brightness distribution. We first convolved the model with the restoring beam kernel, and then determined the best fit S{\'e}rsic model by comparing the convolved model and the data pixel-by-pixel using  Markov chain Monte Carlo (MCMC) method with  the {\footnotesize{emcee}}\footnote{\url{http://dfm.io/emcee/current/}} package \citep{emcee2013}. The best-fit models are shown in the left panels of Figs. \ref{sbcii} (for the [\ion{C}{II}] line), \ref{sbco98} (for the CO\,(9--8) line), \ref{sbciidust} (for the dust continuum underlying the [\ion{C}{II}] emission), and \ref{sbcodust} (for the dust continuum underlying the CO\,(9--8) emission). During the fitting, we considered uniform error maps as we only used the central regions of the observed intensity maps. 
Motivated by the possible existence of a torus surrounding the AGN  in the AGN unification model and from observations of several molecular lines and dust continuum (e.g., \citealt{Honig2019}; \citealt{Combes2019}; \citealt{Garcia-Burillo2021}), we also add an additional unresolved nuclear component (which is concentric with the extended component) to represent the dusty and molecular torus for the CO\,(9--8) line and the dust continuum emission. For the [\ion{C}{II}] line, we do not include this component, as the observed [\ion{C}{II}] emission is suppressed by the high dust opacity, as we discuss in Sect. \ref{sec_dis}. We note that in the following analysis and discussion on the spatially resolved gas and dust content in the quasar host galaxy, we remove the emission associated with  the dusty and molecular AGN torus (i.e., subtract the unresolved nuclear component model from the observed image).
To determine which model (S{\'e}rsic or point source+S{\'e}rsic) can better fit the line and/or the dust continuum data, we first measured the pixel-to-pixel rms within a $1\arcsec\times1\arcsec$ square at the center of the residual map (the measured values are shown in the captions of Figs. \ref{sbco98}, \ref{sbciidust} and \ref{sbcodust}), and then calculate the distribution of residual values inside the square (the top-right histogram of Figs.  \ref{sbco98}, \ref{sbciidust}, and \ref{sbcodust}). In addition, we compared the surface brightness profiles of both the modeled and observed intensity maps with projected rings (corrected by the inclination angle and position angle given in Table \ref{par_kin} determined from the kinematic models in Sect. \ref{sec_kin}).

The  parameters  of the best-fit image decomposition for each emission line and its associated continuum are listed in Table \ref{par_ima}. For the [\ion{C}{II}] line (shown in Fig. \ref{sbcii}), the best-fit S{\'e}rsic index is 0.59, which is smaller than a typical exponential disk with a S{\'e}rsic index of 1. It is also smaller than  that of the CO\,(9--8) line and the dust continuum of our quasar J2310+1855. This discrepancy may be due to the high dust opacity, rather than the intrinsic distribution of the [\ion{C}{II}] line,  as we  discuss in Sect. \ref{sec_dis}. 
For the CO\,(9--8)  line, the best-fit S{\'e}rsic index is 2.01 in the case of a single S{\'e}rsic model. When we add a point component in the center, the  S{\'e}rsic index of the extended component decreases to 1.21. The statistics on the residual maps for the one- and two-component scenarios appear identical.  But the two-component model  better matches the observed CO\,(9--8) within the central $\sim$1 kpc region, as shown by the surface brightness difference distribution in Fig. \ref{sbco98}.  
The [\ion{C}{II}] and CO\,(9--8)  underlying dust continua, similar to the CO\,(9--8)  line, seem better described by a combination of a point component and an extended S{\'e}rsic component, as shown in Figs. \ref{sbciidust} and \ref{sbcodust}. And the S{\'e}rsic index for the dust continuum emission is a little bit larger than 1. 
The dust continuum is more concentrated than the gas emission, as measured by the  half-light radius. This is consistent with what is found in observational and theoretical studies of high-redshift galaxies (e.g., \citealt{Strandet2017}; \citealt{Tadaki2018}; \citealt{Cochrane2019}).  We discuss the spatial distribution and extent of the ISM in Sect. \ref{sec_disext}.
The central positions (see Columns 3--4 in Table \ref{par_ima}) of the gas and the dust continuum are identical.

\begin{figure*}
\centering
\subfigure{\includegraphics[width=0.98\textwidth]{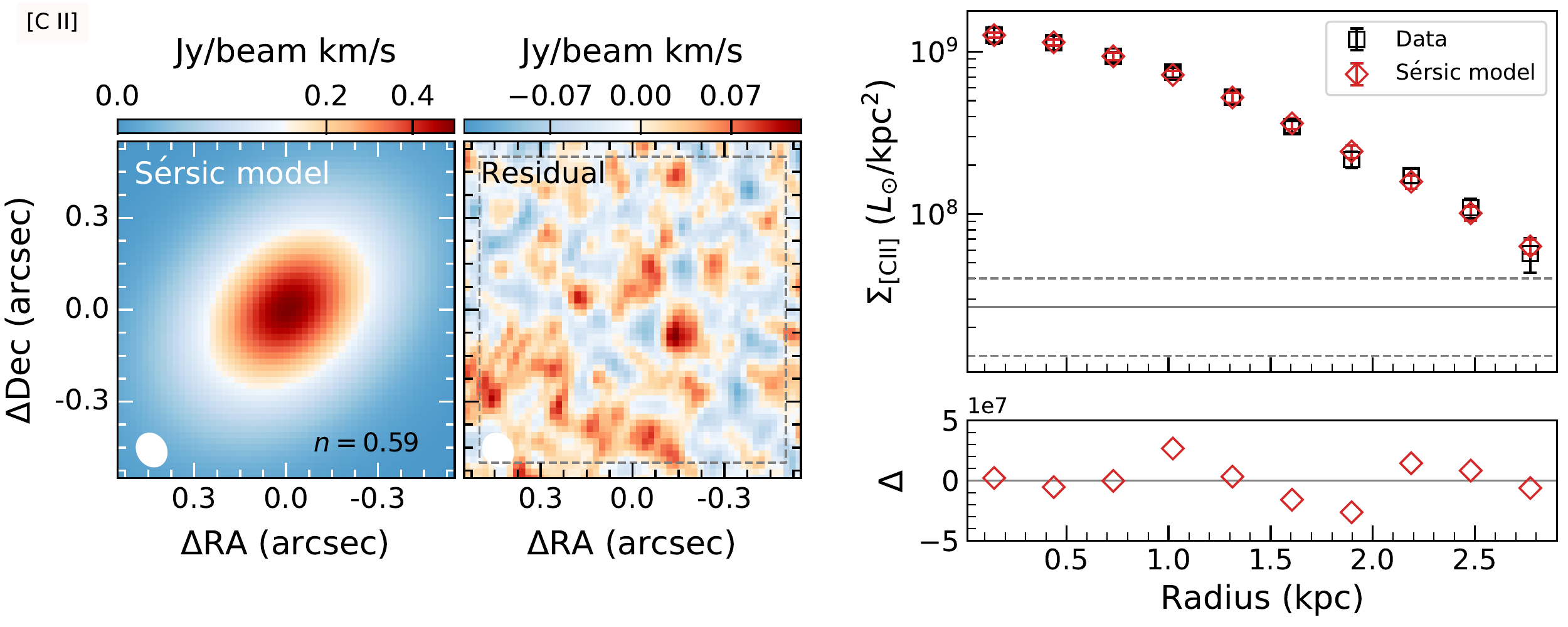}} 
\caption{Image decomposition of the [\ion{C}{II}] line. Left and middle panels: 2D elliptical S{\'e}rsic model and the residual between the observed and modeled [\ion{C}{II}] intensity maps. The shape of the [\ion{C}{II}] synthesized beam  with a FWHM size of $0\farcs111\times0\farcs092$ is plotted in the bottom-left corner of each panel. We measured the residual rms to be 0.037 Jy/beam km/s inside the dashed gray square with a side length of 1$\arcsec$. Top-right panel: [\ion{C}{II}] luminosity surface brightness (black squares and red diamonds with error bars are measurements from the observed intensity map and the 2D elliptical S{\'e}rsic model, respectively) at different radii, measured using elliptical rings with the ring width along the major axis half ($0\farcs05$) that of the [\ion{C}{II}] clean beam size, the rotation angle equal to $\overline{\rm PA}$ (=199$\degr$), and the ratio of semiminor and semimajor axis -- $b/a$ of $\cos(\overline{i})$ ($\overline{i}=42\degr$), where $\overline{\rm PA}$ and $\overline{i}$ come from the [\ion{C}{II}] line kinematic modeling. The error bar represents the standard deviation of the values of all pixels in each ring. The solid and dashed gray  lines are three times the [\ion{C}{II}] luminosity surface brightness limit measured in the emission-free region with an elliptical annulus that is the same with the ones used to measure the [\ion{C}{II}] luminosity surface brightness but with a larger radius, and its corresponding rms. Bottom-right panel: Luminosity surface brightness difference (red diamonds) between the measurements from the observed intensity map and the 2D elliptical S{\'e}rsic model. Note that the vertical scale is linear and in units of $10^{7}L_{\odot}/\rm kpc^{2}$.}
\label{sbcii}
\end{figure*}

\begin{figure*}
\centering
\subfigure{\includegraphics[width=0.98\textwidth]{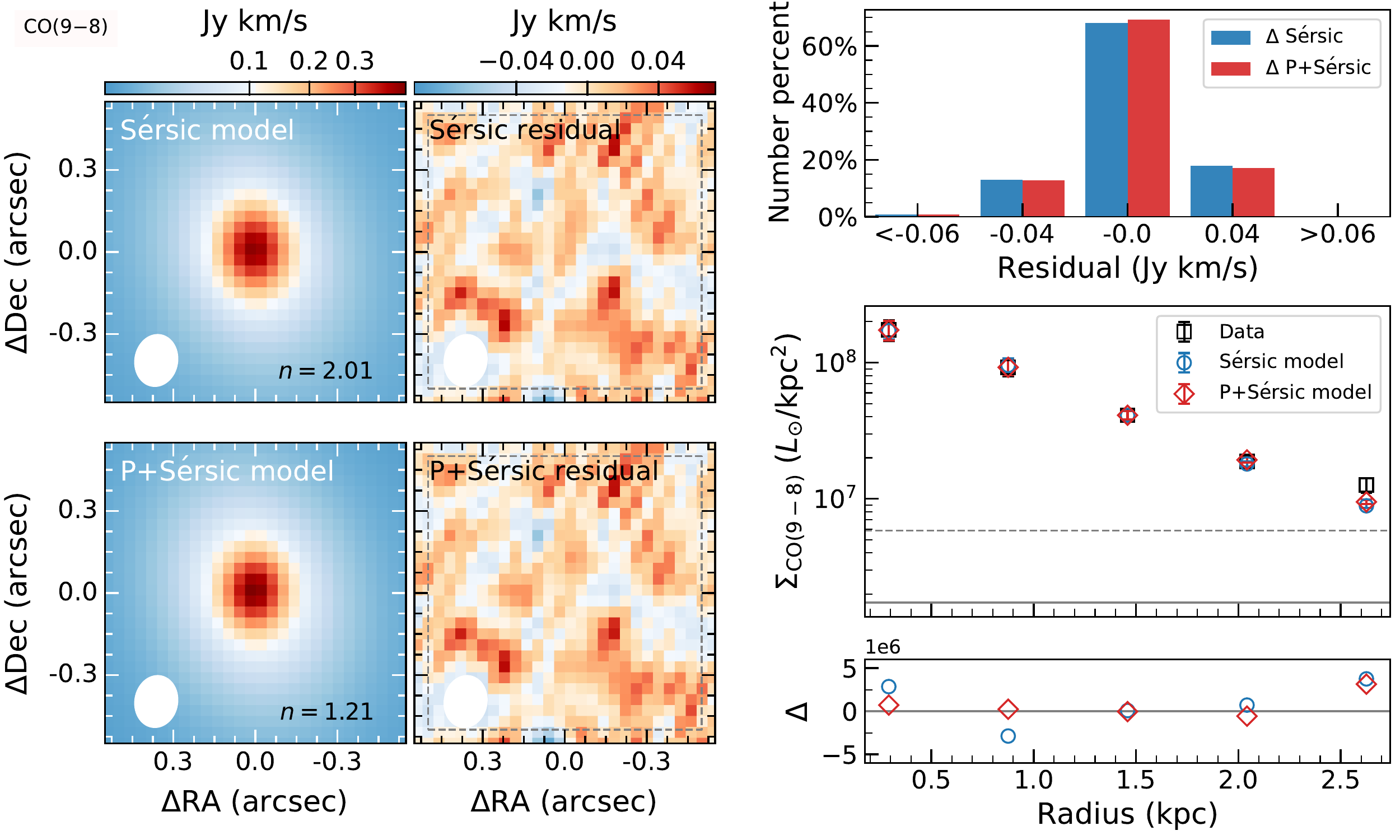}} 
\caption{ Image decomposition of the CO\,(9--8) line. Top-left panels: 2D elliptical S{\'e}rsic model and the residual between the observed and modeled CO\,(9--8) intensity maps, from left to right. We measured the residual rms inside the dashed gray square with a side length of 1$\arcsec$. Bottom-left panels: Same as in the top panels but for the Point+2D elliptical S{\'e}rsic modeling. The peak values and rms within the dashed gray square are 0.061 and 0.021, and 0.059 and 0.020 Jy/beam km/s for the S{\'e}rsic and P+S{\'e}rsic residual maps, respectively. The shape of the CO\,(9--8) synthesized beam  with a FWHM size of $0\farcs187\times0\farcs153$ is plotted in the bottom-left corner of each panel. Top-right panel: Distributions of the residuals for pixels inside the dashed gray squares (S{\'e}rsic modeling: blue histogram; P+S{\'e}rsic modeling: red histogram). Middle-right panel: CO\,(9--8) luminosity surface brightness (black squares, blue circles, and red diamonds with error bars are measurements from the observed intensity map, the 2D elliptical S{\'e}rsic model, and the Point+2D elliptical S{\'e}rsic model, respectively) at different radii, measured using elliptical rings with the ring width along the major axis half ($0\farcs1$) that of the CO\,(9--8) clean beam size, the rotation angle equal to $\overline{\rm PA}$ (=198$\degr$), and the ratio of semiminor and semimajor axis -- $b/a$ of $\cos(\overline{i})$ ($\overline{i}=43\degr$), where $\overline{\rm PA}$ and $\overline{i}$ come from the CO\,(9--8) line kinematic modeling (listed in Table \ref{par_kin}). The error bar represents the deviation of the values of all pixels in each ring. The solid and dashed gray lines are three times the CO\,(9--8) luminosity surface brightness limit measured in the emission-free region with an elliptical annulus that is the same with the ones used to measure the CO\,(9--8) luminosity surface brightness but has a larger radius, and its corresponding rms. Bottom-right panel: Luminosity surface brightness difference between the measurements from the observed intensity map and the ones from the 2D elliptical S{\'e}rsic model (blue circles) and from the Point+2D elliptical S{\'e}rsic model (red diamonds). Note that the vertical scale is linear and in units of $10^{6}L_{\odot}/\rm kpc^{2}$.}
\label{sbco98}
\end{figure*}

\begin{sidewaystable}
\scriptsize
\renewcommand\arraystretch{1.7}
\caption{Parameters derived from best-fit of the image decomposition.}
\label{par_ima}
\centering
\begin{tabular}{lcccccccccccccc}
\hline\hline
Species&Model&center$_{x}$&center$_{y}$&$n_{\rm ISM}$&$r_{{\rm s}\_x}$&$r_{{\rm s}\_y}$&$\theta$&$r_{{\rm e}\_x}$&$r_{{\rm e}\_y}$&$\nu S_{\nu\_\rm point}$&$\nu S_{\nu\_\rm ext}$&$S_{\nu\_\rm point}$&$S_{\nu\_\rm ext}$&$f_{\rm point\_all}$\\
&&(hh:mm:ss)&(dd:mm:ss)&&($\arcsec$)&($\arcsec$)&($\degr$)&(kpc)&(kpc)&(Jy km/s)&(Jy km/s)&(mJy)&(mJy)&($\%$)\\
(1)&(2)&(3)&(4)&(5)&(6)&(7)&(8)&(9)&(10)&(11)&(12)&(13)&(14)&(15)\\
\hline

[\ion{C}{II}]&S&23:10:38.8994\,$\pm$\,0.0001s&18:55:19.7925\,$\pm$\,$0\farcs0011$&$0.59^{+0.01}_{-0.01}$&$0.236^{+0.003}_{-0.003}$&$0.165^{+0.002}_{-0.003}$&$43^{+1}_{-1}$&$1.262^{+0.012}_{-0.011}$&${\bf{0.883^{+0.008}_{-0.007}}}$&--&$6.85^{+0.05}_{-0.06}$&--&--&--\\

CO\,(9--8) &S&23:10:38.9016\,$\pm$\,0.0001s&18:55:19.7781\,$\pm$\,$0\farcs0019$&$2.01^{+0.07}_{-0.06}$&$0.013^{+0.002}_{-0.002}$&$0.010^{+0.002}_{-0.001}$&$109^{+3}_{-4}$&${\bf{1.088^{+0.040}_{-0.035}}}$&$0.82^{+0.03}_{-0.03}$&--&$1.51^{+0.03}_{-0.03}$&--&--&--\\

CO\,(9--8) &P+S&23:10:38.9016\,$\pm$\,0.0001s&18:55:19.7787\,$\pm$\,$0\farcs0019$&$1.21^{+0.09}_{-0.07}$&$0.087^{+0.012}_{-0.015}$&$0.066^{+0.010}_{-0.011}$&$107^{+4}_{-4}$&${\bf{1.259^{+0.054}_{-0.045}}}$&$0.949^{+0.040}_{-0.040}$&$0.16^{+0.02}_{-0.02}$&$1.31^{+0.04}_{-0.04}$&--&--&$11^{+1}_{-1}$\\

$[\ion{C}{II}]_{\rm con}$&S&23:10:38.9003\,$\pm$\,0.0001s&18:55:19.8033\,$\pm$\,$0\farcs0017$&$1.57^{+0.02}_{-0.02}$&$0.022^{+0.001}_{-0.001}$&$0.020^{+0.001}_{-0.001}$&$33^{+1}_{-1}$&$0.652^{+0.003}_{-0.003}$&${\bf{0.587^{+0.003}_{-0.003}}}$&--&--&--&$9.06^{+0.03}_{-0.03}$&--\\

$[\ion{C}{II}]_{\rm con}$&P+S&23:10:38.9003\,$\pm$\,0.0001s&18:55:19.8032\,$\pm$\,$0\farcs0016$&$1.16^{+0.02}_{-0.02}$&$0.053^{+0.002}_{-0.002}$&$0.047^{+0.002}_{-0.002}$&$33^{+1}_{-1}$&$0.687^{+0.004}_{-0.004}$&${\bf{0.621^{+0.004}_{-0.004}}}$&--&--&$0.51^{+0.02}_{-0.02}$&$8.31^{+0.04}_{-0.04}$&$6^{+1}_{-1}$\\

CO\,(9--8) $_{\rm con}$&S&23:10:38.9013\,$\pm$\,0.0001s&18:55:19.7740\,$\pm$\,$0\farcs0005$&$1.18^{+0.03}_{-0.03}$&$0.039^{+0.003}_{-0.002}$&$0.034^{+0.002}_{-0.002}$&$79^{+3}_{-3}$&${\bf{0.522^{+0.006}_{-0.006}}}$&$0.452^{+0.005}_{-0.005}$&--&--&--&$1.52^{+0.01}_{-0.01}$&--\\

CO\,(9--8) $_{\rm con}$&P+S&23:10:38.9013\,$\pm$\,0.0001s&18:55:19.7741\,$\pm$\,$0\farcs0005$&$1.17^{+0.05}_{-0.04}$&$0.042^{+0.003}_{-0.004}$&$0.037^{+0.003}_{-0.003}$&$80^{+2}_{-3}$&${\bf{0.558^{+0.012}_{-0.012}}}$&$0.484^{+0.011}_{-0.010}$&--&--&$0.08^{+0.02}_{-0.02}$&$1.45^{+0.02}_{-0.02}$&$5^{+2}_{-1}$\\

\hline
\end{tabular}
\tablefoot{Column 1: the name of the ISM tracer. Column 2: the best-fit model. ``S" and ``P" present the S{\'e}rsic and point component, respectively. Columns 3--4: source position -- RA and Dec, respectively. Column 5: the S{\'e}rsic index of the ISM (the gas and the dust emission). Columns 6--7: the scale lengths along the major and minor axes, respectively. Column 8: the rotation angle defined in Appendix \ref{sec_fun}. Columns 9--10: the half-light  radii. The values in boldface are measurements near the kinematic major axes. Columns 11--12:  the line flux for the point component and extended component, respectively. Columns 13--14:  the dust continuum flux density for the point component and extended component, respectively. Column 15: the fraction of the nuclear component in the whole emission. Note that the errors of these parameters are all fitting-type errors. The complex correlated noise is not included. As claimed by the task reference of the CASA IMFIT,  the correlated noise can bring 4$\%$ effect when 5$\sigma$ detection of the emission for a 2D elliptical Gaussian fitting.}
\end{sidewaystable}

\subsection{Dust diagnostic}
\label{sec_dd}

With the dust emission size from the image decomposition, and following \citet{Weiss2007} and \citet{Walter2022}, we are able to constrain the dust temperature and the dust mass with a general gray-body formula instead of  using the optically thin approximation as done in our previous work \citep{Shao2019}:
\begin{equation}
S_{\nu} = \Omega_{\rm app}[B_{\nu}(T_{\rm dust})-B_{\nu}(T_{\rm CMB})][1-\exp(-\tau_{\nu})](1+z)^{-3},
\label{graybody1}
\end{equation}

and 

\begin{equation}
\tau_{\nu} = \kappa_{0}(\nu/\nu_{\rm ref})^{\beta}M_{\rm dust,\,app}/(D_{\rm A}^{2}\Omega_{\rm app}),
\label{graybody2}
\end{equation}
where $S_{\nu}$ is the observed flux density for a target at redshift $z$. $B_{\nu}(T_{\rm dust})$ and $B_{\nu}(T_{\rm CMB})$ are the black-body functions with dust temperature $T_{\rm dust}$ and the cosmic microwave background (CMB) temperature $T_{\rm CMB}$, respectively. $\Omega_{\rm app}$ is the apparent solid angle. 
In principle, $\Omega_{\rm app}$ can be different for different wavelengths. Due to the dust continuum sizes remaining constant within $\sim$20$\%$ at observed frame wavelengths from 500 $\mu$m to 2 mm for $z\sim1-5$ main-sequence galaxies from simulations \citep{Popping2022}, and our findings for the similar dust sizes (within $\sim$10$\%$) at wavelengths of 158 and 290 $\mu$m in the rest frame, we do not consider changes in this quantity below.
$\tau_{\nu}$ is the optical depth, which is a function of frequency $\nu$ (rest-frame)  and dust mass surface density $M_{\rm dust,\,app}/(D_{\rm A}^{2}\Omega_{\rm app})$. $M_{\rm dust,\,app}$ is the apparent dust mass. $D_{\rm A}$ is the angular distance. $\beta$ is the emissivity index. We here adopted the absorption coefficient per unit dust mass $\kappa_{0}$ of 13.9 cm$^{2}/$g at the reference frequency $\nu_{\rm ref}$ of 2141 GHz (\citealt{Draine2003}; \citealt{Walter2022}). 

Motivated by the AGN contribution to the strength of the dust continuum emission of the quasar-host system (e.g., \citealt{DiMascia2021}), we decomposed the UV-to-FIR SED of J2310+1855 shown in Fig. \ref{sedfitting}. With the best-fitting UV/optical power law (red lines) and NIR to mid-infrared (MIR) CAT3D AGN torus model (brown lines; \citealt{Honig2017}) in \citet{Shao2019}, and the source size of twice the effective radius of the dust continuum underlying  CO\,(9--8) listed in Table \ref{par_ima}, we  derived an average dust temperature $T_{\rm dust}$ of $53^{+4}_{-4}$ K, a total dust mass $M_{\rm dust,\,app}$ of $1.27^{+0.62}_{-0.35}\times10^{9}M_{\odot}$ and an overall emissivity index $\beta$ of $1.90^{+0.33}_{-0.27}$. 
Our dust temperature is smaller than the value of $71\pm4$ K derived by \citet{Tripodi2022}, and our dust mass is $\sim$3 times higher. This is due to the differences in the AGN dust torus model used, which has a significant contribution in the rest frame wavelength range $\lesssim$100 $\mu$m in our case, whereas the impact of the AGN dust torus on the total dust continuum emission  in \citet{Tripodi2022} is limited $\lesssim$50 $\mu$m. As a result, the peak of our gray-body model is at a longer wavelength, which leads to a lower dust temperature and a higher dust mass according to Eqs. \ref{graybody1} and \ref{graybody2}, as well as a smaller gas-to-dust ratio (GDR) compared with the dust property results in \citet{Tripodi2022}.
The FIR and IR luminosities are $8.84^{+2.68}_{-2.19}\times 10^{12} L_{\odot}$ and $1.19^{+0.40}_{-0.29}\times 10^{13}L_{\odot}$, respectively, by integrating the gray-body model (which can be taken as purely star-forming heated dust emission) from 42.5--122.5 $\mu$m and from 8--1000 $\mu$m. The average SFR is $2055^{+691}_{-499}\, M_{\odot}$/yr calculated from the above-mentioned IR luminosity and the formula in \citet{Kennicutt1998}. 

\begin{figure*}
\centering
\subfigure{\includegraphics[width=0.59\textwidth]{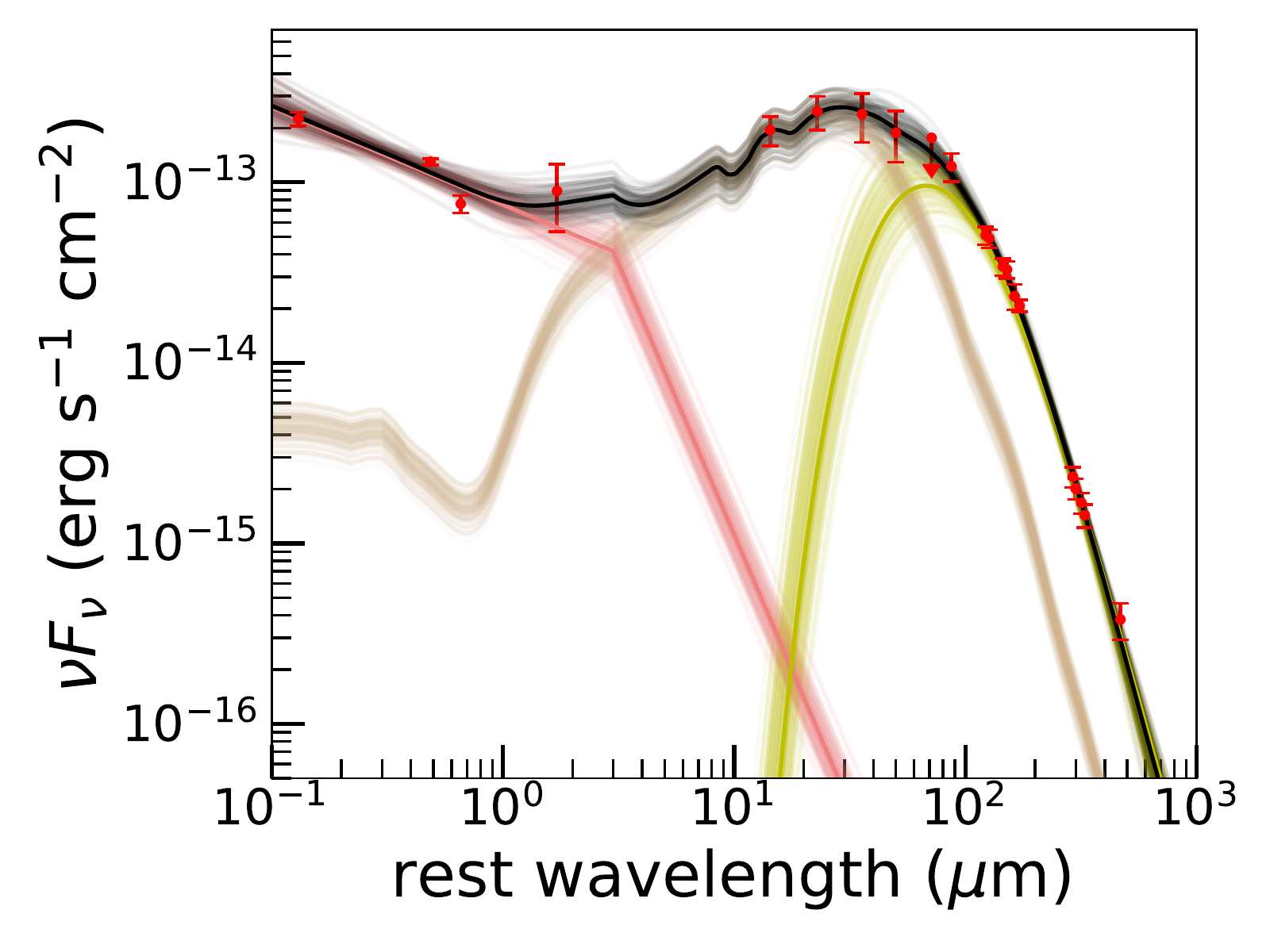}} 
\subfigure{\includegraphics[width=0.4\textwidth]{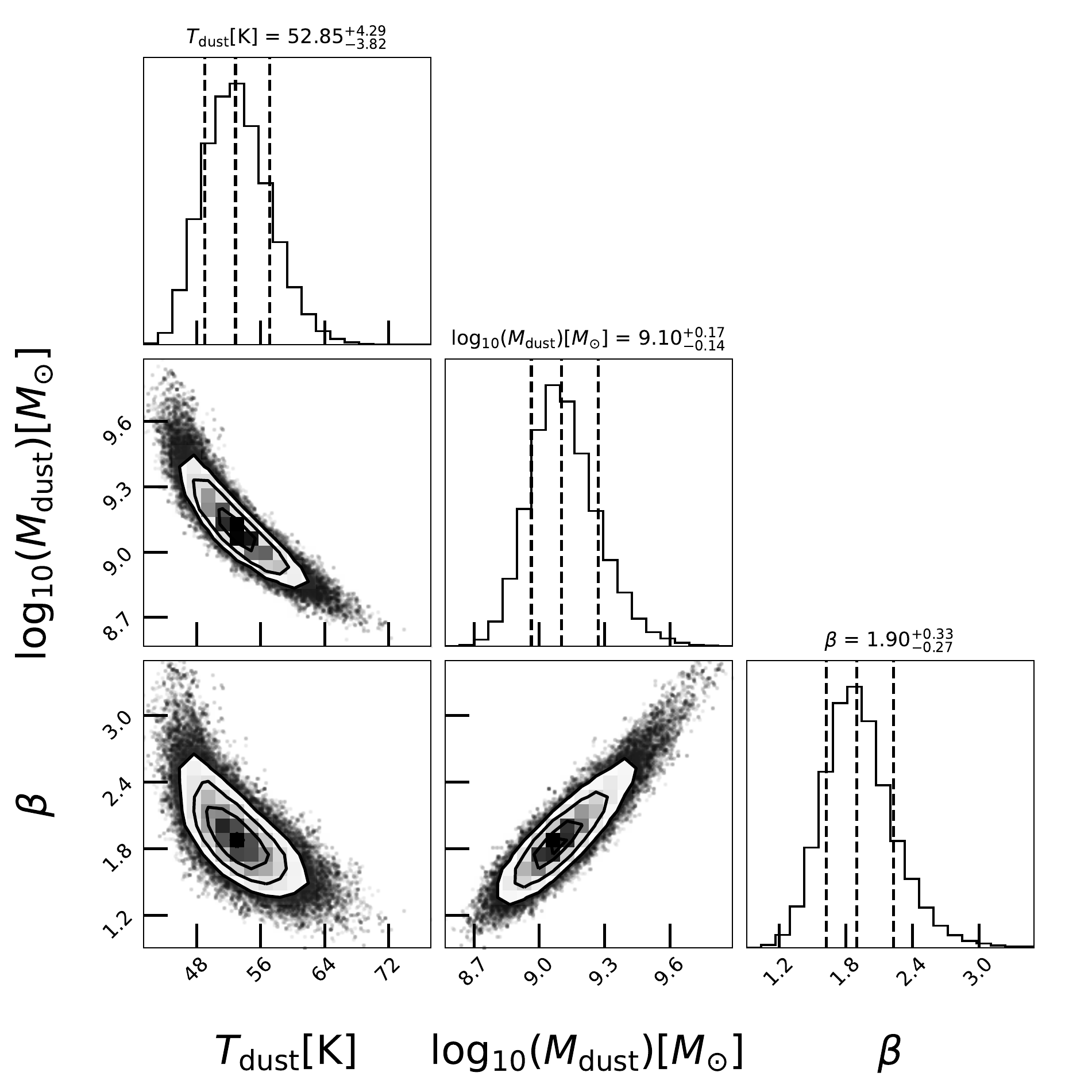}}  
\caption{SED decomposition toward J2310+1855. Left panel: Rest-frame UV-to-FIR SED fitting. The red points with error bars or downward arrows are observed data (see Table 3 in \citealt{Shao2019} for details). The pink lines represent the UV/optical power law from the accretion disk. The brown lines correspond to the CAT3D AGN torus model \citep{Honig2017}. The physical properties derived for these two components are detailed in Table 4 in \citet{Shao2019}.  The green lines correspond to a gray-body profile (Eqs. \ref{graybody1} and \ref{graybody2}) associated with star formation activity in the quasar host galaxy. The black lines are the sum of all components.  The fit employed the {\footnotesize{MCMC}} method with  the {\footnotesize{emcee}} package \citep{emcee2013}. We visualized the model uncertainties with shaded areas by randomly selecting 100 models from the parameter space.
Right panel: Corner map for the parameters (dust temperature, dust mass, and dust emissivity index) associated with the gray-body model for the FIR region SED fitting (green lines in the left panel). The contours are drawn at $1-\exp(-m^{2}/2)$ ($m=0.5, 1, 1.5, 2$) of the volume. The vertical dashed lines show 16th, 50th (median), and 84th percentiles.}
\label{sedfitting}
\end{figure*}

With the resolved dust continuum at two different frequencies, we are able to investigate the radial distributions of the dust temperature, dust mass and SFR. We first used circular annuli to do aperture photometry on both the [\ion{C}{ii}] and CO\,(9--8) underlying dust continuum maps (after removing the central point components that may come from the AGN dust torus) with ring width of $0\farcs1$. The right panel of Fig. \ref{conmap} shows the comparison of the surface brightness of the continuum at observed frame 262 and 147 GHz. The slight differences in the dust continuum profiles at different wavelengths may indicate different distribution of the dust temperature along with distance.  For each ring we fitted the measured flux densities with Eqs. \ref{graybody1} and \ref{graybody2} using the $\beta$ value of 1.90 from our SED fitting, to get the average dust temperature, dust mass and dust optical depth at each radius. Finally, with fixed gray-body formula for each ring, we calculated the IR luminosity and SFR. The results are shown in Fig. \ref{dusteachring}.
The dust temperature drops with increasing radius as seen in the top-left panel. A similar dependence is found in a  star-forming galaxy at $z=7.13$ \citep{Akins2022}. We fit a power law, $T_{\rm dust}\propto r^{-0.41\pm0.20}$.
The radial dependence of $\Sigma_{\rm dust\_mass}$ and $\Sigma_{\rm SFR}$ are consistent with an exponential distribution $\Sigma(r)=\Sigma_{0}\exp^{-r/r_{\rm s}}$, where $r_{\rm s}$ is the exponential scale length and $\Sigma_{0}$ is the central surface density. 
The best-fit relation for  $\Sigma_{\rm dust\_mass}$ is shown in the top-right panel of Fig. \ref{dusteachring} with $r_{\rm s,\,dust\_mass}=0.77\pm0.27$ kpc, corresponding to an effective radius of $1.29\pm0.45$ kpc. It is about two times larger than the dust continuum effective radius (i.e., $\sim$0.6 kpc) shown in Table \ref{par_ima}. This may indicate that the single-band dust continuum emission is not a good tracer of the dust mass, which is consistent with the finding from numerical simulations \citep{Popping2022}, and that the dust emission depends more strongly on dust temperature than on dust mass. Thus, one requires at least two bands of high-resolution imaging to map the dust temperature across the galaxy disks when using the dust continuum emission of galaxies as a reliable tracer of the dust mass distribution. The total dust mass summing over all  rings is $(1.84\pm0.69)\times10^{9} M_{\odot}$, which is consistent with the value measured from the SED fitting. 
An exponential fit for  $\Sigma_{\rm SFR}$ is shown in the bottom-right panel of Fig. \ref{dusteachring} with $r_{\rm s,\,SFR}=0.38\pm0.09$ kpc, corresponding to an effective radius of $0.64\pm0.15$ kpc. It is consistent with the dust effective radius, which means that the resolved dust continuum emission is very closely linked to the SFR distribution.  These are consistent with the conclusions from TNG50 star-forming galaxy simulations that  the single-band dust emission is a less robust tracer of the dust distribution, but is a decent tracer of the  obscured star formation activity in galaxies \citep{Popping2022}. The total SFR summing over all  rings is $3415^{+5746}_{-1777}\ M_{\odot}$/yr, which is consistent with the value measured from the SED fitting. We suggest adopting the total SFR derived from the SED fitting, as the SFR value for each ring is  measured from only two data points at different wavelengths and thus has a lot of uncertainty.
As shown in the bottom-left panel of Fig. \ref{dusteachring}, as the dust mass surface density decreases with the galactocentric radius, the dust emission in both bands becomes less optically thick. 
The rest-frame wavelength of both dust bands ($\sim$290 $\mu$m for the CO\,(9--8) underlying dust continuum, and $\sim$158 $\mu$m for the [\ion{C}{ii}] underlying dust continuum) is shortward of the Rayleigh-Jeans tail (i.e., a rest-frame wavelength of $\sim$350 $\mu$m). 
The dust mass surface density is very high (i.e.,  $\Sigma_{\rm dust\_mass}>10^{8}M_{\odot}$/kpc$^{2}$),  and the optical depth for the dust continuum emission at both frequencies is above 0.1 inside $\sim$1 kpc.

\begin{figure*}
\centering
\subfigure{\includegraphics[width=\textwidth]{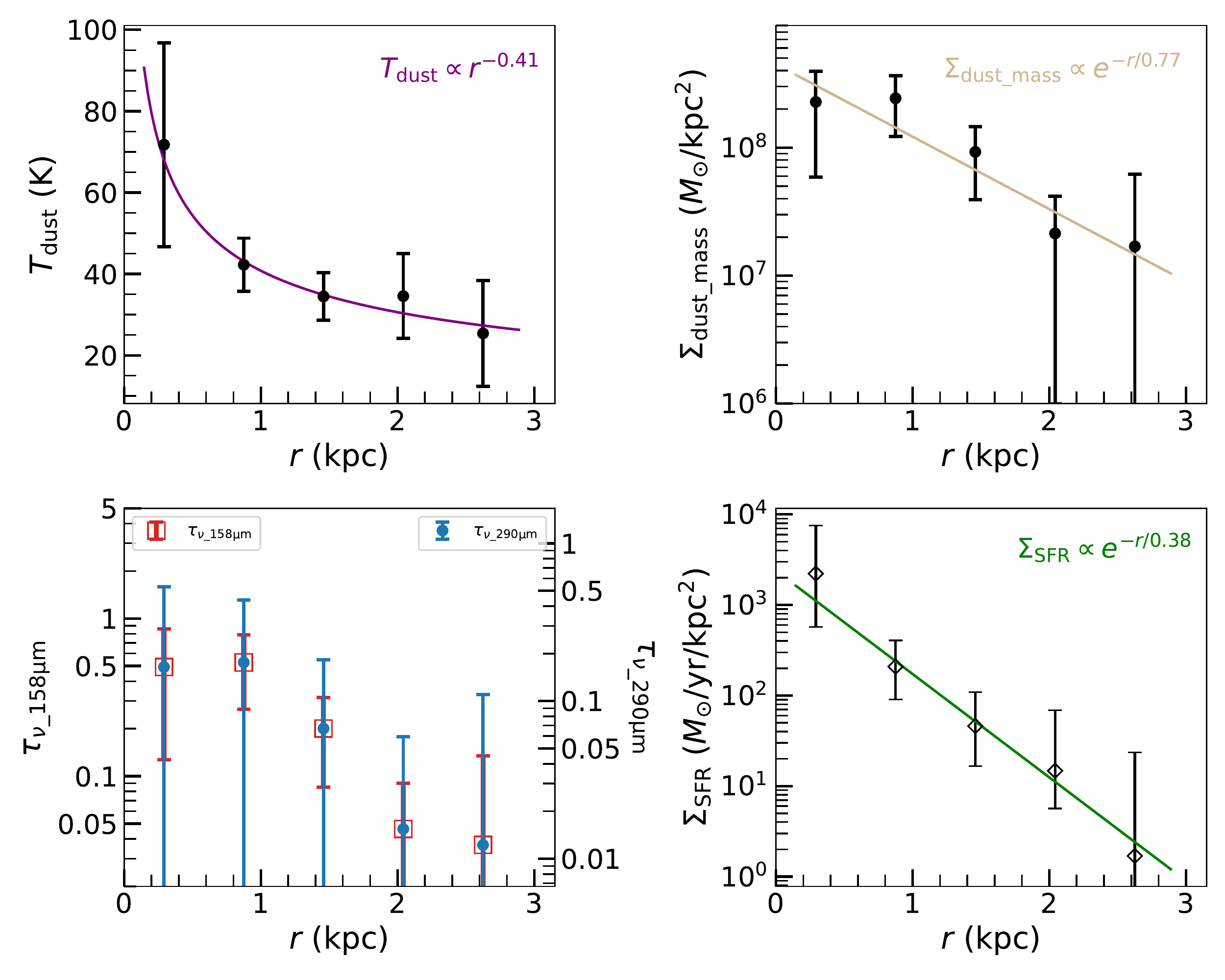}} 
\caption{Radial distributions of the dust temperature (top left), dust mass surface density (top right), optical depth (bottom left) at [\ion{C}{ii}] and CO\,(9--8) underlying dust continuum frequencies (red dots and blue squares with error bars, respectively), and the SFR surface density (bottom right).  The purple, gold, and green lines are best-fit lines. Note that the uncertainties of the optical depth at different wavelengths are same in the bottom-left panel but appear different due to different ranges of the left and right axes. The error bars of the SFR surface density (which is associated with the dust temperature and dust mass) in the bottom-right panel are measured with a Monte Carlo method: we calculate samples by changing the parameter values with random Gaussian draws centered on their best-fit values and deviated by their errors shown in the top-left and top-right panels, and the lower and upper values are taken as the 16th and 84th percentiles.
}
\label{dusteachring}
\end{figure*}

\subsection{Gas kinematic modeling}
\label{sec_kin}

The [\ion{C}{II}] and CO\,(9--8)  lines can be used to trace the kinematic properties of atomic, ionized, and molecular gas in the quasar host galaxy. As shown in Fig. \ref{ciichannel}, the [\ion{C}{II}] emission peak moves in a circular path with frequency. As shown in Figs. \ref{ciidatamodel} and \ref{co98datamodel}, the velocity fields traced by both [\ion{C}{II}] and CO\,(9--8)  show clear velocity gradients. In addition, the position-velocity diagrams along the kinematic major axes have an ``S" shape (especially apparent for the [\ion{C}{II}] line; however, the resolution of the CO data is roughly two times lower than that of the [\ion{C}{II}] data, so the ``S" shape is not obvious). These are consistent with a rotating gas disk. We are able to construct a 3D disk model for these data cubes with the package called 3D-Based Analysis of Rotating Object via Line Observations ($^{\rm 3D}$B{\scriptsize{AROLO}}\footnote{\url{http://editeodoro.github.io/Bbarolo/}}; \citealt{DiTeodoro2015}).

The $^{\rm 3D}$B{\scriptsize{AROLO}} software fits 3D tilted ring models to spectroscopic data cubes. For each ring, the algorithm builds a 3D disk model of the gas distribution in both the spatial and velocity axes, and then convolves it with the restoring beam of the observed data cube. Finally, it compares the convolved data set with the observed one. The geometry of the tilted ring model can be seen in Fig. \ref{tilted} and the corresponding geometrical and kinematic parameters (e.g., the inclination angle $i$; the position angle $\phi$; the rotation velocity $V_{\rm rot}$) are described in Appendix \ref{sec_til}. The 3D tilted ring model performs better than the standard 2D modeling on the velocity fields. For example, a common problem when deriving the kinematic properties from the velocity fields is the beam smearing effect, which  smears the steep velocity gradient especially in the central region. And there exists differences among the velocity fields measured using different methods (e.g., an intensity-weighted velocity field from {\footnotesize{CASA}}, versus a mean gas velocity map based on Gaussian fitting using {\footnotesize{AIPS}}\footnote{\url{http://www.aips.nrao.edu/}}). The $^{\rm 3D}$B{\scriptsize{AROLO}} algorithm avoids the beam smearing with the convolution step and directly conducts the modeling on the data cube. In addition to the rotation velocity for each ring, $^{\rm 3D}$B{\scriptsize{AROLO}} can also measure the asymmetric-drift correction ($V_{\rm A}$), which is caused by the random motions and can be directly measured with the velocity dispersion and the density profile (e.g., \citealt{Iorio2017}). There is no direct way to estimate parameter errors, and $^{\rm 3D}$B{\scriptsize{AROLO}} adopts a Monte Carlo method: it calculates models by changing the parameter values with random Gaussian draws centered on the minimum of the function, once the minimization algorithm has converged.

We took the ring width $W_{\rm ring}$ as half of the clean beam size (i.e., Nyquist sampling): $0\farcs05$  and $0\farcs1$ for the modeling of the [\ion{C}{II}] and CO\,(9--8)  lines, respectively. The central radius of each ring is $W_{\rm ring}\times t$  where $t$ is the $t$-th ring. Eight rings are adopted for the kinematic modeling on our high-resolution [\ion{C}{II}] data. The [\ion{C}{II}] emission extends to a radius of $\sim$2.8 kpc as shown by the [\ion{C}{II}] surface brightness distribution (top-right panel in Fig. \ref{sbcii}). We chose to use data within a radius of $\sim$2.4 kpc, taking advantage of higher S/N data. We should note that as the projected [\ion{C}{II}] kinematic minor axis ($\sim$2.5 kpc) is longer than the kinematic major axis, we will lack some information near the kinematic major axis for the outer two rings. However, since we have plenty of information from near the kinematic minor axis and in the region between these two axes, we still have enough data to carry out the kinematic modeling in the outer two rings. The initial guess of the position angle and the inclination angle are 200$\degr$ and 40$\degr$. The initial position angle are roughly measured from the velocity maps based on the definition of the position angle (see Appendix \ref{sec_til}). One method for roughly determining the inclination angle, assuming an intrinsic round disk, uses the photometric minor and major source size ratio ($r_{\rm size}$): $i = \cos^{-1}(r_{\rm size})$. As the kinematic major axis of the [\ion{C}{II}] line is far from its photometric major axis and [\ion{C}{II}] kinematic major axis is shorter than its kinematic minor axis, but the kinematic major axis of the CO\,(9--8) line is close to its photometric major axis (see Table \ref{par_ima}) and CO\,(9--8) kinematic major axis is longer than its kinematic minor axis, we derived the initial inclination angle  from the CO\,(9--8) minor and major source size ratio. In addition, during the fitting, we adopt  pixel-by-pixel normalization, which allows the code to exclude the parameter of the surface density of the gas from the fit. It means that we force the value of each spatial pixel along the spectral dimension in the model to equal that in the observations, which allows a non-axisymmetric model in density and avoids untypical regions \citep{Lelli2012}. We present the fitted  parameters in Table \ref{par_kin}. The average inclination angle and position angle derived from the [\ion{C}{II}] and CO\,(9--8)  lines are consistent. We plot the intensity, velocity, velocity dispersion, and position-velocity maps and the line spectra measured from the modeled line data cubes in Figs. \ref{ciidatamodel} and \ref{co98datamodel} for the [\ion{C}{II}] and CO\,(9--8)  lines, respectively.  The rotation velocity,  gas velocity  dispersion, and asymmetric-drift correction  for each ring are shown in Fig. \ref{modelciico98}.  The circular rotation velocities (the rotation velocities corrected by the asymmetric drifts; see Eq. \ref{vcvc}) presented in Fig. \ref{rcd}  for [\ion{C}{II}] and CO\,(9--8)  lines  are consistent with each other.

\begin{figure*}
\centering
\subfigure{\includegraphics[width=0.96\textwidth]{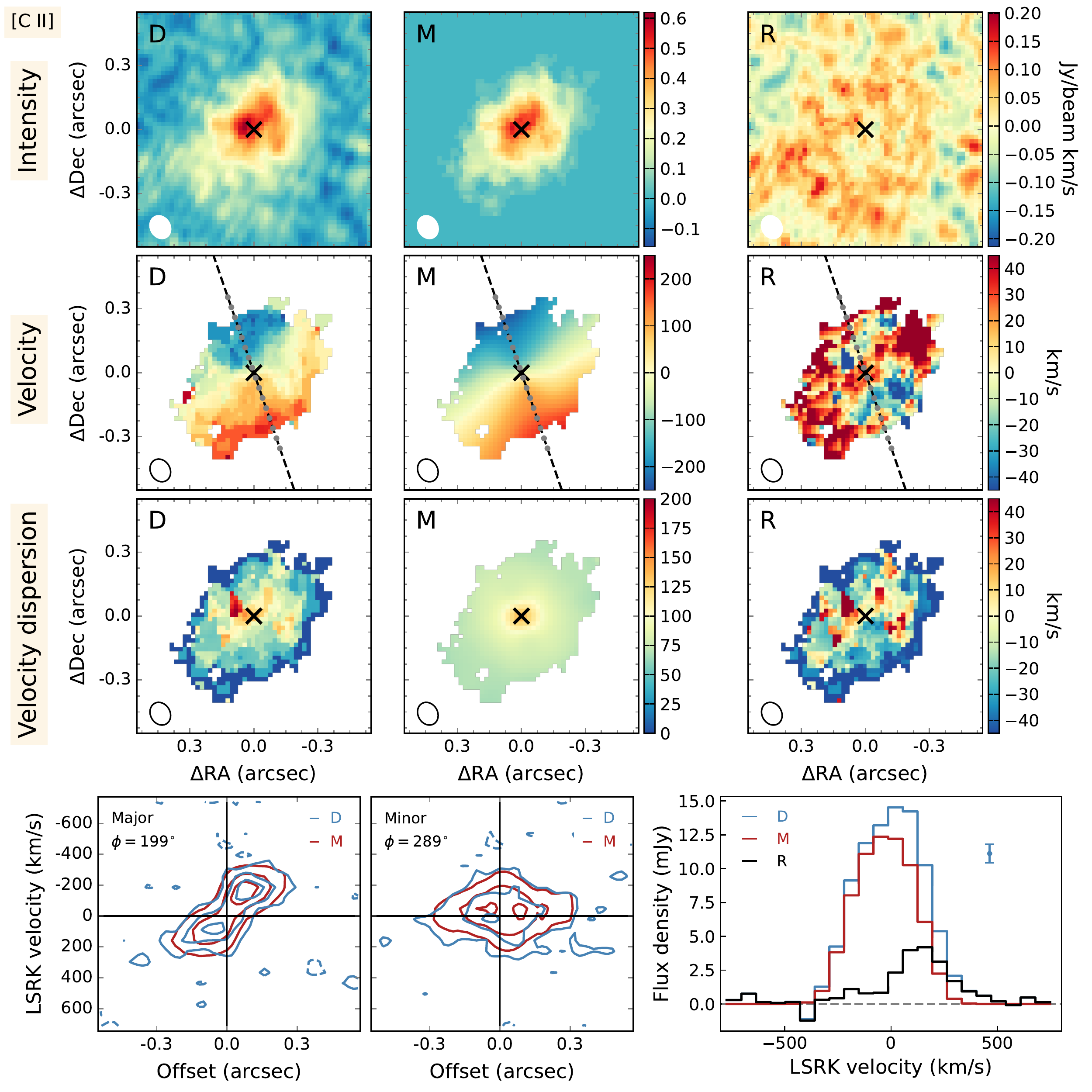}} 
\caption{Observed and modeled [\ion{C}{II}] line of J2310+1855. Upper-left panels: Intensity, velocity, and velocity dispersion maps for the observed data from top to bottom, which are labeled ``D.'' Upper-central panels: Same as the left panels but for the best-fit modeling from $^{\rm 3D}$B{\scriptsize{AROLO}}, labeled ``M.'' Upper-right panels: Same as the left panels but for the residuals between the observed data (``D'') and the modeled ones (``M''), which are labeled ``R'' (note that the dynamical range of these panels are different from others). The black cross in each panel marks the center of the rotating gas disk. The shape of the synthesized beam with a FWHM size of $0\farcs111\times0\farcs092$ is plotted in the bottom-left corner of each panel. In the velocity field panels, the dashed black lines are the kinematic major axis of the gas disk, and the plotted solid gray dots represent the ring positions. Lower-left and lower-middle panels: Position-velocity maps extracted along the kinematic major and minor axes from the observed data cube (blue contours) and the $^{\rm 3D}$B{\scriptsize{AROLO}} model cube (red contours). The contour levels are [--2, 2, 6, 10] $\times$ 0.15 mJy/beam for both the data and the model. Lower-right panel: [\ion{C}{II}] spectra extracted from the observed data cube (blue histogram) and the $^{\rm 3D}$B{\scriptsize{AROLO}} model cube (red histogram). The residual spectrum between the data and the model is shown as the black histogram.  The spectral resolution is 62.5 MHz, corresponding to 68 km/s. In the $^{\rm 3D}$B{\scriptsize{AROLO}} modeling, we used a re-binned (four original channels) data cube in order to optimize the data S/N per frequency and velocity bins and the sampling of the FWHM of the [\ion{C}{II}] line.}
\label{ciidatamodel}
\end{figure*}

\begin{figure*}
\centering
\subfigure{\includegraphics[width=\textwidth]{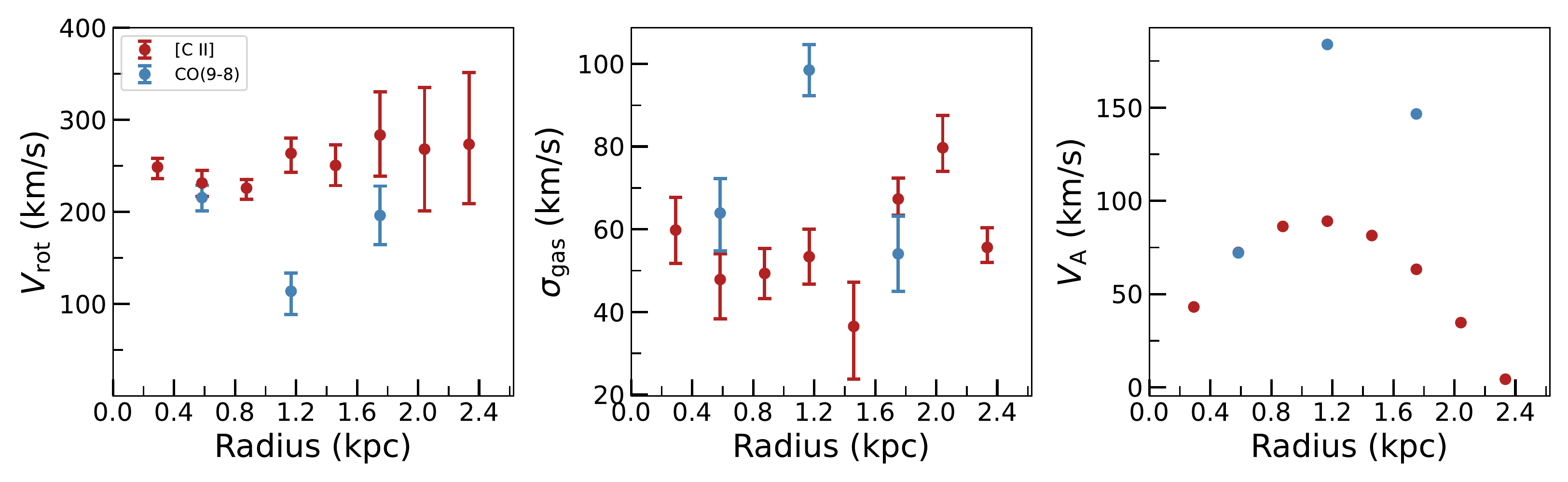}} 
\caption{Measured rotation velocity (left panel), gas velocity dispersion (middle panel), and asymmetric-drift velocity (right panel) as a function of radius (the central radius of each ring) measured from the  $^{\rm 3D}$B{\scriptsize{AROLO}} modeling for the [\ion{C}{II}] (red points with error bars) and CO\,(9--8) (blue points with error bars) lines, respectively.}
\label{modelciico98}
\end{figure*}

\begin{table*}
\renewcommand\arraystretch{1.8}
\caption{Kinematic parameters derived from the $^{\rm 3D}$B{\scriptsize{AROLO}} modeling.}
\label{par_kin}
\centering
\begin{tabular}{lcccccccccc}
\hline\hline
Data cube&$i$&PA&$\overline{i}$&$\overline{\rm PA}$&$V_{\rm flat}$&$V_{\rm max}$&$\sigma_{\rm ext}$&$\sigma_{\rm med}$&$V_{\rm flat}/\sigma_{\rm ext}$&$V_{\rm max}/\sigma_{\rm med}$\\
&($\degr$)&($\degr$)&($\degr$)&($\degr$)&(km/s)&(km/s)&(km/s)&(km/s)&&\\
(1)&(2)&(3)&(4)&(5)&(6)&(7)&(8)&(9)&(10)&(11)\\
\hline
[\ion{C}{II}]&(37--45)$^{+(1-6)}_{-(2-8)}$&(191--214)$^{+(3+26)}_{-(3-23)}$&$42\pm3$&$199\pm8$&$250^{+22}_{-13}$&$283^{+45}_{-47}$&$37^{+13}_{-11}$&$54^{+6}_{-6}$&$7^{+2}_{-2}$&$5^{+1}_{-1}$ \\
CO\,(9--8) &(41--43)$^{+(2-9)}_{-(3-9)}$&(195--202)$^{+(7-9)}_{-(5-9)}$&$43\pm1$&$198\pm3$&--&$215^{+13}_{-14}$&--&$64^{+8}_{-9}$&--&$4^{+1}_{-1}$\\
\hline
\end{tabular}
\tablefoot{Column 1: the data cube name. Columns 2--3: the inclination angle and position angle for each ring. Columns 4--5: the mean values of the inclination angle and position angle shown in columns 2--3. Note that the errors only represent the standard deviations of the measured values for all rings. Column 6: the rotation velocity in the flat part. Column 7:  the maximum rotation velocity. Column 8: the velocity dispersion in the flat part. Column 9:  the median velocity dispersion. Columns 10--11: the ratios  between the  rotation velocity and the  velocity dispersion.}
\end{table*}

\subsection{Rotation curve decomposition}
\label{sec_rcd}

In order to quantify the dynamical contribution of each matter component, we performed a decomposition of the circular rotation curve measured from the high-resolution [\ion{C}{II}] line.

The circular velocity ($V_{\rm c}$) directly traces the galactic gravitational potential ($\Phi$) and can be measured by correcting the rotation curve for the asymmetric drift: 

\begin{equation}
\label{vcvc}
R\frac{\partial\Phi}{\partial R} = V_{\rm c}^{2} = V_{\rm rot}^{2} + V_{\rm A}^{2}
,\end{equation}
where $V_{\rm rot}$ is the rotation velocity, $V_{\rm A}$ is the asymmetric drift correction caused by  random motions and can be modeled given the velocity dispersion and the density profile (e.g., \citealt{Iorio2017}) by $^{\rm 3D}$B{\scriptsize{AROLO}}, and $\Phi$ is the sum of the potentials of the different mass components.

We consider four matter components (black hole, stellar, gas, and dark matter) that  contribute to the total gravitational potential of the quasar-host system. Thus,  the circular velocity can be expressed as
\begin{equation}
V_{\rm c} =  \sqrt{ V_{\rm BH}^{2} + V_{\rm star}^{2} + V_{\rm gas}^{2} +V_{\rm DM}^{2} },
\end{equation}
where $V_{\rm BH}$, $V_{\rm star}$, $V_{\rm gas}$ and $V_{\rm DM}$ are the contributions of the black hole, stellar, gas and dark matter components to the circular velocity. The following applies for each component at a radius $R$:

First, the Keplerian velocity due to the central black hole -- $V_{\rm BH}$ is

\begin{equation}
V_{\rm BH} = \sqrt{ \frac{G M_{\rm BH}}{R} },
\end{equation}
where $G$ and $ M_{\rm BH}$ are the gravitational constant and the black hole mass, respectively. We measured a circular velocity of $249_{-12}^{+9}$ km/s at the innermost ring with a central radius of 0.29 kpc. Only considering the gravitational potential caused by the central black hole at the radius of 0.29 kpc, we calculated a black hole mass upper limit of $4.18^{+0.31}_{-0.39}\times 10^{9} M_{\odot}$. The black hole mass for our target J2310+1855 measured from the \ion{Mg}{II} and \ion{C}{IV} lines is (3.15--5.19)\,$\times 10^{9} M_{\odot}$ \citep{Jiang2016}, which corresponds to a $V_{\rm BH}$ range of 216--277 km/s at $R=0.29$ kpc. The black hole contribution to the total potential is significant in the innermost region, but  some values of $V_{\rm BH}$ exceed the measured value at 0.29 kpc. We set the black hole mass as a free parameter in the dynamical modeling. The black hole sphere of influence ($r_{\rm h}$) is defined as the region within which the gravitational potential of the black hole dominates over that of the surrounding stars, which can be measured with  $r_{\rm h}=GM_{\rm BH}/\sigma_{\rm star}^{2}$ where $\sigma_{\rm star}$ is the velocity dispersion of the surrounding stellar population. When $M_{\rm BH}=2.97\times10^{9}\,M_{\odot}$ (from this dynamical modeling) and $\sigma_{\rm star}=250$ km/s (see Sect. \ref{co-evolution}), $r_{\rm h}=0.21$ kpc. When $M_{\rm BH}=10^{9}\,M_{\odot}$ (the order value of the black hole mass measured from NIR spectral lines with the local scaling relations; \citealt{Jiang2016}; \citealt{Feruglio2018}) and $\sigma_{\rm star}=100$ km/s (assuming it is equal to the maximum velocity dispersion measured from our kinematic modeling of the gas), $r_{\rm h}=0.43$ kpc. Our high-resolution ALMA [\ion{C}{II}] observations can zoom into the sphere of influence considering that our innermost point is at 0.29 kpc and, thus, can be used to derive the dynamical mass of the central black hole. Beyond the sphere of influence, the gravitational dominance of the black hole quickly vanishes, which is shown as the yellow line of the left and middle panels of Fig. \ref{rcd}. We are not spatially resolving the sphere of influence considering the ratio $r_{\rm h}/r_{\rm res}$ (i.e., 0.4--0.8 in this work) between the radius of the black hole sphere of influence and the spatial resolution of the data.  However, gas dynamical studies (e.g., \citealt{Ferrarese2000}; \citealt{Graham2001}; \citealt{Marconi2003}; \citealt{Valluri2004}) have addressed that resolving the sphere of influence is an important but not sufficient factor to dynamically estimate the black hole mass, and the ratio $r_{\rm h}/r_{\rm res}$ can be taken as a rough indicator of the quality of the black hole mass estimate.  \citet{Ferrarese2005} have summarized a list of black hole mass detection based on resolved dynamical studies in their Table II, which shows a $r_{\rm h}/r_{\rm res}$ range of 0.4--1700. Our $r_{\rm h}/r_{\rm res}$ ratio is just within the above-mentioned range. In summary, we are able to derive a dynamical mass of the black hole from our [\ion{C}{II}] data.

Second, the stellar component can be described by a S{\'e}rsic profile (e.g., \citealt{Terzic2005}; \citealt{Rizzo2021}), giving rise to
\begin{equation}
\label{vsersic}
V_{\rm star} = \sqrt{  \frac{GM_{\rm star}}{R} \frac{\gamma(n_{\rm star}(3-p),b(R/R_{\rm e;\,star})^{1/n_{\rm star}})}{\Gamma(n_{\rm star}(3-p))}  },  
\end{equation}where $M_{\rm star}$, $R_{\rm e; \,star}$ and $n_{\rm star}$ are the total stellar mass, the effective radius and the S{\'e}rsic index. $\gamma$ and $\Gamma$ are the incomplete and complete gamma functions. $p$ and $b$ are related to the S{\'e}rsic index by $p=1.0-0.6097/n_{\rm star} + 0.05563/n_{\rm star}^{2}$ (when $0.6<n_{\rm star}<10$, and $10^{-2}\le R/R_{\rm e; \,star}\le10^{3}$) and $b=2n_{\rm star}-1/3+0.009876/n_{\rm star}$ (when $0.5<n_{\rm star}<10$). Due to the limited resolution ($\sim$600 pc) of our ALMA data (giving us few data points), the lack of resolved rest-frame optical or NIR data to constrain the two stellar components (i.e., bulge and disk), and the strong degeneracies between the two subcomponents, following the fitting method from \citet{Rizzo2021}, we adopted a single S{\'e}rsic element to  globally describe the stellar component.

Third, the gas component can be represented by a thin exponential disk \citep{Binney1987}. Thus,

\begin{equation}
V_{\rm gas} = \sqrt{  \frac{2GM_{\rm gas}}{R_{\rm gas}} y^{2} [I_{0}(y)K_{0}(y)-I_{1}(y)K_{1}(y)] },
\end{equation}where $M_{\rm gas}$ and $R_{\rm gas}$ are the total gas mass and the gas  scale radius, $I_{ii}(y)$ and $K_{ii}(y)$ are the first and second kind modified Bessel functions of zeroth ($ii=0$) and first ($ii=1$) orders, and $y$ is given by $y=R/(2R_{\rm gas})$. The CO\,(9--8)  image decomposition in Sect. \ref{sec_sb} suggests an approximately exponential distribution of the molecular gas. We fixed the S{\'e}rsic index to be 1 for the CO\,(9--8) gas distribution, and got a scale radius of $\sim$0.78 kpc, which can be taken as the overall molecular gas scale radius under the assumption of an exponential gas disk. With the detected CO (2--1) line toward J2310+1855 ($\nu S_{\nu}=0.18\pm0.02$ Jy km/s; \citealt{Shao2019}), and assuming $L^{' }_{\rm CO (2-1)} \approx L^{' }_{\rm CO (1-0)}$ \citep{Carilli2013}, we are able to measure the gas mass with an assumed CO-to-gas mass conversion factor  ($\alpha_{\rm CO}$), namely: $M_{\rm gas}=\alpha_{\rm CO} L^{' }_{\rm CO (1-0)}$. 
The only free parameter for the gas component is $\alpha_{\rm CO}$.

Finally, the dark matter component contributes much less to the total gravitational potential than the baryonic components in the innermost regions. However, it is required to fit a flat rotation curve in the outer parts in the Galaxy and other local galaxies (e.g., \citealt{Carignan2006}; \citealt{Sofue2009}). We consider the dark matter halo as a Navarro-Frenk-White (NFW; \citealt{Navarro1996}) spherical halo. These assumptions lead to

\begin{equation}
V_{\rm DM} = \sqrt{ \frac{GM_{\rm DM}}{R_{\rm DM}} \frac{1}{x} \frac{\ln(1+cx)-cx/(1+cx))}{\ln(1+c)-c/(1+c)} },
\end{equation}where $M_{\rm DM}$ and $R_{\rm DM}$ are the virial mass and radius ($M_{\rm 200}$ and $R_{\rm 200}$ in \citealt{Navarro1996}), respectively, $c$ is the concentration parameter, and $x = R/R_{200}$ is the radius in units of the virial radius. The virial mass and radius are correlated with the critical density ($\rho_{\rm crit}$): $M_{\rm 200}=200\rho_{\rm crit}(4\pi/3)R_{\rm 200}^{3}$, where $\rho_{\rm crit}=3H_{z}^{2}/(8\pi G)$ and $H_{z}=H_{0}\sqrt{\Omega_{\rm m}(1+z)^{3}+\Omega_{\Lambda}}$ (where $z$ is the redshift). We calculate the concentration parameter ($c$) from the mass-concentration relation estimated in N-body cosmological simulations by \citet{Ishiyama2021}. The only free parameter for the dark matter component is $M_{\rm DM}$.

In summary, for the dynamical modeling, we have six free parameters ($M_{\rm BH}$, $M_{\rm star}$, $n_{\rm star}$, $R_{\rm e; \,star}$, $\alpha_{\rm CO}$, $M_{\rm DM}$) and eight data points (the circular velocities corrected by the asymmetric drift shown in Fig. \ref{rcd}, which have already been corrected for the inclination angle from $^{\rm 3D}$B{\scriptsize{AROLO}}). During the fitting with the {\footnotesize{emcee}} package \citep{emcee2013}, a loose  prior constraint of $[10^{8}, 10^{10}]M_{\odot}$ is adopted for $M_{\rm BH}$. The quantities of $M_{\rm star}$, $n_{\rm star}$ and $R_{\rm e; \,star}$ are tightly coupled together, and we use prior ranges of $[10^{7}, 10^{11}]M_{\odot}$, [0.5, 10], and [0, 2.5] kpc, respectively.  As for $\alpha_{\rm CO}$, we used a range of [0.2, 14] $M_{\odot}/(\rm K\, km/s\, { pc }^{ 2 })$, which is  measured from a sample of nearby AGN, ultra-luminous infrared galaxies, and starburst galaxies \citep{Mashian2015}. For $M_{\rm DM}$, we consider a uniform prior of $10^{10}$ to $10^{13}M_{\odot}$ dark matter halo. 

We present the best-fit  decomposition for the circular rotation curve in the left panel of Fig. \ref{rcd}, and the physical parameters measured from the dynamical modeling in column (a) of Table \ref{par_dyn}. The black hole mass of $2.97^{+0.51}_{-0.77}\times10^{9}M_{\odot}$ is consistent with the measurements -- $(4.17\pm1.02)\times10^{9}$ and $(3.92\pm0.48)\times10^{9} M_{\odot}$ from the \ion{Mg}{II} and \ion{C}{IV} lines \citep{Jiang2016}, and $(1.8\pm0.5)\times10^{9} M_{\odot}$ from the \ion{C}{IV} line \citep{Feruglio2018}. 
The [\ion{C}{II}] emission can be taken as a molecular gas mass tracer in galaxies (e.g., \citealt{Zanella2018}; \citealt{Madden2020}; \citealt{Vizgan2022}) with a conversion factor ($\alpha_{[\ion{C}{II}]}$) by $M_{\rm gas}=\alpha_{[\ion{C}{II}]} L_{[\ion{C}{II}]}$. The derived gas mass corresponds to a $\alpha_{[\ion{C}{II}]}$ of $3.2^{+0.6}_{-0.8}\ M_{\odot}/L_{\odot}$. Given that there are only two degrees of freedom,  and the strong degeneracies among some parameters (i.e., $M_{\rm star}$, $n_{\rm star}$ and $R_{\rm e;\,star}$), we cannot  constrain most of these parameters well with the  $\sim$600 pc resolution data we have. We tried  another experiment by fixing the black hole mass to a smaller value of $1.8\times10^{9}M_{\odot}$ \citep{Feruglio2018}, and present the dynamical results in the middle panel of Fig. \ref{rcd} and  column (b) of Table \ref{par_dyn}. In both situations, the derived stellar mass is on the order of $10^{9}M_{\odot}$ (however with large uncertainty). And when fixing the black hole component, the stellar component (green line in the middle panel of Fig. \ref{rcd}) has a bump in the inner region. This may indicate that a massive stellar bulge already formed at $z=6$.

\begin{figure*}
\centering
\subfigure{\includegraphics[width=0.32\textwidth]{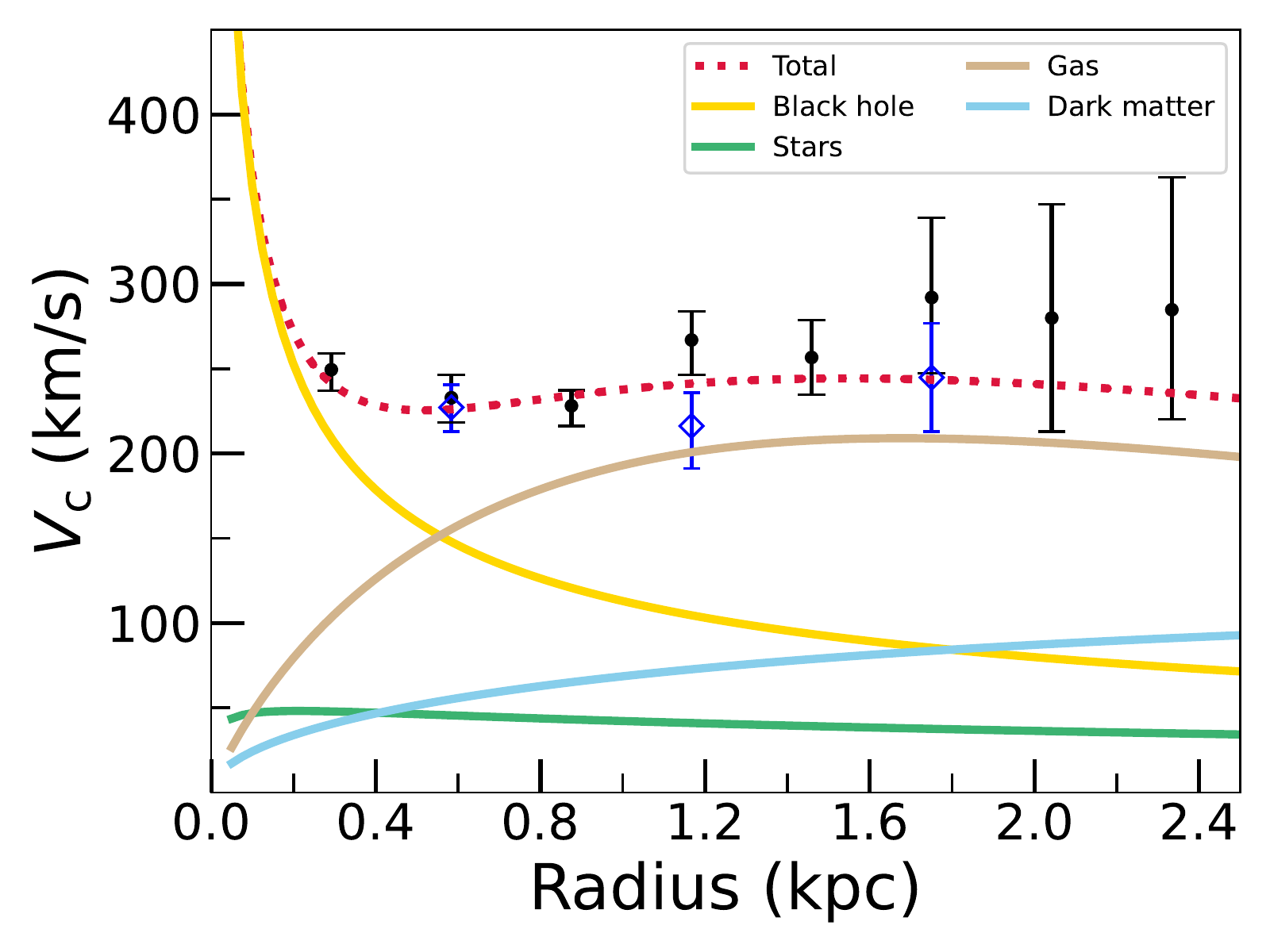}} 
\subfigure{\includegraphics[width=0.32\textwidth]{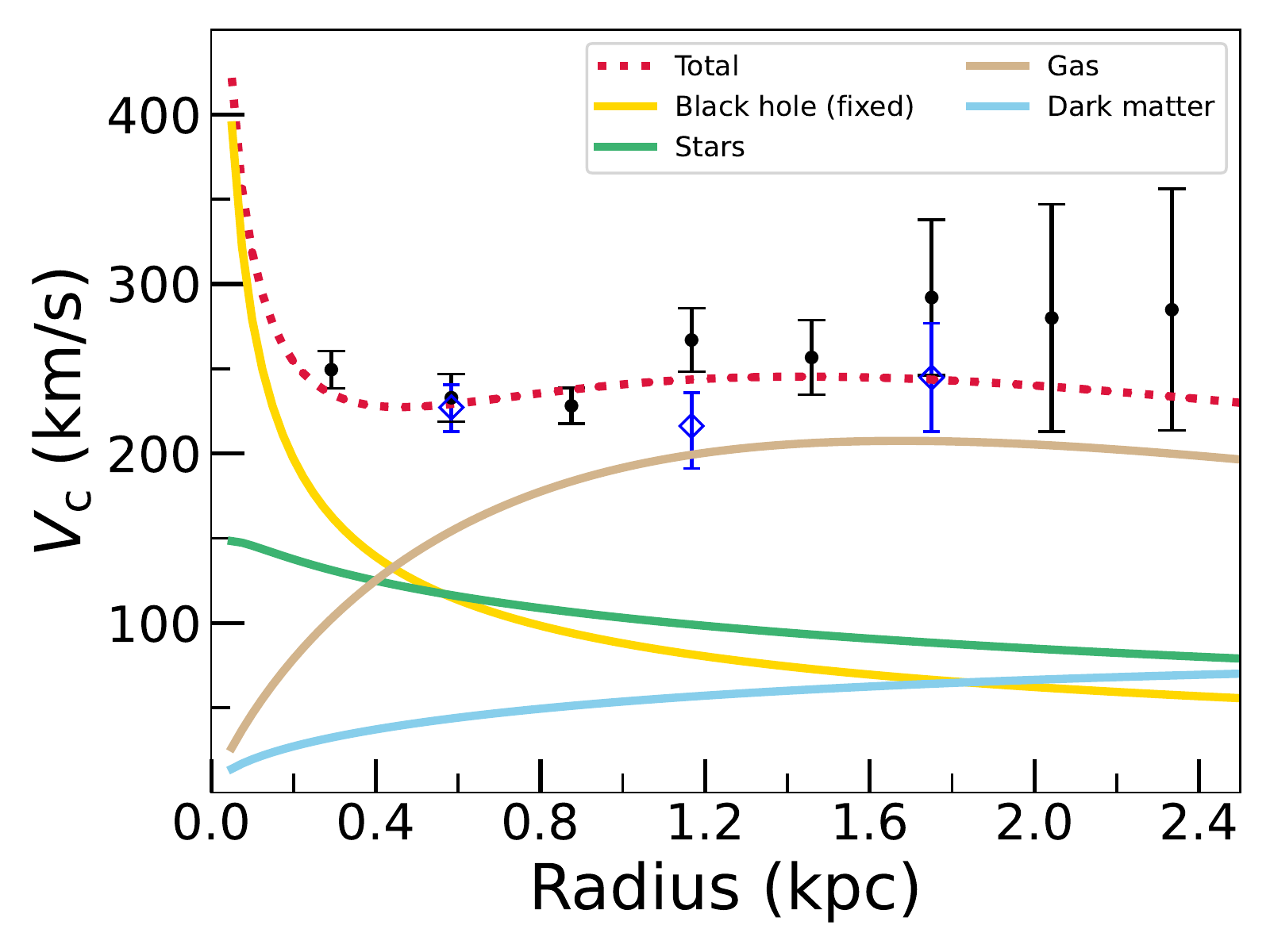}} 
\subfigure{\includegraphics[width=0.32\textwidth]{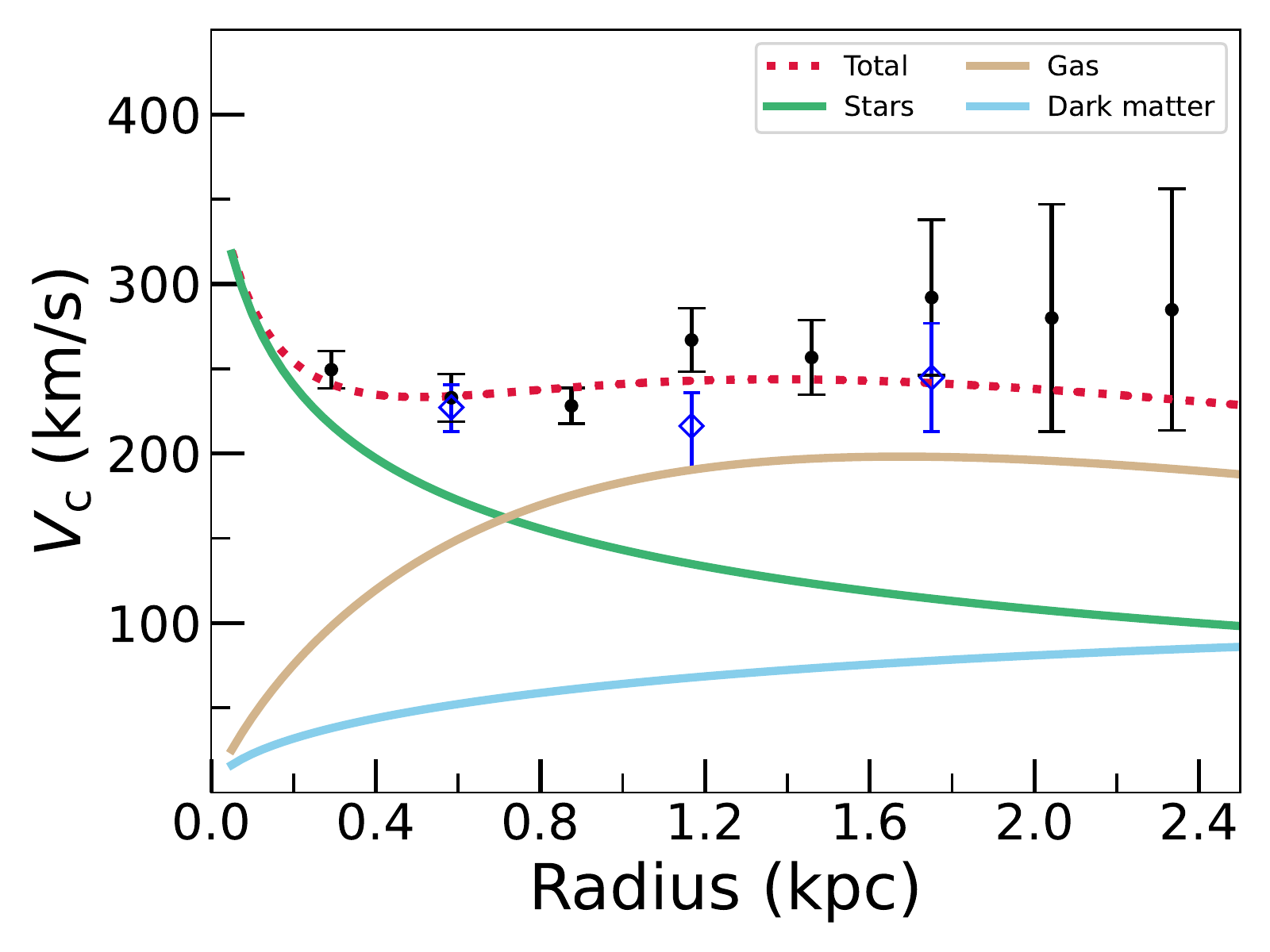}} 
\caption{Best-fit of the decomposition of the circular rotation curve traced by the [\ion{C}{II}] line when allowing the black hole component to be free (left), fixed to $1.8\times10^{9}\,M_{\odot}$ (\citealt{Feruglio2018}; middle) and none (right). Black points with error bars are the circular velocities, these  are the rotation velocities corrected for the asymmetric drift (caused by  gas random motions). The solid yellow, green, brown and blue lines represent the black hole, stellar, gas and dark matter components, respectively. The dashed red lines are the sum of these four components. The blue diamonds with error bars are the circular velocities measured from the  CO\,(9--8) line. These two lines trace the identical gravitational potential within the errors.}
\label{rcd}
\end{figure*}

\begin{table*}
\renewcommand\arraystretch{1.8}
\caption{Derived parameters from the [\ion{C}{II}] dynamical modeling.}
\label{par_dyn}
\begin{center}
\begin{tabular}{lcccccccc}
\hline\hline
Parameter&&(a)&(b)&(c)\\
\hline
$M_{\rm BH}\,(10^{9}M_{\odot})$&(1)&$2.97^{+0.51}_{-0.77}$&$^{aa}1.8$&$-$ \\
$M_{\rm star}\,(10^{9}M_{\odot})$&(2)&$1.16^{+6.51}_{-1.08}$&$5.76^{+5.89}_{-3.53}$&$6.30^{+6.25}_{-2.62}$ \\
$R_{\rm e;\,star}\,(\rm kpc)$&(3)&$1.31^{+0.74}_{-0.70}$&$1.03^{+0.94}_{-0.77}$&$0.21^{+0.55}_{-0.17}$ \\
$n_{\rm star}$&(4)&$4.88^{+3.47}_{-3.18}$&$6.66^{+2.34}_{-2.64}$&$7.17^{+2.02}_{-3.25}$ \\
$M_{\rm DM}\,(10^{10}M_{\odot})$&(5)&$4.55^{+32.28}_{-4.21}$&$1.63^{+26.66}_{-1.58}$&$3.42^{+30.62}_{-3.34}$ \\
$\alpha_{\rm CO}$ [$M_{\odot}/(\rm K\, km/s\, { pc }^{ 2 })$]&(6)&$0.37^{+0.07}_{-0.09}$&$0.37^{+0.09}_{-0.10}$&$0.34^{+0.09}_{-0.09}$ \\
\hline
$M_{\rm gas}\,(10^{10}M_{\odot})$&(7)&$2.05^{+0.40}_{-0.52}$&$2.02^{+0.47}_{-0.55}$&$1.84^{+0.47}_{-0.49}$ \\
$M_{\rm bary}\,(10^{10}M_{\odot})$&(8)&$2.37^{+0.50}_{-0.52}$&$2.74^{+0.45}_{-0.52}$&$2.65^{+0.42}_{-0.52}$ \\
$f_{\rm gas}$&(9)&$0.95^{+0.05}_{-0.25}$&$0.77^{+0.13}_{-0.18}$&$0.74^{+0.11}_{-0.18}$ \\
$V_{200}$\,(km/s)&(10)&$111^{+112}_{-65}$&$79^{+125}_{-54}$&$101^{+116}_{-72}$ \\
$R_{200}$\,(kpc)&(11)&$15.85^{+15.99}_{-9.21}$&$11.25^{+17.91}_{-7.77}$&$14.42^{+16.59}_{-10.24}$ \\
\hline
\end{tabular}
\end{center}
\tablefoot{Columns (a), (b) and (c) are for the dynamical models when allowing the black hole component to be free, fixed and none, respectively. 
Rows 1--6: fitted parameters from the dynamical modeling in Sect. \ref{sec_rcd}. They are the black hole mass, the stellar mass, the effective radius of the stellar component, the S{\'e}rsic index of the stellar component, the dark matter mass, and the CO-to-gas mass conversion factor, respectively. Row 7: the gas mass ($M_{\rm gas}=\alpha_{\rm CO} L^{' }_{\rm CO (1-0)}$). Row 8: the baryonic mass ($M_{\rm bary}=M_{\rm star}+M_{\rm gas}$). Row 9: gas mass fraction ($f_{\rm gas}=M_{\rm gas}/M_{\rm bary}$). Rows 10--11: the virial velocity and virial radius of the dark matter halo. 
}
\tablefoottext{$aa$}{the black hole mass is fixed to  the value of $1.8\times10^{9}\,M_{\odot}$ \citep{Feruglio2018}.}
\end{table*}

\section{Discussion} 
\label{sec_disc}

\subsection{The spatial distribution and extent of the ISM}
\label{sec_disext}

\subsubsection{Comparisons among different ISM tracers}
\label{sec_dis}

The different distribution and extent of various ISM tracers may be due to dissociation regions heated by starbursts and/or AGN in the quasar-host system of J2310+1855. We model the spatial distribution and extent (e.g., quantified by the half-light radii and the S{\'e}rsic index) of the [\ion{C}{II}] and the CO\,(9--8) lines and the dust continuum emission, using a 2D elliptical S{\'e}rsic function in Sect. \ref{sec_sb}.

The extended FIR dust continuum and the CO high-$J$ line emission follow a S{\'e}rsic distribution with the S{\'e}rsic index $n_{\rm ISM}$ a bit larger than 1. Exponential dust-continuum profiles are also observed for nearby (e.g., \citealt{Haas1998}; \citealt{Bianchi2011}) and high-redshift galaxies (\citealt{Hodge2016}; \citealt{CalistroRivera2018}; \citealt{Wang2019}; \citealt{Novak2020}). However, the [\ion{C}{II}] emission profile is slightly flatter,  with a S{\'e}rsic index of 0.59.

\citet{Walter2022} used simple RADEX \citep{vanderTak2007} modeling on a $z\sim7$ quasar host galaxy,  and find that the [\ion{C}{ii}] surface brightness in the very central part (i.e., $<$0.4 kpc) is significantly suppressed by the high dust opacity, which increases the FIR background radiation field on the [\ion{C}{ii}] line,   thereby making the [\ion{C}{ii}] fainter than one would observe without the FIR background. As shown in Fig. \ref{dusteachring}, both the dust mass surface density ($\Sigma_{\rm dust\_mass}>10^{8}M_{\odot}$/kpc$^{2}$) and the [\ion{C}{ii}] underlying dust optical depth ($\tau\sim0.5$) are very high inside the central 1 kpc region in our quasar host galaxy. 
In order to test the high dust opacity effect on the spatial distribution of [\ion{C}{ii}], following \citet{Walter2022}, we model the [\ion{C}{ii}] line intensities as a function of radius with and without the FIR background determined using the dust emission in the host galaxy (i.e., a gray-body radiation of $B_{\nu}(T_{\rm dust})(1-\exp^{-\tau_{\nu}})$). For both situations, we also consider the CMB as a background (i.e., a black-body radiation of $B_{\nu}(T_{\rm CMB})$; $T_{\rm CMB}=2.73\times(1+z)$ K). It should be noted that we only model collisional excitation with molecular hydrogen H$_{2}$, and ignore collisions with electrons and neutral hydrogen \ion{H}{i}. We model the [\ion{C}{ii}] line intensities for the whole galaxy by dividing the galaxy disk into nine concentric rings with ring width of $0\farcs05$. The number density of H$_{2}$ is fixed to be $10^{5}\rm cm^{-3}$. The column density of [\ion{C}{ii}] for each ring can be calculated with a fixed gas-to-dust mass ratio of $\sim$10 (from our measured total dust mass in Sect. \ref{sec_dd} and total gas mass in Sect. \ref{sec_rcd}) and a fixed [\ion{C}{ii}] abundance relative to hydrogen ($7\times10^{-6}$ is the best value in our fitting).  We further assume that the average kinetic temperature is equal to the dust temperature at the central radius of each ring, and the dust temperature can be predicted from the $T_{\rm dust}$--$r$ relation measured in Sect. \ref{sec_dd}. The average optical depth in each ring is associated with the dust mass surface density, which can be predicted from the $\Sigma_{\rm dust\_mass}$--$r$ relation measured in Sect. \ref{sec_dd}. In order to match the observed surface brightness of the [\ion{C}{ii}] line, we reduce the scale length of the $\Sigma_{\rm dust\_mass}$--$r$ relation by a factor of 2 for the inner 0.5 kpc region, which is consistent with the observed surface density profiles of the dust mass for some nearby galaxies (e.g., \citealt{Casasola2017}). The full width at half maximum (FWHM) of the [\ion{C}{ii}] line emission for each ring is measured from our observed [\ion{C}{ii}] line spectrum, which is extracted from each circular annulus.

The RADEX modeling results are shown in Fig. \ref{radexmodel}. The high FIR background associated with the dust radiation, which has dust optical depth $\sim$0.5 inside 1 kpc, indeed  lowers the [\ion{C}{ii}] surface brightness, thus changing the observed spatial distribution of [\ion{C}{ii}], reducing the [\ion{C}{ii}] equivalent width, and lowering the surface brightness ratio between  [\ion{C}{ii}] and FIR emission. The effect becomes weaker in the outer parts of the galaxy, as the dust optical depth decreases.   
As shown in the right panel of Fig. \ref{dusteachring}, the CO\,(9--8) underlying dust continuum optical depths are below 0.2. And importantly, the CO\,(9--8) emits at longer wavelength than [\ion{C}{ii}], further away from the dust emission peak. Thus, the FIR background radiation is not strong enough to significantly modify the observed CO\,(9--8) line intensity.
In summary, the difference of the observed surface brightness distribution between [\ion{C}{ii}] (with a S{\'e}rsic index of 0.59) and CO\,(9--8) (with a S{\'e}rsic index of 1.21 or 2.01) may be just the effect of  high dust opacity (see also \citealt{Riechers2014}), which diminishes the [\ion{C}{ii}] emission much more strongly in the center than in the outer region.

\begin{figure*}
\centering
\subfigure{\includegraphics[width=\textwidth]{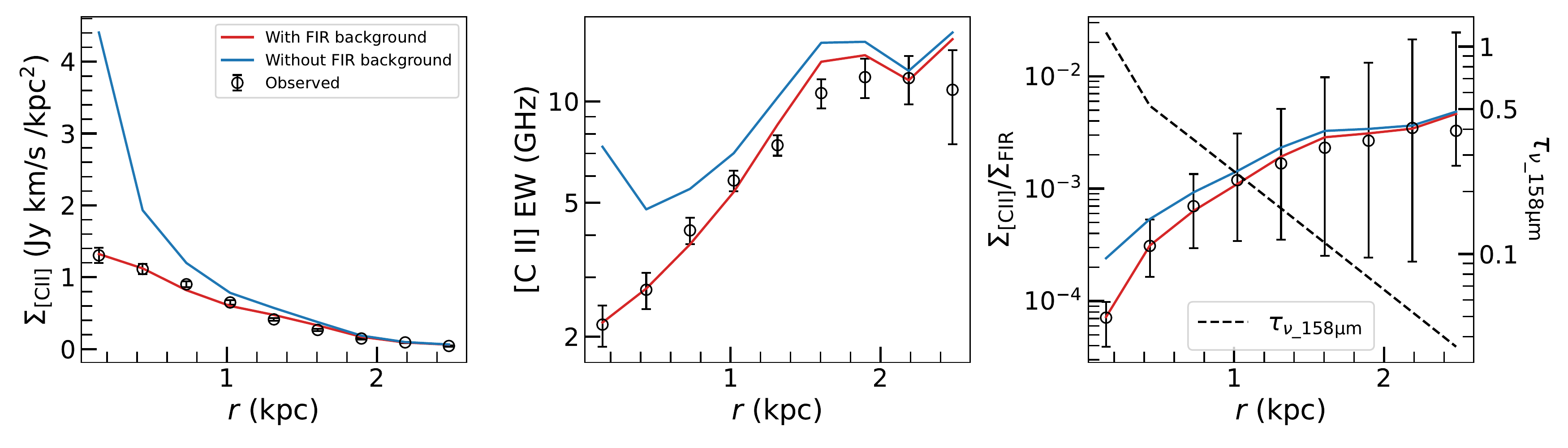}} 
\caption{RADEX modeling of the [\ion{C}{ii}] line emission with (red lines) and without (blue lines) FIR background caused by high dust opacity in the quasar host galaxy, and compared with the observed [\ion{C}{ii}] line emission (black circles with error bars): the  [\ion{C}{ii}] surface brightness profile (left panel), the [\ion{C}{ii}] line equivalent width (middle panel), and the ratio between $\Sigma_{\rm [\ion{C}{ii}]}$ and $\Sigma_{\rm FIR}$ (right panel) as a function of radius. The dashed black line in the right panel represents the radial distribution of the optical depth at the frequency of the [\ion{C}{ii}] underlying dust continuum.}
\label{radexmodel}
\end{figure*}

We next compare the rotation angles (column 8 in Table \ref{par_ima}) of the major axes of the tilted brightness distribution of the ISM tracers with the position angles (column 5 in Table \ref{par_kin}) from the kinematic modeling. The photometric major axis of the [\ion{C}{II}] line is close to that of its underlying dust continuum, but is almost perpendicular to that of the CO\,(9--8) line. The photometric major axis of the CO\,(9--8) line is coincident with the kinematic major axis of the [\ion{C}{II}]/CO\,(9--8) shown in Sect. \ref{sec_kin}, but makes an angle of $\sim$30$\degr$ with respect to that of its underlying dust continuum. Such a misalignment between kinematic and photometric position angles is also reported for the hosts of Palomar-Green quasars by \citet{Molina2021}, which may indicate kinematic twisting.  We compare the effective radii (in boldface in columns 10--11 of Table \ref{par_ima}) of the emission near the kinematic major axis, as they are less influenced  by  projection effects and can be interpreted as the matter distribution along the kinematic major axis. The half-light radius of the [\ion{C}{II}] underlying dust continuum at observed frame 262 GHz is larger by a factor of 1.1 than that of the CO\,(9--8) underlying dust continuum at observed frame 147 GHz. The half-light radii of the dust continuum at both frequencies are smaller than those of their nearby lines, with ratios of $\sim$0.7 and $\sim$0.5 of [\ion{C}{II}] underlying dust continuum to the [\ion{C}{II}] line, and CO\,(9--8) underlying dust continuum to the CO\,(9--8) line, respectively. The half-light radius of the [\ion{C}{II}] line is smaller than that of the CO\,(9--8) line. The more concentrated dust continuum than the [\ion{C}{II}] and the CO\,(9--8) emission is also found in other $z\sim6$ quasars (e.g., \citealt{Wang2019}; \citealt{Novak2020}), high-redshift galaxies (e.g.,  \citealt{Riechers2013, Riechers2014}; \citealt{Chen2017}; \citealt{Gullberg2019}), and numerical simulations (e.g., \citealt{Cochrane2019}; \citealt{Popping2022}).

The higher compactness for the continuum than the lines reflects the temperature dependence on radius; the dust surface brightness depends on both column density and temperature (Eq. \ref{graybody1}), while the lines mainly scale with temperature (e.g., \citealt{CalistroRivera2018}). 
The dust continuum sizes at the rest frame wavelengths of 158 and 290 $\mu$m are identical within $\sim$10$\%$, which is consistent with simulations conducted by \citet{Popping2022}, who found that the dust continuum sizes remain constant within $\sim$20$\%$ at observed frame wavelengths from 500 $\mu$m to 2 mm for $z\sim$1--5 main-sequence galaxies. 
As these two wavelengths are very close, they are both dominated by dust at similar temperatures.

Our RADEX modeling  of the effect of the dust opacity suggest that the actual half-light radius of the [\ion{C}{ii}] gas is much smaller than that measured from the observed $n=0.59$ S{\'e}rsic distribution listed in Table \ref{par_ima}. 
The intensity of  the CO\,(9--8) line  depends strongly on the ISM density that drives the excitation of high-$J$ CO lines. The [\ion{C}{ii}] line is mainly collisionally excited by electron and hydrogen, which are heated by UV photons. The different effective radii of the [\ion{C}{ii}] and CO\,(9--8) lines may be a reflection of the radial dependence of the gas excitation  and the radiation field.

\subsubsection{The nuclear and extended components of the ISM}
\label{sec_ismorigin}

The long-wavelength dust continuum emission appears to trace the bulk of the star formation activity in  quasar-starburst systems at $z\sim6$ (e.g., \citealt{Venemans2018}; \citealt{LiQ2020}). Detailed SED decomposition from rest-frame UV to FIR allows one to constrain the relative contribution to the continuum of different components (i.e., accretion disk, torus, host galaxy; e.g., \citealt{Leipski2013, Leipski2014}; \citealt{Shao2019}). We find (Fig. \ref{sedfitting}) that the AGN dust torus contributes $\sim$40$\%$ of the integrated FIR  luminosity of J2310+1855, and the star-formation heated dust continuum emission dominates the FIR luminosity above rest-frame 50 $\mu$m (\citealt{Shao2019}; Sect. \ref{sec_dd} in this paper). The dust continuum  decomposition results presented in Table \ref{par_ima} and Figs. \ref{sbciidust}--\ref{sbcodust} reveal both a compact central component and an extended component, which may stem from the AGN dust torus and the quasar host galaxy associated with star formation, respectively.
The AGN dusty molecular torus is a key element in the AGN unification model, which  obscures the accretion disk and the broad line region in type-2 AGN (e.g., \citealt{Honig2019}). The AGN torus might be unstable, warped and unaligned with the host galaxy orientation, and has been revealed by observations of several molecular lines and dust continuum in low-$z$ AGN (i.e., 10 pc scale; e.g., \citealt{Garcia-Burillo2014, Garcia-Burillo2021}; \citealt{Combes2019}).
Our image decomposition assigns $\sim$5$\%$ of the dust continuum near both [\ion{C}{ii}] and CO\,(9--8) to a point source, which we interpret as emission from the AGN torus.

The diameters of the dusty molecular tori measured by high-resolution ALMA observations of nearby Seyfert galaxies  range from $\sim$25 pc to $\sim$130 pc, with a median value of $\sim$42 pc \citep{Garcia-Burillo2021}. We here consider a radius of 25 pc of the torus in our quasar, and with the point emission at the two wavelengths listed in Table \ref{par_ima} and Eqs. \ref{graybody1} and \ref{graybody2} ($\beta$ fixed to be 1.90), we calculate a torus dust temperature of $1600\pm500$ K and a torus dust mass of ($1.46\pm1.14$) $\times10^{6}M_{\odot}$. 
\citet{Esparza-Arredondo2021} explored the torus GDR for a sample of 36 nearby AGN with \textit{NuSTAR} and \textit{Spitzer} spectra, and found that it lies between 1 and 100 times the Galactic ratio (i.e., 100; \citealt{Bohlin1978}). From our dynamical modeling, we get an average GDR of $\sim$10 for the quasar host galaxy. We assume a GDR value of 100 for the torus alone, as this is consistent with the existence of gas located in the dust-free inner region of the torus.  Thus, the molecular gas mass derived from the dust emission in the torus is ($1.46\pm1.14$) $\times10^{8}M_{\odot}$.

We present the best-fit  CO\,(9--8)  image decomposition results in Fig. \ref{sbco98} and Table \ref{par_ima} with both a nuclear and extended components,  representing the CO\,(9--8)  emission from the AGN molecular torus and the AGN host galaxy, respectively. The nuclear component comprises $\sim$11$\%$ of the whole CO\,(9--8) emission.
The nuclear CO\,(9--8)  emission can be converted to a molecular gas mass with a flux ratio between CO\,(9--8)  and CO (2--1) lines (i.e., $\sim$8; \citealt{Shao2019}) and a CO-to-H$_{2}$ conversion factor - $\alpha_{\rm CO}$, assuming  $L^{'}_{\rm CO(2-1)}  \approx L^{'}_{\rm CO(1-0)}$ (\citealt{Carilli2013}). We derive an average $\alpha_{\rm CO}$ value of $\sim$0.37 from our dynamical study, but this may be smaller in the core. For example the conversion factor in the  circumnuclear disk can be  3--10 lower than that of the ISM (i.e., \citealt{Usero2004}; \citealt{ Garcia-Burillo2014}). Assuming a $\alpha_{\rm CO}$ that is 10 times smaller than the overall value, the molecular gas mass is derived to be ($2.44\pm0.28$) $\times10^{8}M_{\odot}$, which is roughly consistent with that measured from the dust emission from the AGN dusty molecular torus. However, the decomposed unresolved emission from the AGN dusty molecular torus is quite uncertain given our current angular resolution and sensitivity. And the estimate of the molecular gas mass in the dusty molecular torus depends on many uncertain quantities: the torus size, the dust emissivity index, the GDR, the CO line ratio and the $\alpha_{\rm CO}$ in the unresolved central region.

\subsection{The surface density of the gas and the star formation}
\label{sec_ism_dist}

Next,  we investigated the surface density of the gas and star formation. We measured the surface densities of gas and dust as a function of distance from the quasar center from the intensity maps using  circular rings. These rings have widths of $0\farcs05$ and $0\farcs1$ for the [\ion{C}{II}] and CO\,(9--8) lines, respectively, about half the restoring beam sizes.
For each ring, we used Eqs. \ref{graybody1} and \ref{graybody2} with a fixed dust emissivity index of 1.90 (from the UV to FIR SED decomposition in Sect. \ref{sec_dd}) and with the dust temperature and dust mass surface density following the relations from our dust diagnostics in Sect. \ref{sec_dd}. Then we measured the FIR luminosity  by integrating from 42.5 to 122.5 $\mu$m and the IR luminosity by integrating from 8 to 1000 $\mu$m.

\subsubsection{$\Sigma_{\rm [\ion{C}{II}]}/\Sigma_{\rm FIR}$}

Local spiral galaxies have a typical [\ion{C}{II}]-FIR luminosity ratio of $\sim$$5\times10^{-3}$ when $\Sigma_{\rm FIR}<10^{11}\,L_{\odot}\rm /kpc^{2}$. However, the ratio is substantially smaller at higher FIR luminosity surface densities (e.g.,  \citealt{Diaz-Santos2013, Diaz-Santos2017}). This so-called [\ion{C}{II}]-FIR deficit  is shown in the left panel of Fig. \ref{ciifir}, together with results from other spatially resolved $z\sim6$ quasars (e.g., \citealt{Wang2019}) and $z\sim3$  submillimeter galaxies (SMGs; e.g., \citealt{Rybak2019}). As shown in the right panel of Fig. \ref{radexmodel}, the ratio of  $\Sigma_{\rm [\ion{C}{II}]}$ and $\Sigma_{\rm FIR}$ of our target increases with increasing radii, appearing to asymptote at larger radii. Our [\ion{C}{II}]-FIR ratios are a few times higher than those of spatially resolved star-forming associations at similar FIR surface brightness. Our measurements push to higher $\Sigma_{\rm FIR}$ than do the comparison samples, but do not probe deeper into the faint end (i.e., $\Sigma_{\rm FIR}<10^{10}\,L_{\odot}\rm /kpc^{2}$) occupied by  Lyman break galaxies (LBGs; e.g., \citealt{Hashimoto2019lbg}) and the local luminous infrared galaxies in the GOALS sample \citep{Diaz-Santos2013, Diaz-Santos2017}.

The origin of the overall  [\ion{C}{II}]-FIR deficit of galaxies has been investigated in detail (e.g., \citealt{Luhman1998}; \citealt{Stacey2010}; \citealt{Diaz-Santos2013}; \citealt{Narayanan2017}; \citealt{Lagache2018}), yet remains under debate. The [\ion{C}{II}] line is mainly excited through  collisions with electrons, neutral hydrogen \ion{H}{I}, and molecular hydrogen H$_{2}$. Close to the quasar host galaxy center (corresponding to the region with the highest $\Sigma_{\rm FIR}$), the UV photons are absorbed by large dust grains (which may be reemitted at FIR wavelengths, thus  boosting the FIR luminosity). Therefore, the hydrogen and electrons will be less heated, giving rise to few collisions to excite ionized carbon. In addition, the dust absorption reduces the number of ionizing photons, decreasing the extent to which carbon becomes ionized (e.g., \citealt{Luhman2003}). Other popular explanations for the [\ion{C}{II}]-FIR deficit are the saturation of [\ion{C}{II}] emission at very high temperatures (e.g., $T_{\rm gas}>T_{[\ion{C}{II}]}$; e.g., \citealt{Muno2016}) or in extremely dense environments (in which more carbon is in the form of CO than C$^{+}$; e.g., \citealt{Narayanan2017}). 

The observed larger [\ion{C}{II}]-FIR deficit in the center of our quasar host galaxy may be a reflection of the high dust temperature at small radius, and the different dependence of $\Sigma_{\rm FIR}$ and $\Sigma_{\rm [C II]}$ on $T_{\rm dust}$. As shown in the right panel of Fig. \ref{ciifir}, $\Sigma_{\rm FIR}$ is an integral over the gray-body relating to the dust temperature with a power-law index ($\sim T_{\rm dust}^{5.0\pm0.4}$; black line), larger than that ($\sim T^{4}$) of the Stefan-Boltzmann law. However,  $\Sigma_{\rm [C II]}$ scales as $\sim T_{\rm dust}^{2.4\pm0.7}$ (green line). Thus, higher dust temperatures in the galaxy center produce a larger [\ion{C}{II}]-FIR deficit (see also \citealt{Gullberg2015}; \citealt{Walter2022}). 
In addition, as demonstrated by \citet{Walter2022} and our RADEX test in Sect. \ref{sec_dis}, the high dust opacity, which increases the FIR background radiation field on the [\ion{C}{ii}] line (see also \citealt{Riechers2014}), suppresses the observed [\ion{C}{ii}] line, especially in the central region, thus enhancing the  [\ion{C}{II}]-FIR deficit.

\begin{figure*}
\centering
\subfigure{\includegraphics[width=0.49\textwidth]{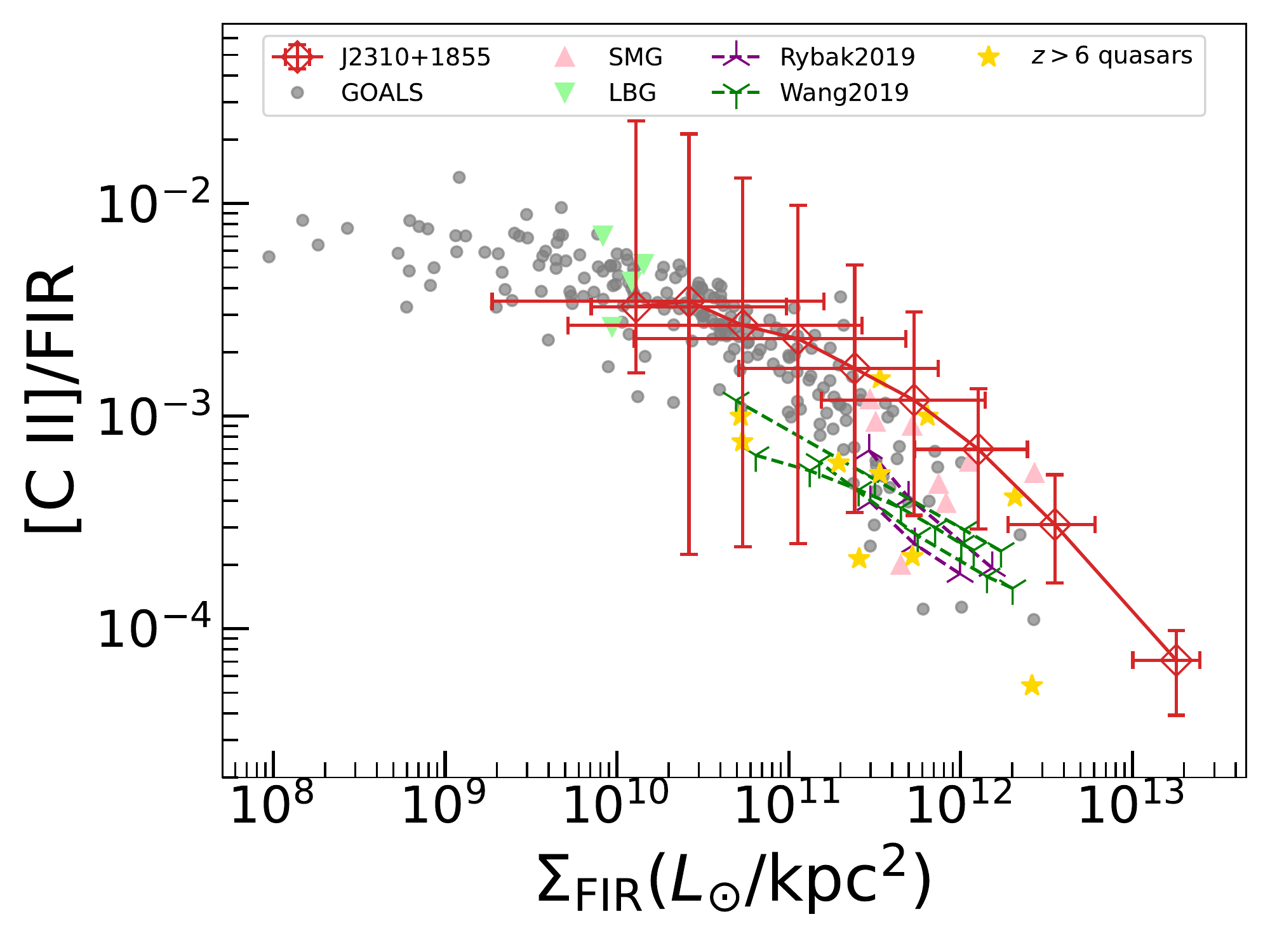}} 
\subfigure{\includegraphics[width=0.49\textwidth]{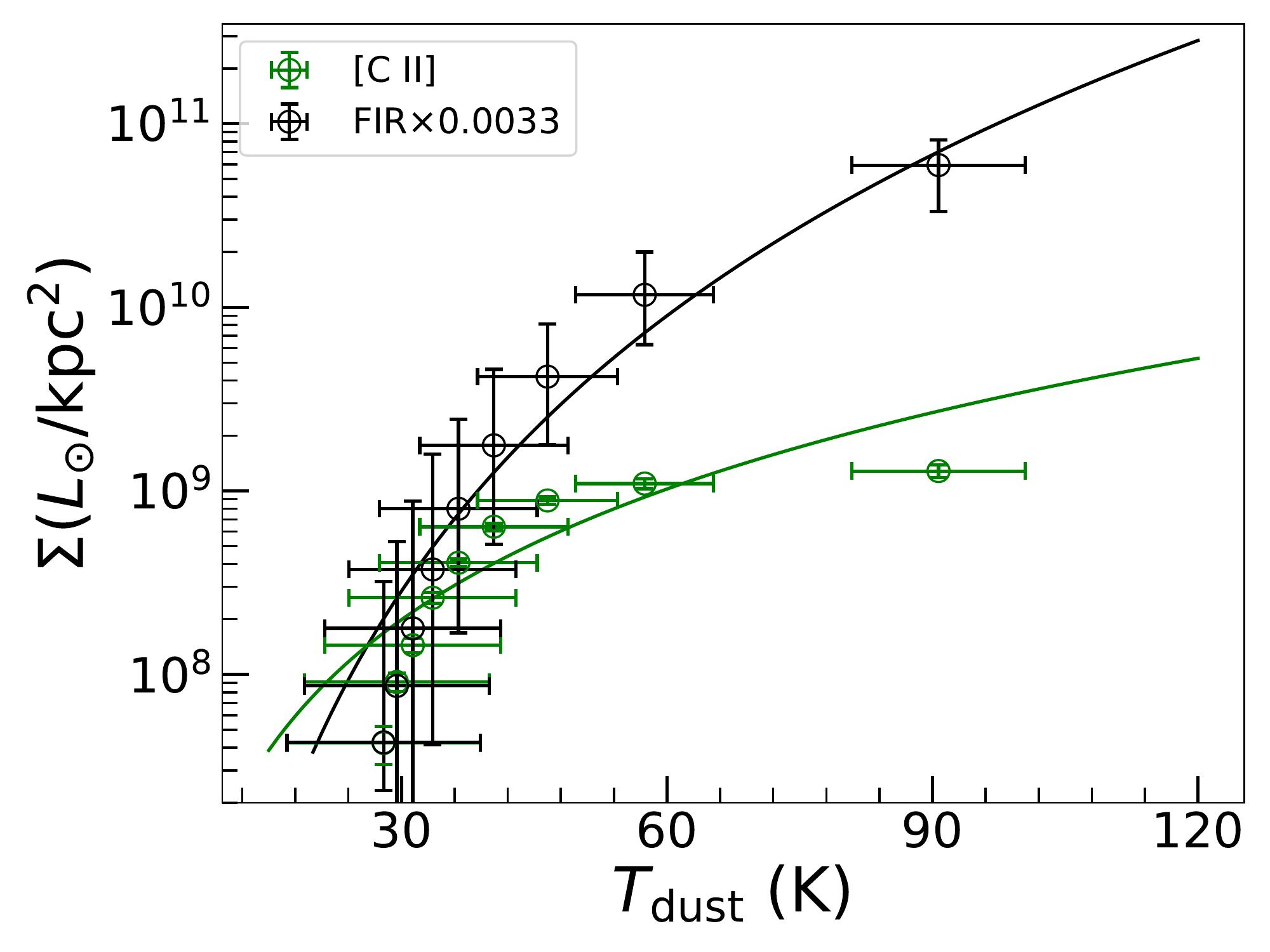}}  
\caption{
[\ion{C}{II}]-FIR deficit. Left panel: [\ion{C}{II}]/FIR luminosity ratio as a function of FIR surface brightness. The red open diamonds with error bars connected by the solid red line represent measurements for J2310+1855. The comparison samples are spatially resolved $z\sim6$ quasars (green triangles connected by dashed lines; \citealt{Wang2019}) and $z\sim3$ SMGs (purple triangles connected by dashed lines; \citealt{Rybak2019}), other $z>6$ quasars (yellow stars; \citealt{Wang2013}; \citealt{Venemans2016, Venemans2017837}; \citealt{Decarli2018}), other high-$z$ SMGs (pink triangles; \citealt{Riechers2013}; \citealt{Neri2014}; \citealt{Gullberg2018}; \citealt{Rybak2019}),  the GOALS sample of local luminous infrared galaxies (gray circles; \citealt{Diaz-Santos2013, Diaz-Santos2017}), and LBGs (green triangles; \citealt{Capak2015}; \citealt{Jones2017}; \citealt{Hashimoto2019lbg}).
Right panel: Comparison between the trends of the surface brightnesses of [\ion{C}{II}] (green  circles with error bars) and the scaled FIR (black circles with error bars) luminosities,  as a function of dust temperature. The green and black lines are the power-law fits to the $\Sigma_{\rm [C II]}$ and $\Sigma_{\rm FIR}$ as a function of $T_{\rm dust}$, respectively.}
\label{ciifir}
\end{figure*}

\subsubsection{$\Sigma_{M_{\rm gas}}$  and $\Sigma_{\rm SFR}$}

The Kennicutt-Schmidt (KS) power-law relation between the gas mass and the SFR surface density, $\Sigma_{\rm SFR}\sim\Sigma_{M_{\rm gas}}^{N}$, describes how efficiently galaxies turn their gas into stars, which enables us to understand  galaxy formation and evolution across cosmic time. The KS relation is nearly linear ($N\sim$1--1.5) in galaxies ranging from the local Universe to high redshifts ($z<4$; e.g., \citealt{Kennicutt1998}; \citealt{Bothwell2010}; \citealt{Genzel2010, Genzel2013}), indicating that the star formation processes may be independent of  cosmic time.  It is still unknown whether it holds for higher-redshift galaxies up to the reionization epoch.

In order to determine whether the spatially resolved ISM of J2310+1855 at $z=6.00$ deviates from the star formation law observed in local galaxies and high-$z$ star-forming galaxies, Fig. \ref{gsd} plots the SFR surface density as a function of the molecular gas mass surface density of the resolved ISM of J2310+1855 (black open circles and diamonds with error bars), and  of other populations from local to high-$z$ galaxies in the literature (references in the caption).
The molecular clouds where stars form are primarily composed of molecular hydrogen (H$_{2}$) and  helium. Given cosmological abundances, we consider the mass of the star-forming gas to be the mass of the molecular hydrogen times 1.36.  
We measure the gas mass surface density $\Sigma_{M_{\rm gas}}$ in each ring from the CO\,(9--8)  intensity map with a flux ratio between CO\,(9--8)  and CO (2--1) lines of $\sim$8 \citep{Shao2019} and assume  $L^{'}_{\rm CO(2-1)}  \approx L^{'}_{\rm CO(1-0)}$ (\citealt{Carilli2013}) and a conversion factor $\alpha_{\rm CO} \sim 0.37 \ M_{\odot}/(\rm K\, km/s\, { pc }^{ 2 })$ from the dynamical modeling in Sect. \ref{sec_rcd}. In addition, we also measure $\Sigma_{M_{\rm gas}}$ from the [\ion{C}{II}] intensity map with $\alpha_{\rm [\ion{C}{II}]}\sim3.2 \ M_{\odot}/L_{\odot}$ from the dynamical modeling in Sect. \ref{sec_rcd}. We similarly measure the SFR  surface density $\Sigma_{\rm SFR}$ from the resolved dust continuum  by converting the  IR luminosity to SFR using the formula in \citet{Kennicutt1998}.  Note that we subtracted the  IR luminosity contributed by the AGN torus as a point source (see Sects. \ref{sec_sb} and \ref{sec_ismorigin}).

Our SFR and gas surface densities are at the high end of  the  samples shown in Fig. \ref{gsd}, comparable to the SMGs at $2<z<4$.   The best-fit KS relation with $\Sigma_{M_{\rm gas}}$ measured from the CO\,(9--8) line for J2310+1855 is
\begin{equation}
\frac{\log_{10}\Sigma_{\rm SFR}}{M_{\odot}/\rm yr/kpc^{2}}=-20.50(\pm1.87) + 2.50(\pm0.21) \times \frac{\log_{10}\Sigma_{M_{\rm gas}}}{M_{\odot}/\rm kpc^{2}},
\end{equation}
and the best-fit KS relation with $\Sigma_{M_{\rm gas}}$ measured from the [\ion{C}{II}] line  is
\begin{equation}
\frac{\log_{10}\Sigma_{\rm SFR}}{M_{\odot}/\rm yr/kpc^{2}}=-10.59(\pm0.64) + 1.37(\pm0.07) \times \frac{\log_{10}\Sigma_{M_{\rm gas}}}{M_{\odot}/\rm kpc^{2}},
\end{equation}
for which we do not fit with the inner most measurement, as the observed [\ion{C}{II}] emission in the central of the galaxy is highly suppressed due to the high dust opacity (see Sect. \ref{sec_dis}).

The  power-law index $N=2.50\pm0.21$ measured from the CO\,(9--8) line is higher than that of local disk galaxies (e.g., \citealt{Kennicutt1998, Kennicutt2007}) and  high-redshift star-forming populations (e.g., \citealt{Genzel2013}; \citealt{Sharon2019}). This may indicate that this quasar host galaxy at $z=6.00$ is undergoing much faster star formation  than do galaxies in the local Universe and at lower redshifts.  
However, we should note that, as mentioned in other studies (e.g., \citealt{Daddi2010}; \citealt{Thomson2015})   the CO-to-H$_{2}$ conversion factor $\alpha_{\rm CO}$ can vary for different galaxy types, position scales, and  metallicities. We adopted the $\alpha_{\rm CO}$ value from our dynamical studies in Sect. \ref{sec_rcd}, which only represents the global value, and ignored any possible variation within the galaxy. This will bias the slope of the KS law we found in Fig. \ref{gsd}. In addition, the indirect measurements of the low-$J$ CO line transitions requires assuming a CO excitation model, which brings additional uncertainties when comparing  KS relations from different studies.  We simply use a bulk ratio between the ALMA  CO\,(9--8) emission and the VLA  CO\,(2--1) emission, which in reality will decrease with radius. This leads to an over-estimation of the slope of the KS law. Low-$J$ CO observations with higher spatial resolution will be needed to improve on this analysis.
And there appears to be an excess in CO\,(9--8) luminosities in distant starbursts due to for example shock excitation (e.g., \citealt{Riechers2021GADOT}), which would make the CO\,(9--8) line a poorer tracer of gas mass. 

The  power-law index  $N=1.37\pm0.07$ measured from the  [\ion{C}{II}] line is consistent with both the local and other high-redshift samples (e.g., \citealt{Bothwell2010}; \citealt{Carilli2010}; \citealt{Genzel2010}; \citealt{Thomson2015}). This may indicate that this quasar host galaxy at $z = 6$ in our work follows similar star formation process with that of the local Universe and lower redshifts. However, the value of the $\alpha_{\rm [\ion{C}{II}]}$ might be variable within the galaxy. The depletion times defined by $\Sigma_{M_{\rm gas}}/\Sigma_{\rm SFR}$ for the ISM of our quasar host galaxy are 1--100 Myr,  increasing from the inner region (corresponding to the highest $\Sigma_{M_{\rm gas}}$ and $\Sigma_{\rm SFR}$) to outer. The short gas consumption timescales suggest a rapid starburst mode for our quasar host galaxy.

\begin{figure*}
\centering
\subfigure{\includegraphics[width=0.8\textwidth]{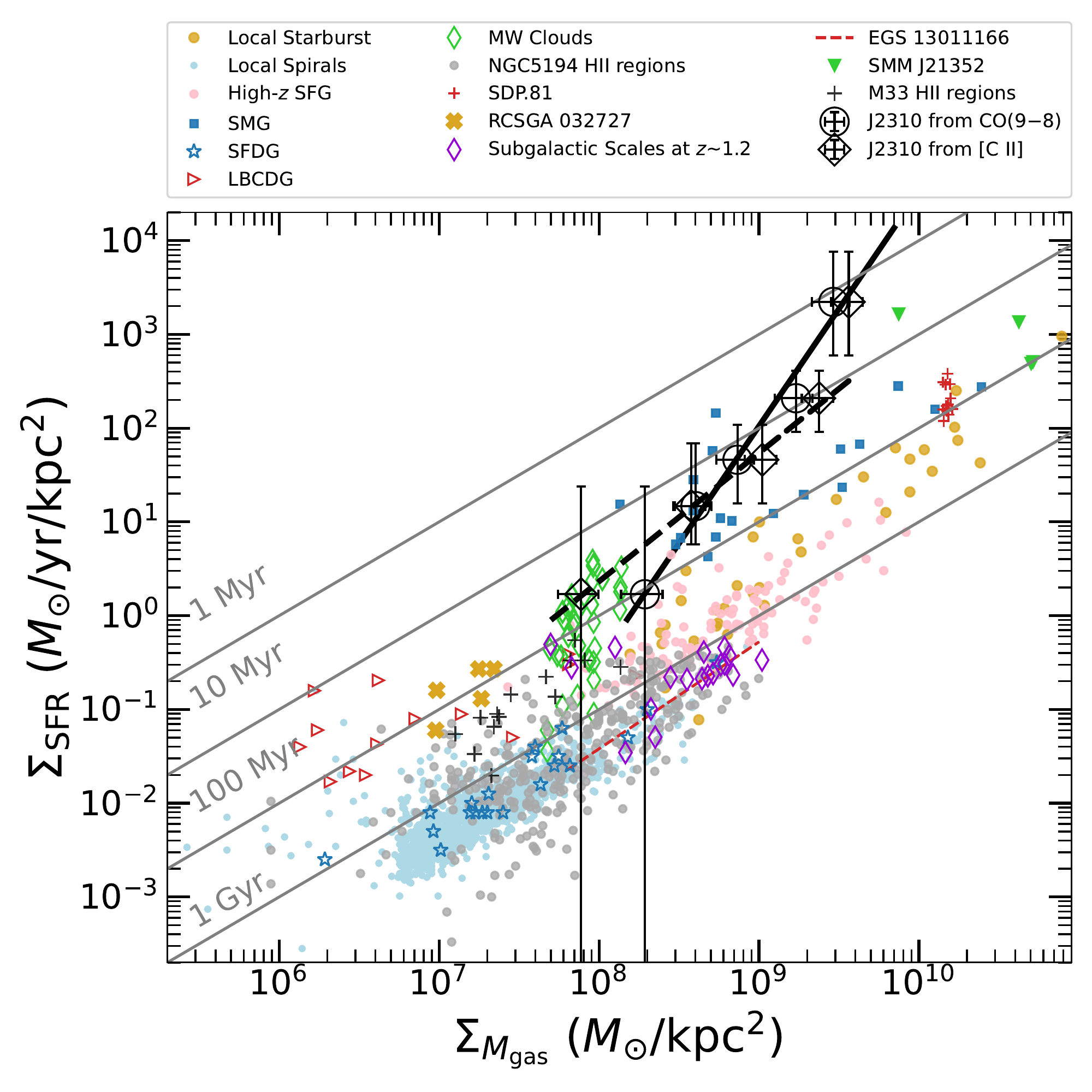}} 
\caption{Spatially resolved KS plot for J2310+1855 (black open circles with error bars, and its best-fit power-law shown as black solid line when calculating $\Sigma_{M_{\rm gas}}$ from the CO\,(9--8) line; black open diamonds with error bars, and its best-fit power-law shown as  dashed black  line when calculating $\Sigma_{M_{\rm gas}}$ from the [\ion{C}{II}] line), compared with the surface density KS relations of other galaxy samples including local starburst galaxies (brown filled circles; \citealt{Kennicutt1998}), local spiral galaxies (blue filled circles; \citealt{Kennicutt1998}; \citealt{Bigiel2010}), $z\sim$1--3 star-forming galaxies (pink filled circles; \citealt{Genzel2010}; \citealt{Freundlich2013}; \citealt{Tacconi2013}), $z\sim$2--4 SMGs (blue filled squares; \citealt{Bothwell2010}; \citealt{Genzel2010}; \citealt{Carilli2010}), star-forming dusty galaxies (blue open stars; \citealt{Villanueva2017}), local blue compact dwarf galaxies (red open triangles; \citealt{Amorin2016}), and some spatially resolved galaxies --  Milky Way clumps (green open diamonds; \citealt{Heiderman2010}; \citealt{Evans2014}), star-forming regions of  the spiral galaxy NGC 5194 (gray filled circles; \citealt{Kennicutt2007}), the lensed SMG SDP.81 at $z = 3.042$ (red pluses; \citealt{Hatsukade2015}), the young low-metallicity starburst galaxy RCSGA 032727 at $z=1.7$ (brown crosses; \citealt{Gonzalez-Lopez2017}), massive star-forming galaxies around $z = 1.2$ (purple open diamonds; \citealt{Freundlich2013}), the $z = 1.5$ star-forming galaxy EGS13011166 (dashed red line; \citealt{Genzel2013}), the lensed SMG SMMJ21352 at $z=2.3$ (green filled triangles; \citealt{Thomson2015}), and the \ion{H}{II} regions in the nearby spiral galaxy M33 (black crosses; \citealt{Miura2014}). The solid gray lines represent gas depletion timescales of 1Myr, 10 Myr, 100 Myr, and 1 Gyr. We applied a 1.36 correction factor to the molecular hydrogen mass for the presence of helium for the comparison samples. For our target, the conversion factor to the gas mass is from the dynamical modeling, which should include the contributions from all other elements (e.g., helium). When deriving the index of the KS power-law with the $\Sigma_{M_{\rm gas}}$ derived from the [\ion{C}{II}] line, we do not use the innermost measurement, as the observed [\ion{C}{II}] emission in the central of the galaxy is highly suppressed due to the high dust opacity (see Sect. \ref{sec_dis}).}
\label{gsd}
\end{figure*}

\subsection{The asymmetric line profile}
\label{sec_dip}

The integrated spectra of the gaseous disks of many galaxies both in the local Universe and at high redshift often have a characteristic double-horned shape (e.g., \citealt{Roberts1978}; \citealt{Walter2008}; \citealt{Shao2017}). Higher velocity dispersion and steeper density profiles of the gas tend to decrease the depth of the valley between the horns, and can even reduce the global profile  to flat-topped Gaussian shapes (e.g., \citealt{deBlok2014}; \citealt{Stewart2014}). In addition, the gaseous disk can be disturbed by interactions of galaxies with their neighbors and environments (e.g., \citealt{vanEymeren2011}; \citealt{Bok2019}; \citealt{Watts2020}; \citealt{Reynolds2020}), ram-pressure stripping (e.g., \citealt{Gunn1972}; \citealt{Kenney2015}) and gas accretion from the cosmic web (e.g., \citealt{Bournaud2005}). These distortions may drive morphological (e.g., the gas distribution) and kinematic (e.g., the velocity field, and the global spectral profile) asymmetries. We now turn to the spectral profile of J2310+1855 to understand it in the context of these ideas.

The [\ion{C}{II}] and CO\,(9--8) spectral profiles of J2310+1855 are asymmetric as shown in Figs. \ref{ciimap} and \ref{co98map}. The spectra are slightly enhanced on the red side. The enhancements ($[S_{\nu,\,\rm redpeak}-S_{\nu,\,\rm bluepeak}]/S_{\nu,\,\rm bluepeak}$; from the double-Gaussian fitting) are about $26\%\pm10\%$ and $18\%\pm14\%$ of the [\ion{C}{II}] and CO\,(9--8) lines, respectively. The two possible peaks are separated by  $213\pm9$ and $215\pm14$ km/s for the [\ion{C}{II}] and CO\,(9--8) lines, respectively. From our kinematic modeling of the [\ion{C}{II}] and CO\,(9--8) data cube in Sect. \ref{sec_kin}, we found a median gas dispersion of $54^{+6}_{-6}$ and $64^{+8}_{-9}$ km/s for the [\ion{C}{II}] and CO\,(9--8) lines, respectively (Table \ref{par_kin}). The ratios between the separation of the two peaks in the red and blue parts and the median velocity dispersion are $3.9^{+0.5}_{-0.5}$ and $3.3^{+0.5}_{-0.5}$ for the [\ion{C}{II}] and CO\,(9--8) lines, respectively. In addition, as shown in Sect. \ref{sec_kin}, the radial distribution of [\ion{C}{II}] and CO\,(9--8) lines are not steep (e.g., the S{\'e}rsic index is around 1). Thus, we are able to detect the double-horned profiles for both the [\ion{C}{II}] and CO\,(9--8) lines (we consider the two peaks to be resolved if the separation of the two peaks is twice the velocity dispersion). 
The redward enhancement of the spectra may be due to more material on one side of the galaxy, or temperature differentials from one side to another.

\subsection{Ionized and molecular gas kinematics}
\label{sec_gas_kin}

We studied the gas kinematics with both [\ion{C}{II}] and CO\,(9--8) lines using $^{\rm 3D}$B{\scriptsize{AROLO}} in Sect. \ref{sec_kin}. These two lines present consistent kinematic properties, including the gas disk geometry and the gas motions.  

\subsubsection{[\ion{C}{II}] versus CO\,(9--8)}
\label{sec_gas_kin_cii_co98}

The [\ion{C}{II}] and CO\,(9--8) emission trace similar gas disk geometries in our quasar J2310+1855 host galaxy. As shown in Table \ref{par_kin}, the inferred inclination angle and position angle are identical for the two lines. The modest inclination angle ($\sim$40$\degr$) makes J2310+1855 an ideal target to investigate the gas kinematics in the early Universe. The rotation speed as well as the velocity dispersion (discussed below) are also consistent between the [\ion{C}{II}] and CO\,(9--8) lines. This suggests that the [\ion{C}{II}] and CO\,(9--8) gas are coplanar and corotating in the host galaxy of quasar J2310+1855.

The rotation speed of the [\ion{C}{II}] line presented in the left panel of Fig. \ref{modelciico98} is roughly constant  ($\sim$250 km/s) with radius. However, the rotation speed of the CO\,(9--8) line at $\sim$1.2 kpc is  somewhat lower than that of the other two rings. 
The measured rotation speed of the [\ion{C}{II}] line is overall consistent with that of the CO\,(9--8) line, given that the errors in Fig. \ref{modelciico98} do not include the covariance with other fitting parameters.   
The  [\ion{C}{II}] and CO\,(9--8) data cubes used in the kinematic modeling in Sect. \ref{sec_kin} are from the same telescope and follow an identical data reduction process. They have a similar velocity resolution ($\sim$65 km/s), which is sufficient to sample the intrinsic width of both lines, and the line sensitivities are also comparable (i.e., $\rm S/N_{\rm peak}>10$). The only difference  is the angular resolution: that of the [\ion{C}{II}] data is roughly two times better than that of the CO\,(9--8) data. Considering that we measured the gas rotation speed in rings with widths that are half  the angular resolution for both lines, and that  $^{\rm 3D}$B{\scriptsize{AROLO}} performs well even when the galaxy is resolved with a small number of beams \citep{DiTeodoro2015}, the spatial resolution difference may not affect the inferred rotation curve significantly. Thus, the lower rotational speed of the $\sim$1.2 kpc ring  from CO\,(9--8) kinematical modeling may be due to  higher random motions of the CO\,(9--8) gas in that ring. 
The maximum rotation velocities ($\sim$250 km/s) of J2310+1855 in this work are smaller than that ($\sim$400 km/s) of quasar ULAS J131911.29+095051.4 at $z = 6.13$ in our previous work (\citealt{Shao2017}), but are at the high end of the rotation velocities of a sample of 27 $z\ge6$ quasars \citep{Neeleman2021}.
The velocity dispersions of the [\ion{C}{II}] and CO\,(9--8) lines have consistent median values of $54_{-6}^{+6}$ and $64_{-9}^{+8}$ km/s, respectively, as shown in the middle panel of Fig. \ref{modelciico98} and Table \ref{par_kin}. This is different from what is normally observed in nearby galaxies where the ionized gas (i.e., \ion{H}{$\alpha$}) velocity dispersions are substantially higher than the molecular gas (i.e., CO) velocity dispersions (e.g., \citealt{Fukui2009}; \citealt{Epinat2010}). However, because the angular resolution of these two lines are different, the velocity dispersions are measured over different volumes, and in the Milky Way, the velocity dispersion in molecular clouds is proportional to the size and mass of the clouds (e.g., \citealt{Heyer2015}). In addition, our velocity dispersions are much smaller than the average velocity dispersion (i.e., $129\pm10$ km/s) estimated using ALMA [\ion{C}{II}] data at a resolution of $\sim$$0\farcs25$ in a sample of  27 $z\ge 6$ undisturbed quasar host galaxies \citep{Neeleman2021}.

There is a general trend that the gas velocity dispersion increases with redshift or cosmic time (e.g., \citealt{Glazebrook2013}). The predicted velocity dispersion for $z=6$ galaxies is about $\sim$80 km/s following the empirical relationship for $0<z<4$ galaxies with measured velocity dispersions from either ionized or molecular gas \citep{Ubler2019}. The predicted velocity dispersion is consistent with our measured ones -- $36^{+11}_{-13}$--$80^{+8}_{-6}$ km/s for the [\ion{C}{II}] line and $54^{+9}_{-9}$--$98^{+6}_{-6}$ km/s for the CO\,(9--8) line. This may indicate that the luminous AGN in the center has little influence on  the velocity dispersion of the quasar host galaxy. The physical process to drive and maintain the velocity dispersion over cosmic time is still not clear. But  a constant energy input is required to retain the turbulence in the ISM, or it will decay within a few megayears as proposed by theoretical works (e.g., \citealt{Stone1998}; \citealt{MacLow1998}). This energy supply may be associated with either stellar feedback (i.e., winds; expanding \ion{H}{II} regions; supernovae) or other instability processes (i.e., clump formation; radial flows; accretion; galaxy interactions). The gas outflow along the line of sight traced by the blueshifted absorption of the OH$^{+}$ line would be very much in line with helping the upkeep of turbulence.

The rotation-to-dispersion ratio $V_{\rm rot}/\sigma$ is a measure of the kinematic support from ordered rotation (i.e., circular motions) versus random motions (i.e., noncircular; turbulence) in a system.  
We make two kinds of measurements of $V_{\rm rot}/\sigma$ in Table \ref{par_kin} -- the ratio between the rotation velocity and velocity dispersion in the flat part of the rotation curve (column 10) and the ratio between the maximum rotation velocity and the median velocity dispersion (column 11). The ratios are above 4 for both the [\ion{C}{II}] and CO\,(9--8) lines, which are higher than the cutoffs of a rotating system (i.e., a cutoff of 1 from \citealt{ForsterSchreiber2009}; \citealt{Epinat2009}; a cutoff of 3 from \citealt{Burkert2010}; \citealt{ForsterSchreiber2018}), indicating that the [\ion{C}{II}]  and CO\,(9--8)  gas  motion is dominated by rotation.  These $V_{\rm rot}/\sigma$ values are typical of other $z\sim6$ quasar host galaxies traced by resolved [\ion{C}{II}] in \citet{Neeleman2021}. Similarly, our measured $V_{\rm rot}/\sigma$ ratios are comparable to those of $z\sim2$ star-forming galaxies (e.g., \citealt{Genzel2008}; \citealt{ForsterSchreiber2009, ForsterSchreiber2018}; \citealt{Wisnioski2015}), and $z\sim4.5$ gravitational lensed dusty star-forming galaxies (e.g., \citealt{Rizzo2021}).

The inclination angle measured from our kinematic modeling is larger than the one (25$\degr$) derived by \citet{Tripodi2022}, who also used $^{\rm 3D}$B{\scriptsize{AROLO}}. They constrain the inclination angle $<$30$\degr$ considering that $M_{\rm dyn}$ should be larger than $M_{\rm H_{2}}$, which is derived from the CO\,(2--1) emission line \citep{Shao2019} and adopting an $\alpha_{\rm co}$ value of 0.8 $M_{\odot}/(\rm K\, km/s\, { pc }^{ 2 })$. As noted in Sect. \ref{sec_rcd}, the values of $\alpha_{\rm co}$ may differ from source to source (e.g., \citealt{Mashian2015}), and based on our dynamical model, we derive an $\alpha_{\rm co}$ value of 0.37 for J2310+1855. The loose constraint on $\alpha_{\rm co}$ (i.e., we adopt a wide value range of 0.2--14; \citealt{Mashian2015}) allows for a more flexible fitting of the inclination angle in the $^{\rm 3D}$B{\scriptsize{AROLO}} modeling. The initial guess of the inclination angle of 40$\degr$ in the $^{\rm 3D}$B{\scriptsize{AROLO}} modeling is from the ratio between the photometric minor and major source sizes (which are roughly along the kinematic minor and major axes, respectively) of the CO\,(9--8) emission assuming an intrinsic round gas disk (before inclination). However, the observed kinematic minor source size is larger than the kinematic major source size (after inclination) for the [\ion{C}{II}] line, which indicates that the intrinsic kinematic minor source size of [\ion{C}{II}] must be larger than its intrinsic kinematic major source size (before inclination). As we discuss above, the [\ion{C}{II}] and CO\,(9--8) gas are coplanar and corotating in the host galaxy of quasar J2310+1855. Thus, [\ion{C}{II}] and CO\,(9--8) gas are well mixed. Therefore, the  intrinsic kinematic minor source size of CO\,(9--8) might be larger than its intrinsic kinematic major source size. As a result, from the kinematical aspect, the inclination angle should be larger than 40$\degr$, or conservatively the inclination angle cannot be much below 40$\degr$ considering a 5$\%$ uncertainty of the initial guess of the inclination angle.

\subsubsection{Outflow traced by OH$^{+}$}
\label{outflow}

The  OH$^{+}$ line has been detected in emission, absorption and  P-Cygni-shaped profiles in galaxies (e.g.,  \citealt{vanderWerf2010}; \citealt{Rangwala2011}; \citealt{Spinoglio2012}; \citealt{Gonzalez-Alfonso2013}; \citealt{Riechers2013, Riechers2021GADOT, Riechers2021}; \citealt{Berta2021}; \citealt{Butler2021}; \citealt{Stanley2021}).  The production of OH$^{+}$ is mainly driven by $\rm H^{+} + O \rightarrow O^{+} + H$ followed by $\rm O^{+} + H_{2} \rightarrow OH^{+} + H$ and $\rm H^{+} + OH \rightarrow OH^{+} + H$ (e.g.,  \citealt{vanderWerf2010}). The  OH$^{+}$ is in a unstable state, and it reacts rapidly with H$_{2}$ molecules following the main chemical reactions: $\rm OH^{+}+H_{2}\rightarrow H_{2}O^{+}+H$ and then $\rm H_{2}O^{+} + H_{2} \rightarrow H_{3}O^{+}+H$ (e.g., \citealt{Gerin2010}). The key initiation for this simple chemical sequence is the production of H$^{+}$, and it is argued that the atomic H ionization is most likely dominated by  X-rays and/or  cosmic rays (e.g., \citealt{Gonzalez-Alfonso2013}).  
The OH$^{+}$ emission shows double-horned profile with comparable intensities of $\sim0.30\pm0.05$ mJy of the two peaks. Our measured luminosities of the OH$^{+}$ emission and FIR follow the weak correlation between these two parameters found by \citet{Riechers2021GADOT}. 
At $-384\pm128$ km/s (relative to the [\ion{C}{II}] redshift from our ALMA Cycle 0 observations; \citealt{Wang2013}), we detect a $\sim$3$\sigma$ dip with a peak value of $-0.18\pm0.05$ mJy. This absorption velocity shift is at the high end of  the OH$^{+}$ absorption velocity shifts  in the range of $\sim$130--360 km/s of $z$=2--6 starburst galaxies \citep{Riechers2021GADOT}, and is comparable to the widths of the [\ion{C}{II}] and CO\,(9--8) lines of J2310+1855. This P-Cygni profile (i.e., blueshifted absorption and redshifted emission) may indicate the outflow of gas along the line of sight. The outflow signature of broad blue-shifted absorption line of J2310+1855 is recently reported by \citet{Bischetti2022} through VLT/X-shooter observations. The OH$^{+}$ absorption and emission components are co-spatial (left panel of Fig. \ref{ohmap}). This is different from, for example, a hyperluminous dusty starbursting major merger ADFS-27 in \citet{Riechers2021}, where the offset is $>$1 kpc. 

In order to further understand the outflow traced by OH$^{+}$ absorption, we estimate the outflow mass with the observed OH$^{+}$ absorption spectrum and its best-fit Gaussian model. The optical depth of an unsaturated absorption line can be calculated with $\tau_{\rm line}=-\ln(f_{\rm trans})$, where $f_{\rm trans}$ is the fraction of the continuum emission that is still transmitted, which can be expressed as $(S_{\nu_{\rm cont}}+S_{\nu_{\rm line}})/S_{\nu_{\rm cont}}$, where $S_{\nu_{\rm line}}$ is the flux density of the continuum subtracted OH$^{+}$ absorption, and $S_{\nu_{\rm cont}}$ is the flux density of the continuum emission for which we use a constant value of 1.53 mJy for the whole line. We measure the peak optical depth $\tau_{\rm OH^{+}}=0.13\pm0.04$. The velocity-integrated optical depths $\tau_{\rm OH^{+}}$d$v$ are $35\pm7$ and $44\pm31$ km/s by integrating the OH$^{+}$ absorption region  and the best-fit Gaussian region of the  OH$^{+}$ absorption, respectively. The OH$^{+}$ column density $N_{\rm OH^{+}}$ is related to the  velocity-integrated optical depth $\tau_{\rm OH^{+}}$d$v$ through $\tau_{\rm OH^{+}}{\rm d}v=\lambda^{3}AN_{\rm OH^{+}}/8/\pi$, where $A$ is the Einstein coefficient with a value of $2.11\times10^{-2}$ s$^{-1}$ (e.g., \citealt{Butler2021}), and $\lambda$ is the rest frame wavelength of the OH$^{+}$ line (i.e., 0.029 cm). We thus derive $N_{\rm OH^{+}}$ to be $(1.7\pm0.3)\times10^{14}$ and  $(2.2\pm1.5)\times10^{14}$ cm$^{-2}$, respectively. Then assuming a relative abundance of $\log_{10}(N_{\rm OH^{+}}/N_{\rm H})=-7.8$ (e.g., \citealt{Bialy2019}), we measure the  neutral hydrogen column density $N_{\rm H}$ of $(1.1\pm0.2)\times 10^{22}$ and $(1.4\pm0.9)\times 10^{22}$ cm$^{-2}$, respectively. With a source radius of 1.3 kpc (the effective radius of CO\,(9--8) line in Table \ref{par_ima}), we thus derive the neutral hydrogen mass and neutral gas mass (i.e., with a correction factor of 1.36 for the presence of helium) to be $(4.6\pm0.9)\times10^{8}$ and $(6.2\pm1.2)\times10^{8}$, and $(5.8\pm4.0)\times10^{8}$ and $(7.9\pm5.5)\times10^{8}$ $M_{\odot}$, respectively. The neutral gas mass of the outflow is $3\pm1\%$ and $4\pm3\%$ of the molecular gas mass in the host galaxy, respectively.
Our neutral hydrogen mass is a few times smaller than  the median value of $\sim2\times10^{9} M_{\odot}$ for a sample of 18 $z$=2--6 starburst galaxies that are also traced by OH$^{+}$ absorption \citep{Riechers2021GADOT}.

Finally, we do not find evidence of gas outflow in the spectra of [\ion{C}{II}] or CO\,(9--8) emission lines in the form of high-velocity wings (see Figs. \ref{ciimap} and \ref{co98map}). However, \citet{Tripodi2022} reported the presence of high-velocity components on both the red and blue sides of the  [\ion{C}{II}] emission line in J2310+1855. This discrepancy might be due to differences in the sensitivities and/or angular resolution of the data used in the analysis. We have better angular resolution ($0\farcs11\times0\farcs09$ versus $0\farcs17\times0\farcs15$), although the sensitivities are comparable (0.17 mJy/beam per 17 km/s versus 0.23 mJy/beam per 8.5 km/s). To understand this discrepancy, we used a nature weighting (robust=2.0) to {\footnotesize{TCLEAN}} our [\ion{C}{II}] data, which give more weights on short baselines. The resulted clean beam size is $0\farcs16\times0\farcs13$, which is similar to the one of \citet{Tripodi2022}, and a  sensitivity of 0.15 mJy/beam per 17 km/s. The residual spectrum between the observed one and the best-fit double-Gaussian model is only noise-like, with an rms of 0.70 mJy.  The reality of the high-velocity wings reported by \citet{Tripodi2022} is therefore doubtful. As shown in previous studies, the [\ion{C}{II}] emission line has proven to be a difficult tracer of outflow activity (e.g., \citealt{Meyer2022}) and, as shown by  \citet{Spilker2020}, strongly lensed dusty star-forming galaxies at $z>4$ that show clear OH outflow do not display high-velocity wings in their [\ion{C}{II}] emission lines. 
In addition, our CO\,(9--8) data does not have enough S/N to look for broad line wings.

\subsection{The dynamics of the quasar-host system}
\label{sec_mbhmdyn}

The dynamical modeling of the rotation curves can yield detailed knowledge of the composition and distribution of matter in the quasar-host system. In Sect. \ref{sec_rcd}, we decomposed the J2310+1855 quasar-host system into four dynamical components (black hole, stars, gas, and dark matter) and modeled the potential contribution of each component to the measured circular velocities of the high-resolution [\ion{C}{II}] data cube. 

\subsubsection{The dynamical structures}
\label{dsds}

The black hole mass is a physical parameter that is critical for understanding the coevolution of the central SMBHs and their host galaxies from the local to the high-redshift Universe (e.g., \citealt{Ho2013}; \citealt{Shen2013}).
The mass of the black hole of the quasar J2310+1855 at $z=6$ is on the order of $10^{9}\,M_{\odot}$ (\citealt{Jiang2016}; \citealt{Feruglio2018}), which is measured using the width of NIR emission lines and relationships calibrated with low-redshift AGN. The black hole component is important to shape the gravitational potential in the central region of the host, contributing $\sim$120 km/s (for $M_{\rm BH}=10^{9}\,M_{\odot}$) at a radius of 0.29 kpc (the innermost position we can reach with our current resolution), where we measured a circular velocity of $\sim$250 km/s. 
For our target, the value of $R_{\rm e;\,star}$ is strongly covariant with the black hole mass and the inferred physical parameters of the stellar component. As shown in Fig. \ref{rcd} and Table \ref{par_dyn}, a factor of 2 change in the black hole mass causes a factor of 4 change in the inferred stellar mass. 
If we consider an extreme case without a central black hole, the stellar profile needs to become strongly centrally concentrated, with a stellar effective radius of $0.21^{+0.55}_{-0.17}$ kpc to make up for the black hole (as shown in the right panel of Fig. \ref{rcd} and column (c) of Table \ref{par_dyn}). There is no direct observations to constrain the distribution of the stellar component in this quasar host galaxy. A highly concentrated stellar component might be possible, and thus we cannot definitely remove such an extreme case without a central SMBH. When making the black hole mass a free parameter, we confirm that the black hole mass is on the order of $10^{9}\,M_{\odot}$; this is the first time that a black hole mass has been dynamically measured at $z\sim6$.  The evidence of the existence of a central SMBH is marginal, and we need more and better data to verify (i.e., \textit{James Webb} Space Telescope observations to constrain the stellar property, and higher-resolution ALMA [\ion{C}{II}] observations).  However, we can built a strong case for the existence of a $>10^{9}\,M_{\odot}$ black hole from our dynamical modeling, unless the stellar distribution in the case without a black hole is strange (but not impossible), as do the other line width metrics (e.g., \citealt{Jiang2016}; \citealt{Feruglio2018}). 
In either scenario (making the black hole mass free or fixed) the gas mass stays stable, as the effective radius of the gas component is fixed based on our ALMA CO\,(9--8) observations and the gas mass is mainly determined by the circular velocities in the outer regions.

The mass and size of the stellar component can provide essential insights into galaxy evolution  (e.g., \citealt{Hodge2016}). The stellar mass and size correlation of galaxies across  cosmic time provides an important view of the assembly history of galaxies (e.g., \citealt{Kawinwanichakij2021}). However, for quasar host galaxies at $z\sim6$, it is very difficult to determine these stellar properties through  rest frame optical and NIR imaging, as the starlight is overwhelmed by the central quasar. Thus, the comparison with the $z\sim6$ objects on the stellar mass-size plane is currently not feasible. This also leads to limitations on our dynamical modeling including a stellar component of the quasar-host system in Sect. \ref{sec_rcd}. Our modeled stellar mass (around $10^{9}M_{\odot}$) has a large uncertainty.  
It has been found that not all stellar components are necessarily co-spatial with the gas/dust for SMGs. Thus, the total $M_{\rm star}$ may be under-predicted with the dynamical modeling based on the kinematics of the gas (e.g., \citealt{Gomez-Guijarro2018}).
High-resolution stellar continuum emission observations toward our target and similar luminous quasars in the early Universe by for example the  \textit{James Webb} Space Telescope (JWST) are critical to investigate the $M_{\rm star}-R_{\rm e; \, star}$ relation and the role of these $z\sim6$ quasars in galaxy evolution. 
The stellar effective radius $R_{\rm e;\,star}$ (in Table \ref{par_dyn}) is roughly similar to our measured CO and [\ion{C}{ii}] effective radii, but larger than the dust continuum effective radius, albeit with large uncertainties (in Table \ref{par_ima}). This is consistent with the simulated TNG50 star-forming galaxies \citep{Popping2022}.  
Figure \ref{rcd} shows dynamical evidence for a massive ($\sim$$10^{9}M_{\odot}$) stellar component, especially when we fix the black hole mass to be $1.8\times10^{9}M_{\odot}$. In this case, the contribution to $V_{\rm c}$ from stars rises monotonically to the center of the galaxy, as is typical for  nearby massive spiral galaxies (e.g., \citealt{Simard2011}; \citealt{Gao2019, Gao2020}), and such a stellar bulge signature has already been proven to exist using a dynamical method in some dusty star-forming galaxies at $z\sim4-5$  (e.g., \citealt{Rizzo2020, Rizzo2021}; \citealt{Tsukui2021}).

The gas component is well constrained by our ALMA CO\,(9--8) image decomposition and the dynamical modeling  (see Sects. \ref{sec_sb} and \ref{sec_rcd}). The gas fraction ($f_{\rm gas}=M_{\rm gas}/M_{\rm bary}$ and $M_{\rm bary}=M_{\rm star}+M_{\rm gas}$) is $>$70$\%$, and thus it dominates the quasar-host system. The large reservoir of  molecular gas ($M_{\rm gas} \sim 10^{10}M_{\odot}$) is responsible for fueling the bulk of the star formation in the central region of the quasar host galaxy. We should point out that the scale radius of the gas component in the dynamical modeling is from the CO\,(9--8) intensity map assuming an exponential gas distribution. However, the overall molecular gas (e.g., traced by CO\,(1--0)) might be more extended than that traced by high-$J$ CO. We adopt a 1.5 times larger gas scale radius to do the dynamical modeling, and get an $\alpha_{\rm CO}$ value of $0.73^{+0.16}_{-0.24}$.

The dark matter component is less constrained. The mass of the dark matter halo  is 16$^{+26}_{-12}\%$ and 9$^{+28}_{-8}\%$ of the total dynamical mass of the quasar-starburst system  within a radius of two times the gas half-light radius, when making the black hole mass free and fixed, respectively. The virial velocity and radius of the dark matter halo both have large uncertainties as shown in Table \ref{par_dyn}. The virial radius of the dark matter halo is much more extended than the gas distribution as shown in Table  \ref{par_ima}, which is consistent with numerical simulations (e.g., \citealt{Khandai2012}).

\subsubsection{The coevolution of the quasar and its host galaxy}
\label{co-evolution}

To study the coevolution of  central black holes and their host galaxies, we typically investigate the evolution of the $M_{\rm BH}-\sigma$ and the $M_{\rm BH}-M_{\rm bulge}$ relations across cosmic time (e.g., \citealt{Ho2013}), where  $\sigma$ and $M_{\rm bulge}$ are the velocity dispersion and total mass of the bulge, respectively.

Considering that  the stars and molecular gas trace the same underlying potential, and following \citet{Davis2013}, we assume that the second moments of the stars and  molecular gas are equal. Thus, we can derive a simple zeroth-order estimate of the velocity dispersion:

\begin{equation}
V_{\rm stars}^{2} + \sigma_{\rm stars}^{2} = V_{\rm gas}^{2} + \sigma_{\rm gas}^{2},
\label{stellarvd}
\end{equation}
where $V_{\rm gas}$ and $ \sigma_{\rm gas}$ are the velocity and velocity dispersion of the molecular gas, for which we use the maximum circular velocity ($\sim$245 km/s; the maximum rotation velocity corrected by the asymmetric drift) and the median velocity dispersion ($\sim$64 km/s) of the CO\,(9--8) line shown in Table \ref{par_kin} as proxies. $V_{\rm stars}$ is the stellar velocity, for which we use the modeled stellar velocity (40 and 99 km/s when making the black hole mass free and fixed, respectively) at the half-mass radius of the stellar component (1.76 and 1.39 kpc when making the black hole mass free and fixed, respectively). The predicted stellar velocity dispersions ($\sigma_{\rm stars}$) are 250 and 233 km/s where we free or fix the black hole mass, respectively. Based on the local relationship for nearby galaxies from \citet{Ho2013}, we predict a bulge velocity dispersion of $\sim$$335\pm20$ and $\sim$$300\pm16$ km/s for a central black hole of mass  $2.97\times10^{9}$ and $1.8\times10^{9}M_{\odot}$ (fixed value in the dynamical modeling), respectively. These bulge velocity dispersions from the local relationship are higher than the predicted stellar velocity dispersions including the scatter in the local relationship, also when we adopt the [\ion{C}{II}] kinematic parameters. This may indicate that the black hole evolves faster than the host galaxy of J2310+1855. 

We present the baryonic mass $M_{\rm bary}$ from the best-fit dynamical modeling and the ratio of $M_{\rm BH}/M_{\rm bary}$ between the black hole mass and the baryonic mass as a function of radius in Fig. \ref{ratio}. Inside a radius that is two times the half-light radius  of the CO emission ($\sim$2.6 kpc), the measured ratio of $M_{\rm BH}/M_{\rm bary}$ is $\sim$$28^{+12}_{-9}$ and $\sim$$16^{+4}_{-2}$ times higher than the local relationship when making the black hole mass free and fixed in the dynamical modeling, respectively. If we consider a stellar-spheroidal component (i.e., the stellar component in our dynamical modeling in Sect. \ref{sec_rcd}), the $M_{\rm BH}/M_{\rm star}$ will be $>$50 times higher than the local relationship \citep{Ho2013}. This may indicate that the central SMBH grows the bulk of its mass before the formation of most of the stellar mass in this quasar host galaxy in the early Universe.  \citet{Wang2013} and \citet{Tripodi2022} also found the same result for  quasar J2310+1855. This is consistent with what has been found for other high-luminosity  $z\sim6$ quasars (e.g., \citealt{Wang2019}).

However, as we discussed in Sect. \ref{dsds}, the black hole mass and the parameters (e.g., the mass, the effective radius, and the S{\'e}rsic index) of the stellar component are somewhat degenerate in the dynamical modeling. Thus, the modeled black hole mass,  stellar mass and  stellar velocity adopted in the discussion above have substantial and coupled uncertainties to the investigation of the quasar-host coevolution. Higher angular resolution and sensitivity observations toward the gas and stellar components are needed to better constrain these parameters with dynamical modeling.  In addition,  Eq. \ref{stellarvd}, which we used to predict the stellar velocity dispersion comes with additional uncertainties. \citet{Davis2013} have compared the predicted stellar velocity dispersion using this equation with the measured values. Some data points with stellar velocity dispersion similar to the circular velocity (which is fairly common in elliptical galaxies) appear as obvious outliers, which may indicate that the zeroth order approximations may break down in this situation. 

\begin{figure}
\centering
\subfigure{\includegraphics[width=0.5\textwidth]{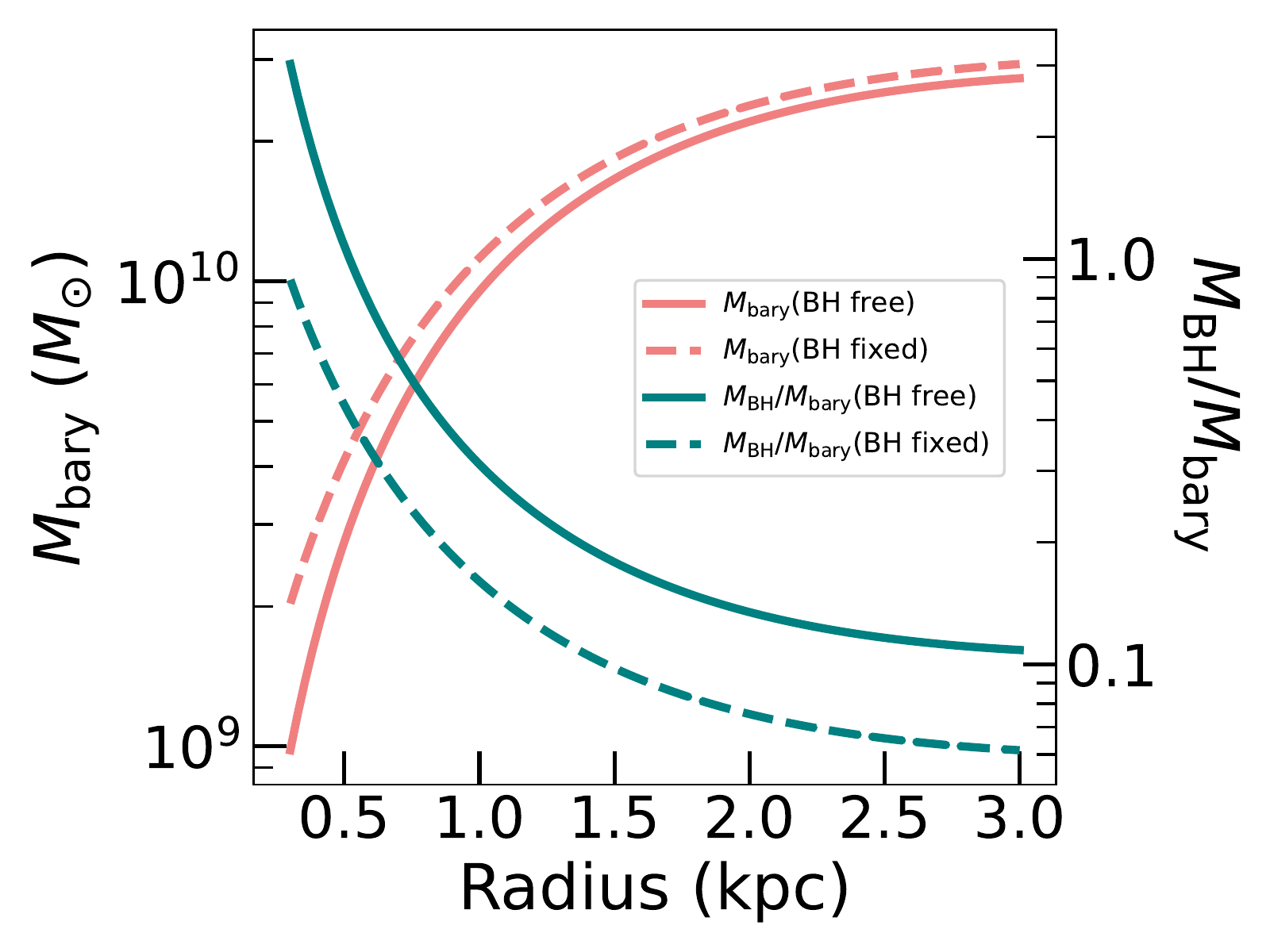}} 
\caption{Baryonic  mass $M_{\rm bary}$ (red lines) measured from the dynamical modeling in Sect. \ref{sec_rcd} as a function of radius, and the corresponding ratio (green lines) between the black hole mass  $M_{\rm BH}$ and the baryonic  mass $M_{\rm bary}$. The solid and dashed lines represent the scenarios in which the black hole mass is free and fixed during the dynamical modeling, respectively.}
\label{ratio}
\end{figure}

\section{Summary} 
\label{sec_sum}
We have reported on ALMA high-resolution (sub-kiloparsec- to kiloparsec-scale) observations of the [\ion{C}{II}], CO\,(9--8), and OH$^{+}$\,($1_{1}$--$0_{1}$) lines and their underlying dust continuum toward the FIR luminous quasar SDSS J2310+1855 at $z = 6.0031$. We have investigated the ISM distributions, the gas kinematics, and the quasar-host system dynamics with various models. Below are the main results: 
 
\begin{itemize}

\item The [\ion{C}{II}] and CO\,(9--8) lines and their underlying dust continuum are all spatially resolved. The OH$^{+}$ emission is a point source, and the emission peak is only  at a level of $\sim$5$\sigma$. The line widths of the [\ion{C}{II}] and CO\,(9--8) emission are consistent, which suggests that these two lines  trace the same gravitational potential. We observed asymmetric spectral line profiles for the [\ion{C}{II}] and CO\,(9--8) lines. The enhanced intensity of the lines toward the red part of the spectrum may be due to an asymmetric distribution of gas, or because the temperatures are higher on one side of the galaxy.

\item We used 2D elliptical S{\'e}rsic models to reproduce the observed intensity maps of the [\ion{C}{II}] and CO\,(9--8) lines and the dust continuum. The extended FIR dust continuum and the CO\,(9--8) emission can be fitted with a S{\'e}rsic distribution with a S{\'e}rsic index of $\sim$1 as well as an additional central point component. The point and extended components may represent emission from an AGN dusty molecular torus and the quasar host galaxy, respectively. The derived flux ratio from the AGN dust torus and the star-forming activity for the dust emission  are consistent with our rest-frame UV-to-FIR SED decomposition result. The [\ion{C}{II}] emission follows a flatter distribution (S{\'e}rsic index of 0.59), which may be due to high dust opacity increasing the FIR background of the [\ion{C}{II}] line and reducing the [\ion{C}{II}] emission at smaller radii. 
The dust temperature drops with distance from the center. The effective radius of the dust continuum is smaller than that of the line emission and dust mass surface density, but is consistent with that of the SFR surface density. This may indicate that dust emission is a less robust tracer of the dust and gas distribution but is a decent tracer of the obscured star formation activity. 
The effective radius of the [\ion{C}{II}] underlying dust continuum is consistent within $\sim$10$\%$ with that of the CO\,(9--8) underlying dust continuum. The two wavelengths are very close and shortward of the Rayleigh-Jeans tail; therefore, both are dominated by dust at similar temperatures. The half-light radius of the [\ion{C}{II}] line is smaller than that of the CO\,(9--8) line,  which may be  a reflection of the radial dependence of the gas excitation and the radiation field.  
A larger [\ion{C}{II}]-FIR deficit is observed in the center compared to the outer region, which is likely due to a higher dust temperature and higher dust opacity in the galaxy center. 
The spatially resolved KS relation ($\Sigma_{\rm SFR}\sim\Sigma_{M_{\rm gas}}^{2.50\pm0.21}$) with $\Sigma_{M_{\rm gas}}$ measured from the CO\,(9--8) line of J2310+1855 is steeper than in the local Universe and $z\sim2-4$ starburst samples. However, when $\Sigma_{M_{\rm gas}}$ is measured from the [\ion{C}{II}] line, the relation ($\Sigma_{\rm SFR}\sim\Sigma_{M_{\rm gas}}^{1.37\pm0.07}$) is consistent with the local and low-redshift samples.

\item A rotating gas disk is observed in both the [\ion{C}{II}] and CO\,(9--8) lines,  as seen from the obvious gradients in the velocity maps and the ``S'' shaped position-velocity diagrams. We applied 3D tilted ring models to both line data cubes with $^{\rm 3D}$B{\scriptsize{AROLO}}. These two lines show similar kinematical signatures: the gas disk geometry (i.e., consistent inclination angle and position angle of the gas disk) and the gas motions (i.e., rotation dominated). The rotation velocities and velocity dispersions traced by [\ion{C}{II}] and CO\,(9--8) lines are consistent. The circular velocities of [\ion{C}{II}] and CO\,(9--8) lines are in excellent agreement, which is consistent with their identical line width, indicating that they trace the same gravitational potential. We suggest that the [\ion{C}{II}] and CO\,(9--8) gas are coplanar and corotating in this quasar host galaxy. The velocity dispersions and the $V_{\rm rot}/\sigma$ ratio of our quasar host galaxy are comparable with other high-$z$ populations, for example  $z\ge6$ quasar hosts and  $z\sim2$ star-forming galaxies. The OH$^{+}$ line shows a P-Cygni profile with an absorption at $\sim$--400 km/s, which may indicate an outflow with a neutral gas mass of $(6.2\pm1.2)\times10^{8} M_{\odot}$ along the line of sight.

\item For the purpose of quantifying the dynamical contributions from different matter components, we decomposed the circular rotation curve measured from the high-resolution [\ion{C}{II}] line into four components (black hole, stars, gas, and dark matter). The whole quasar-host system appears baryonic matter dominated. The gas component is well constrained, which reveals a large quantity of gas with a mass on the order of $10^{10}M_{\odot}$. The stellar and dark matter components are not well constrained and depend on whether we allow the black hole mass to be a free parameter. We measured the black hole mass to be $2.97^{+0.51}_{-0.77}\times 10^{9}\,M_{\odot}$. This is the first time that a SMBH mass has been measured dynamically at $z\sim6$. This result is consistent with that measured from \ion{Mg}{II} and \ion{C}{IV} lines with the local scaling relations. The stellar component has a dynamical mass measured on the order of $10^{9}\,M_{\odot}$ when the Universe was only $\sim$0.93 Gyr old. This result is more robust when we fix the black hole mass to be $1.8\times 10^{9}\,M_{\odot}$ from the quasar spectrum. We note that we may miss the  stellar components outside the gas and dust-emitting regions. The relations between the black hole mass and  the stellar velocity dispersion and the baryonic mass of this quasar indicate that the central SMBH evolved earlier than its host galaxy.
\end{itemize}

In the future we plan to improve our understanding of the ISM distribution, gas kinematics, and system dynamics of quasars at the earliest epoch, with higher-resolution observations toward various rest-frame FIR ISM tracers with ALMA and toward the rest-frame optical and NIR stellar emission with the JWST.

\begin{acknowledgements}
We would like to dedicate this paper to the memory of Dr. Yu Gao. Dr. Gao passed away on 21 May 2022. During his career, Dr. Gao made numerous key contributions to the physics of star formation and the interstellar medium in galaxies; in particular, his work  established the star formation law in terms of dense molecular gas content. Dr. Gao left a lasting legacy as a brilliant scientist, as a most respected colleague and caring mentor, and above all, a true gentleman. Dr. Gao will always be in our hearts and thoughts as a truly remarkable human being.
We would like to thank the anonymous reviewer for all valuable comments and suggestions, which helped us to improve the quality of the manuscript.
This work makes use of the following ALMA data: ADS$/$JAO.ALMA$\#$ 2018.1.00597.S; 2015.1.01265.S; 2015.1.00997.S and 2011.0.00206.S. ALMA is a partnership of ESO (representing its member states), NSF (USA) and NINS (Japan), together with NRC (Canada), MOST and ASIAA (Taiwan), and KASI (Republic of Korea), in cooperation with the Republic of Chile. The Joint ALMA Observatory is operated by ESO, AUI$/$NRAO and NAOJ. The National Radio Astronomy Observatory is a facility of the National Science Foundation operated under cooperative agreement by Associated Universities, Inc. 
R.W. acknowledges supports from the National Science Foundation of China (NSFC) grants No. 11991052, 11721303, and 11533001. Y.S. acknowledges discussions with Francesca Rizzo, Gang Wu, Aiyuan Yang and Sudeep Neupane.

\end{acknowledgements}

\clearpage
\bibliographystyle{aa} 
\bibliography{mybib}

\onecolumn
\begin{appendix}

\section{The [\ion{C}{II}] channel maps}
\label{sec_channel}
We present the [\ion{C}{II}] and CO\,(9--8) channel maps of J2310+1855 in Figs. \ref{ciichannel} and \ref{co98channel}. The [\ion{C}{II}]  emission peak moves in a circular, counterclockwise path from 219 to $-$281 km/s, a signature of a rotating gas disk.

\begin{figure*}[h]
\centering
\subfigure{\includegraphics[width=0.99\textwidth]{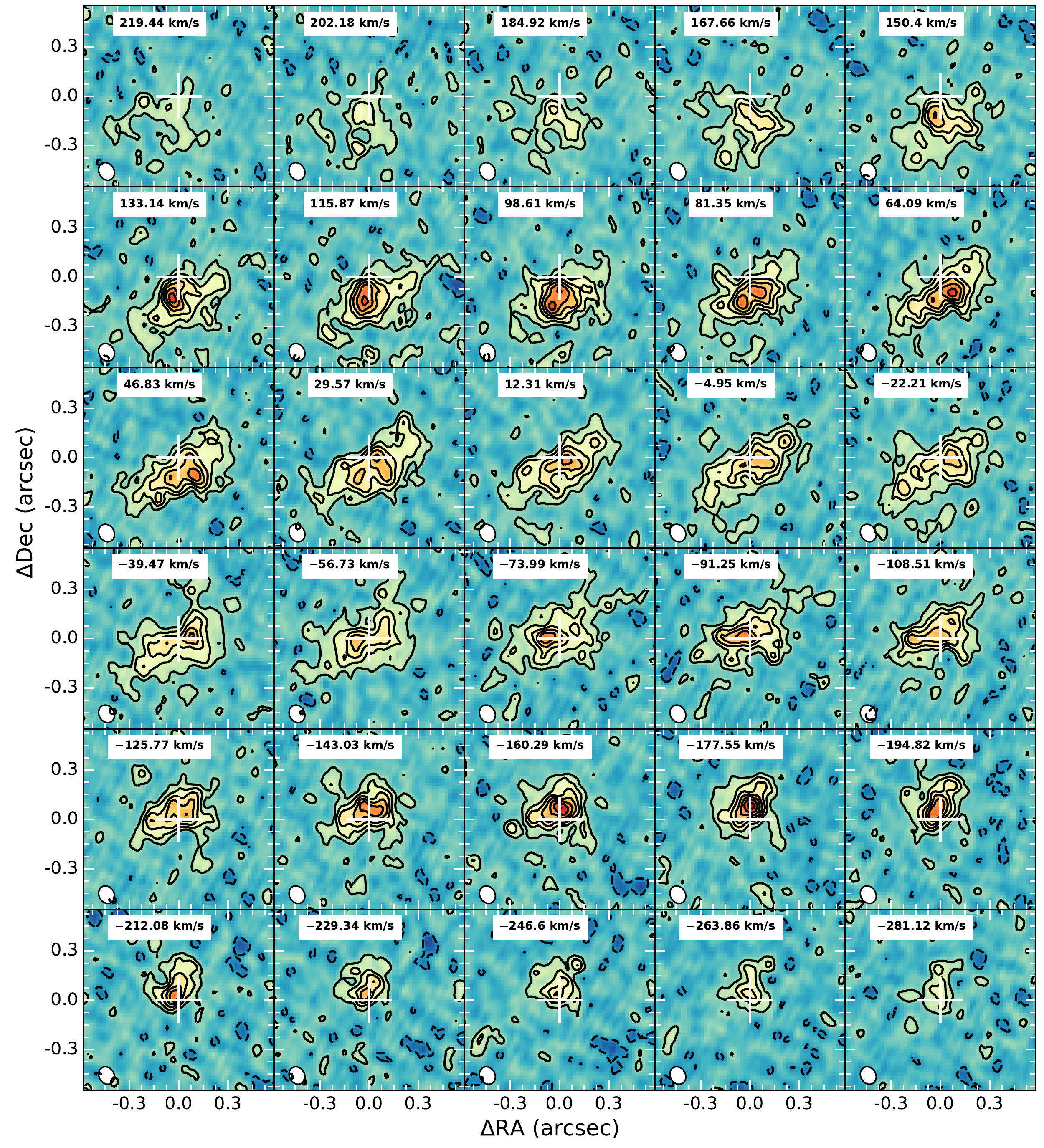}} 
\caption{ [\ion{C}{II}] channel maps of J2310+1855. The velocity takes the [\ion{C}{II}] redshift from \citet{Wang2013} as a reference. The channel width is $\sim$ 17 km/s (corresponding to 15.625 MHz). The contour levels are [$-$2, 2, 4, 6, 8, 10, 12, 14] $\times$ rms, where rms = 0.17 mJy/beam. The white plus sign represents the HST quasar position. The [\ion{C}{II}]  emission peak moves in a circular, counterclockwise path from 219 to $-$281 km/s.}
\label{ciichannel}
\end{figure*}

\begin{figure*}[h]
\centering
\subfigure{\includegraphics[width=0.99\textwidth]{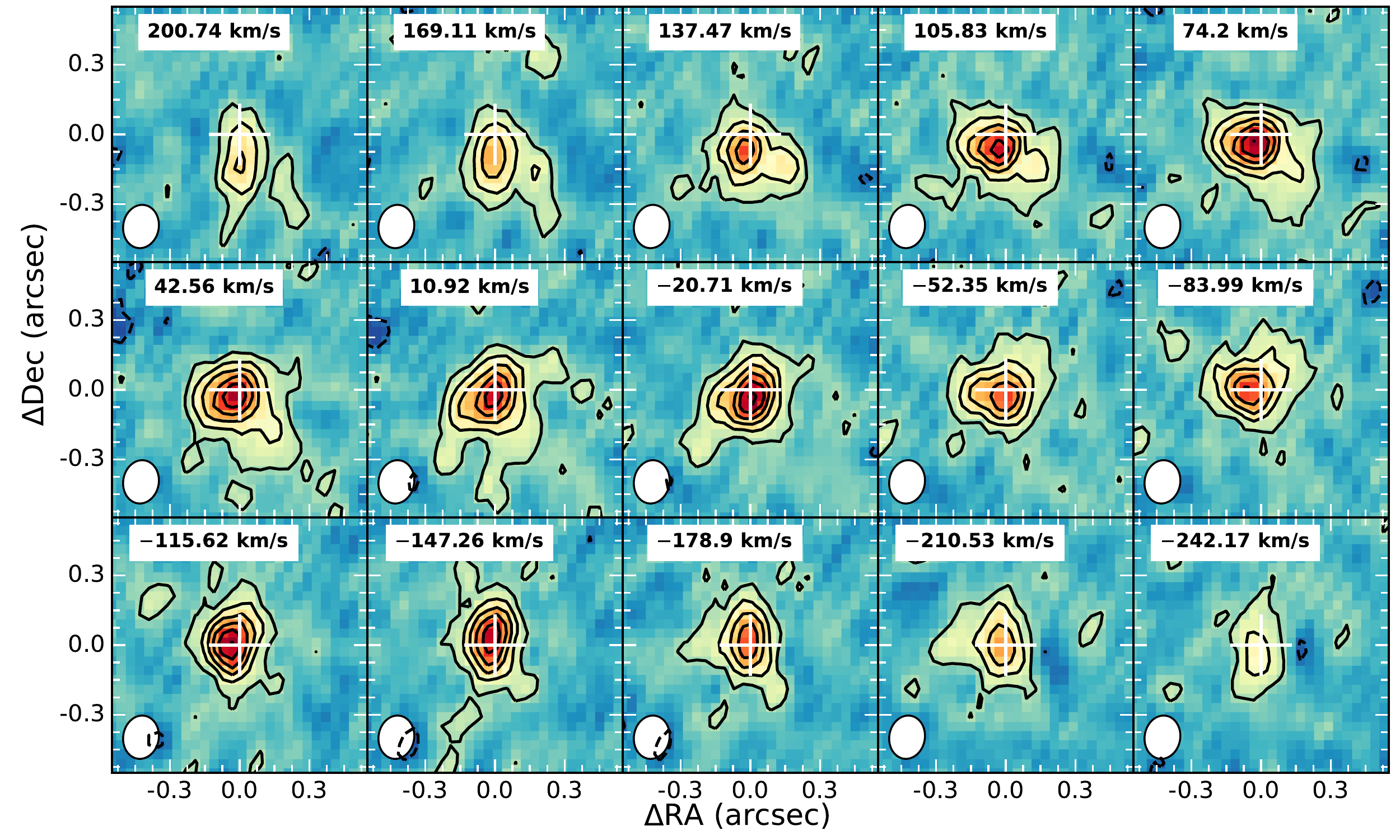}} 
\caption{ CO\,(9--8) channel maps of J2310+1855. The velocity takes the [\ion{C}{II}] redshift from \citet{Wang2013} as a reference. The channel width is $\sim$ 32 km/s (corresponding to 15.625 MHz). The contour levels are [$-$2, 2, 4, 6, 8, 10, 12, 14] $\times$ rms, where rms = 0.10 mJy/beam. The white plus sign represents the HST quasar position. }
\label{co98channel}
\end{figure*}

\section{Two-dimensional elliptical S{\'e}rsic function}
\label{sec_fun}

The 1D S{\'e}rsic profile as a function of radius, $r$, is given by \citet{Sersic1963}:

\begin{equation}
\label{sersicfunc}
\Sigma(r) = \Sigma_{0}\exp^{- (r/r_{s})^{1/n} },
\end{equation}
where $\Sigma_{0}$ is the central surface brightness, $r_{s}$ is the scale length, which is the radius at which the surface brightness $\Sigma(r)$  drops by $e^{-1}$, and $n$ is the shape parameter that controls the degree of curvature of the profile, the so-called S{\'e}rsic index. 
When $n$ is large (i.e., $>4$), $r_{s}$ is too small to measure; therefore, one often redefines the
profile in terms of the half-light radius, $r_{\rm e}$, also known as the effective radius:
 
\begin{equation}
\Sigma(r) = \Sigma_{0}\exp^{- k(r/r_{\rm e})^{1/n} },
\end{equation}
where the dependent variable $k$ is coupled to $n$, and $r_{\rm e}=k^{n}r_{s}$. $k$ satisfies $\frac{1}{2}\Gamma(2n) = \gamma(2n; k)$, where $\Gamma$ and  $\gamma$ are the Gamma function and lower incomplete Gamma function, respectively. $k$ can be approximated as $2n-\frac{1}{3}+\frac{4}{405n}+\frac{46}{25515n^{2}}+\frac{131}{1148175n^{3}}+\frac{2194697}{30690717750n^{4}}$ for $n>0.36$ \citep{Ciotti1999}.

The 2D elliptical S{\'e}rsic function can be expressed as
\begin{equation}
\label{2dsersicfunc}
f(x, y) = A\exp^{- [a(x-x_{0})^{2}+2b(x-x_{0})(y-y_{0})+c(y-y_{0})^{2}]^{1/(2n)} }
\end{equation}

\begin{equation}
a=\frac{\cos^{2}\theta}{h^{2}_{x}}+\frac{\sin^{2}\theta}{h^{2}_{y}}
\end{equation}

\begin{equation}
b=\frac{\sin2\theta}{2h^{2}_{x}}-\frac{\sin2\theta}{2h^{2}_{y}}
\end{equation}

\begin{equation}
c=\frac{\sin^{2}\theta}{h^{2}_{x}}+\frac{\cos^{2}\theta}{h^{2}_{y}}
,\end{equation}
where the coefficient $A$ is the height of the peak, ($x_{0},\, y_{0}$) is the center, $h_{x}$, $h_{y}$ are the  scale heights along the x and y axis before rotating by $\theta$, and $\theta$ is the rotation angle  in radians, which increases counterclockwise counting from the positive direction of the x axis, and $n$ is the S{\'e}rsic index. 
We show surface plots, projected filled contour plots and surface brightness distributions in Fig. \ref{expblob} for 2D elliptical S{\'e}rsic images with different rotation angles and S{\'e}rsic indices.

\begin{figure*}[h]
\centering
\subfigure{\includegraphics[width=0.97\textwidth]{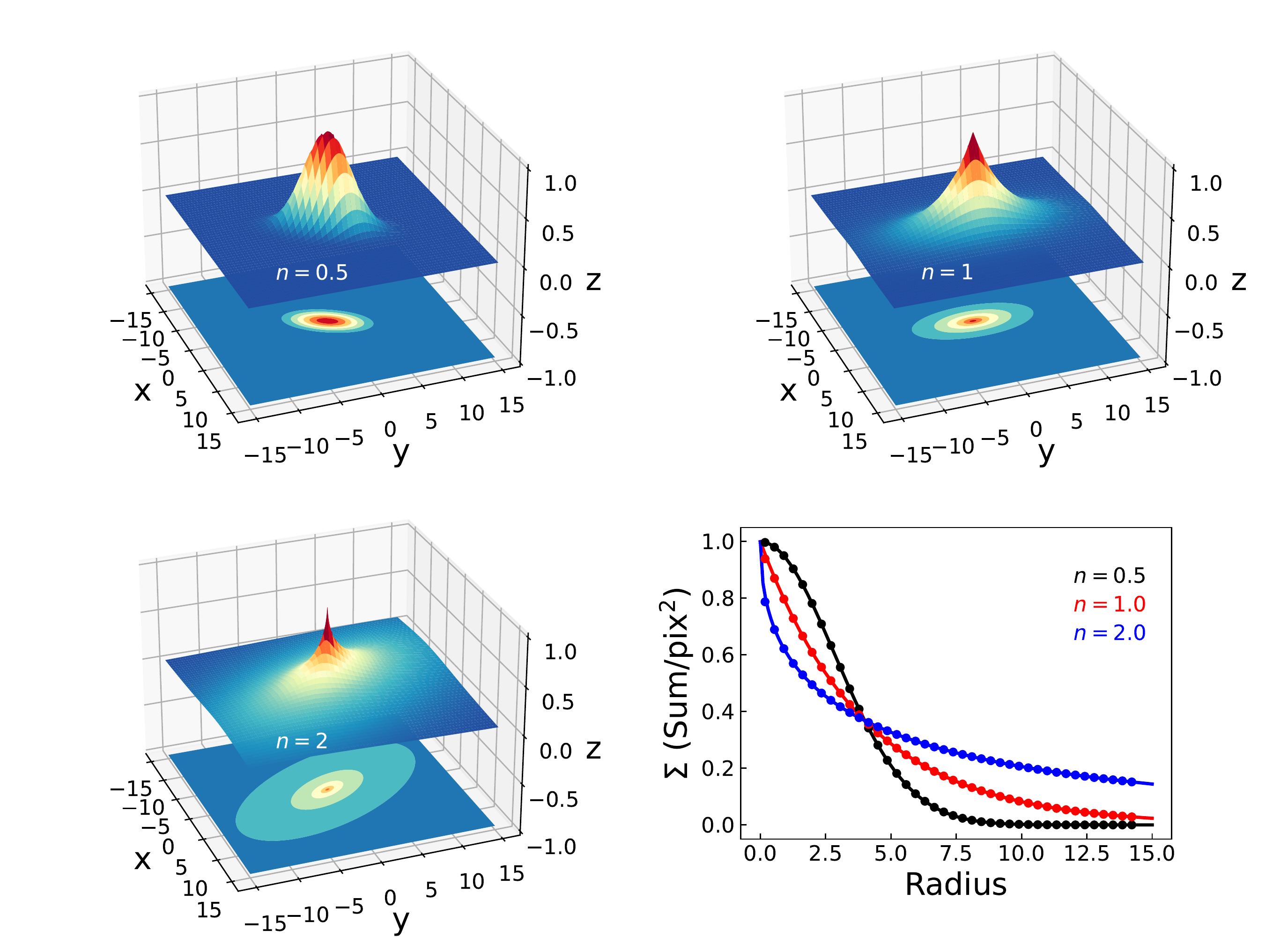}} 
\caption{Surface plots, projected filled contour plots, and surface brightness distributions for 2D elliptical S{\'e}rsic images with ($x_{0},\, y_{0}$) = (0, 0), $h_{x}=4$, and $h_{y}=2$. $\theta$ and $n$ are $\pi/3$ and 0.5 (top left), $\pi/2$ and 1 (top right),  $2\pi/3$ and 2 (bottom left).  The surface brightness as a function of radius shown in the bottom-right panel is measured using elliptical annuli with the same rotation angle of these images. The radius is the one along the major axis. The black, red and blue dots are measured from the top-left, top-right, and bottom-left images, respectively. The black, red, and blue lines are the fits to these surface brightness distributions with 1D S{\'e}rsic function in Eq.  \ref{sersicfunc}. The best-fitted S{\'e}rsic indices are labeled in the same colors as the fitting lines. 
}
\label{expblob}
\end{figure*}

We use a 2D elliptical S{\'e}rsic model to fit the CO\,(9--8)  line and the [\ion{C}{II}] and CO\,(9--8)  underlying dust continuum. And as motivated by the existence of AGN torus, we also fit these intensity maps with an additional unresolved nuclear component to represent the dusty and molecular torus.  The results are shown in Figs. \ref{sbco98}, \ref{sbciidust}, and \ref{sbcodust}.

\begin{figure*}
\centering
\subfigure{\includegraphics[width=0.97\textwidth]{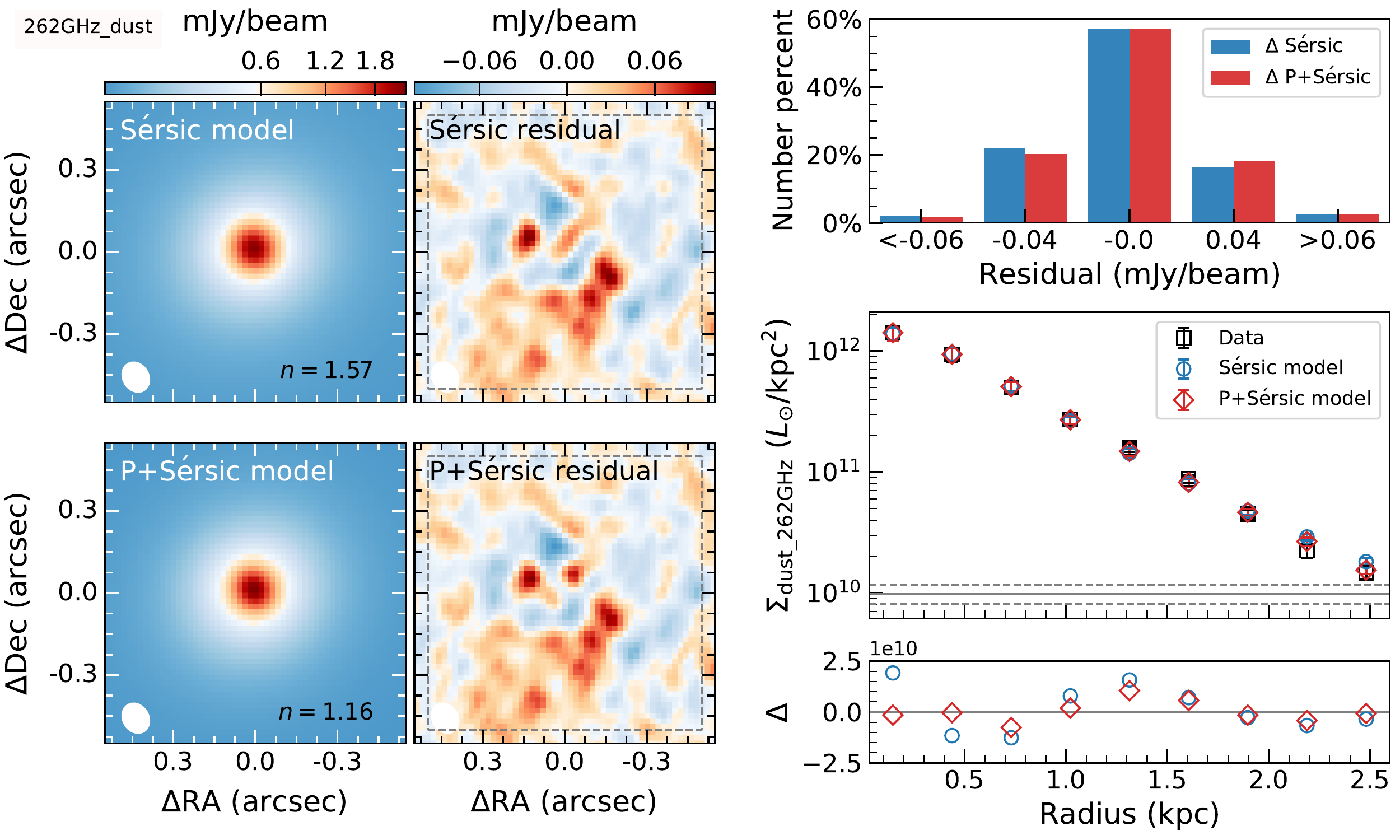}} 
\caption{As in Fig. \ref{sbco98} but for the [\ion{C}{II}] underlying dust continuum. The peak values and rms within the dashed gray square are 0.101 and 0.028, and 0.099 and 0.028 Jy/beam km/s for the S{\'e}rsic and P+S{\'e}rsic residual maps, respectively.  The shape of the synthesized beam with a FWHM size of $0\farcs113\times0\farcs092$ is plotted  as white ellipses. We measured the [\ion{C}{II}] underlying dust continuum luminosity surface density using elliptical rings with the ring width along the major axis of half ($0\farcs05$) that of the major axis of the [\ion{C}{II}] clean beam size, the rotation angle equal to $\overline{\rm PA}$ (=199$\degr$) and the ratio of semiminor and semimajor axis -- $b/a$ of $\cos(\overline{i})$ ($\overline{i}=42\degr$), where $\overline{\rm PA}$ and $\overline{i}$ come from the [\ion{C}{II}] line kinematic modeling (listed in Table \ref{par_kin}).}
\label{sbciidust}
\end{figure*}

\begin{figure*}
\centering
\subfigure{\includegraphics[width=0.97\textwidth]{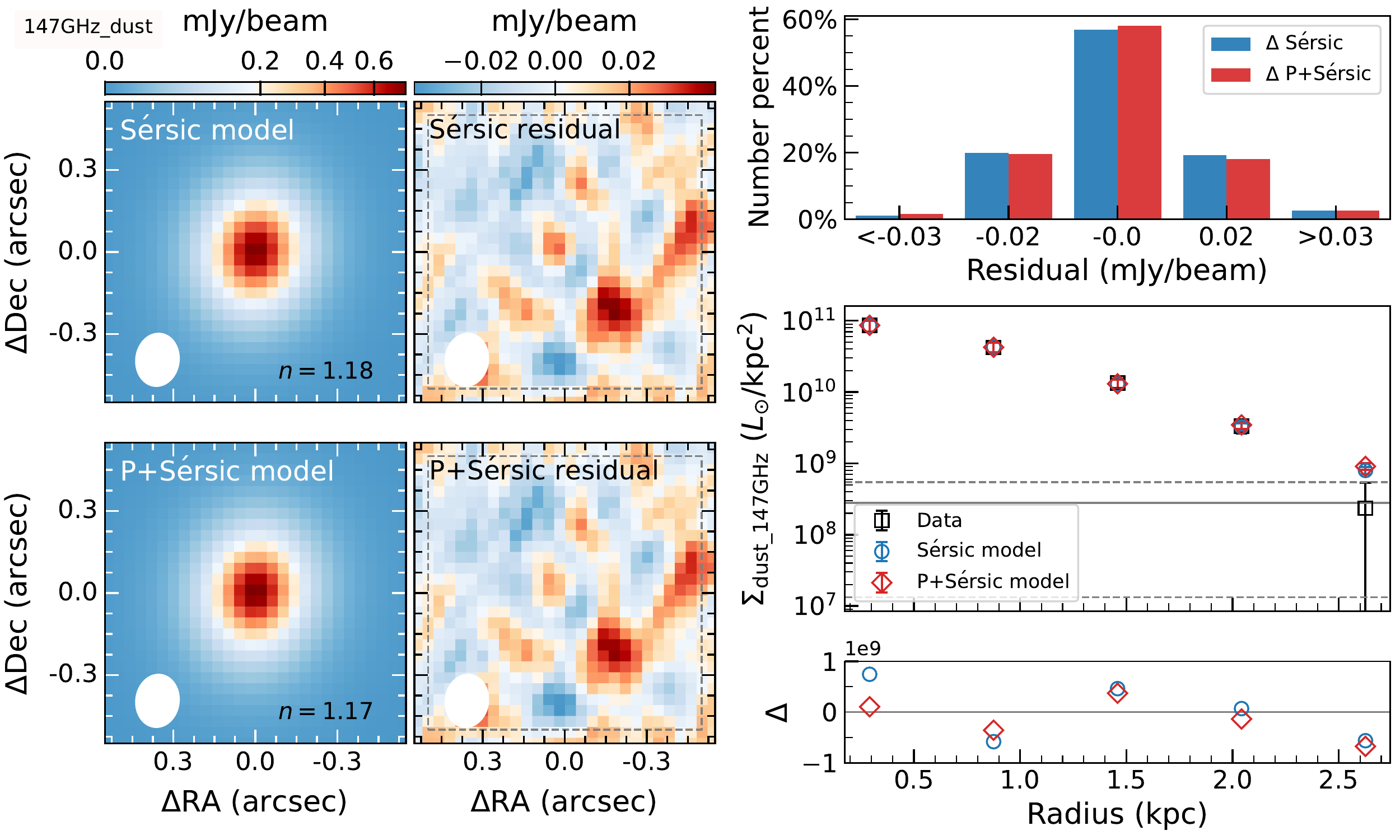}} 
\caption{As in Fig. \ref{sbco98} but for the CO\,(9--8)  underlying dust continuum. The peak values and rms within the dashed gray square are 0.043 and 0.014, and 0.041 and 0.014 Jy/beam km/s for the S{\'e}rsic and P+S{\'e}rsic residual maps, respectively.  The shape of the synthesized beam  with a FWHM size of $0\farcs192\times0\farcs156$ is plotted  as a white ellipse in each panel. }
\label{sbcodust}
\end{figure*}

\section{Tilted ring model}
\label{sec_til}
In this section, we illustrate the parameters used in the tilted ring model. We only observe the line-of-sight velocities as a function of position on the sky, which cannot be directly used to investigate the mass distributions in galaxies. We need to measure the rotation velocity $V_{\rm rot}$ as a function of distance $R$ to the center (the rotation curve). If we assume that the gas is rotating symmetrically in a disk, a number of tilted ring model parameters can define the observed velocity field of a galaxy. The tilted ring model (see Fig. \ref{tilted}) consists of a set of concentric rings characterized by various geometrical and kinematical components. The geometrical components are: 

$\bullet$ $x_{0}, \,y_{0}$: sky coordinates of the rotation center of the galaxy.

$\bullet$ $i$ ($R$): the inclination (i.e., the angle between the normal to the plane of the galaxy and the line-of-sight.

$\bullet$ $\phi$ ($R$): the position angle of the major axis of a ring projected onto the sky (i.e., an ellipse). This is an angle taken in a counterclockwise direction from the north of the sky to the major axis of the receding half of the galaxy.

$\bullet$  $\zeta$ ($R$): the azimuthal angle measured  between the major axis on the sky plane and the line connecting the galaxy center ($x_{0}, \,y_{0}$) and the data point ($X,\, Y$)  in the galaxy plane. It is related to $i$ ($R$), $\phi$ ($R$), ($x_{0}, \,y_{0}$).\\

 \noindent The kinematical components are:

$\bullet$ $V_{\rm sys}$: the velocity of the center of the galaxy with respect to the Sun, the so-called systemic velocity.

$\bullet$ $V_{\rm rot}$($R$): the rotation velocity at distance $R$ from the center. This velocity is assumed to be constant in a ring.\\

\noindent The $V_{\rm rot}$ is perpendicular to the ring and has components in the X and Y directions. $V_{\rm X}$ ($= V_{\rm rot}\sin\zeta$) is parallel to the xy plane and has  no component in the z direction. However, $V_{\rm Y}$ ($= V_{\rm rot}\cos\zeta$) has a component  in the z direction: $V_{\rm z}$ = $V_{\rm Y} \sin i$. This is the radial velocity ($V_{\rm los}$) that we observe at a point ($x,\, y$). It must be corrected for the systemic (line-of-sight) velocity $V_{\rm sys}$. Thus,

\begin{equation}
V_{\rm los}(x, y) = V_{\rm sys} + V_{\rm z} = V_{\rm sys} + V_{\rm rot}\cos\zeta\sin i
\end{equation}

\begin{equation}
\cos\zeta =\frac{X}{R}= \frac{-(x-x_{0})\sin\phi+(y-y_{0})\cos\phi}{R}
\end{equation}

\begin{equation}
\sin\zeta = \frac{Y}{R}= \frac{-(x-x_{0})\cos\phi-(y-y_{0})\sin\phi}{R\cos i}
.\end{equation}

\begin{figure}[h]
\centering
\subfigure{\includegraphics[width=0.45\textwidth]{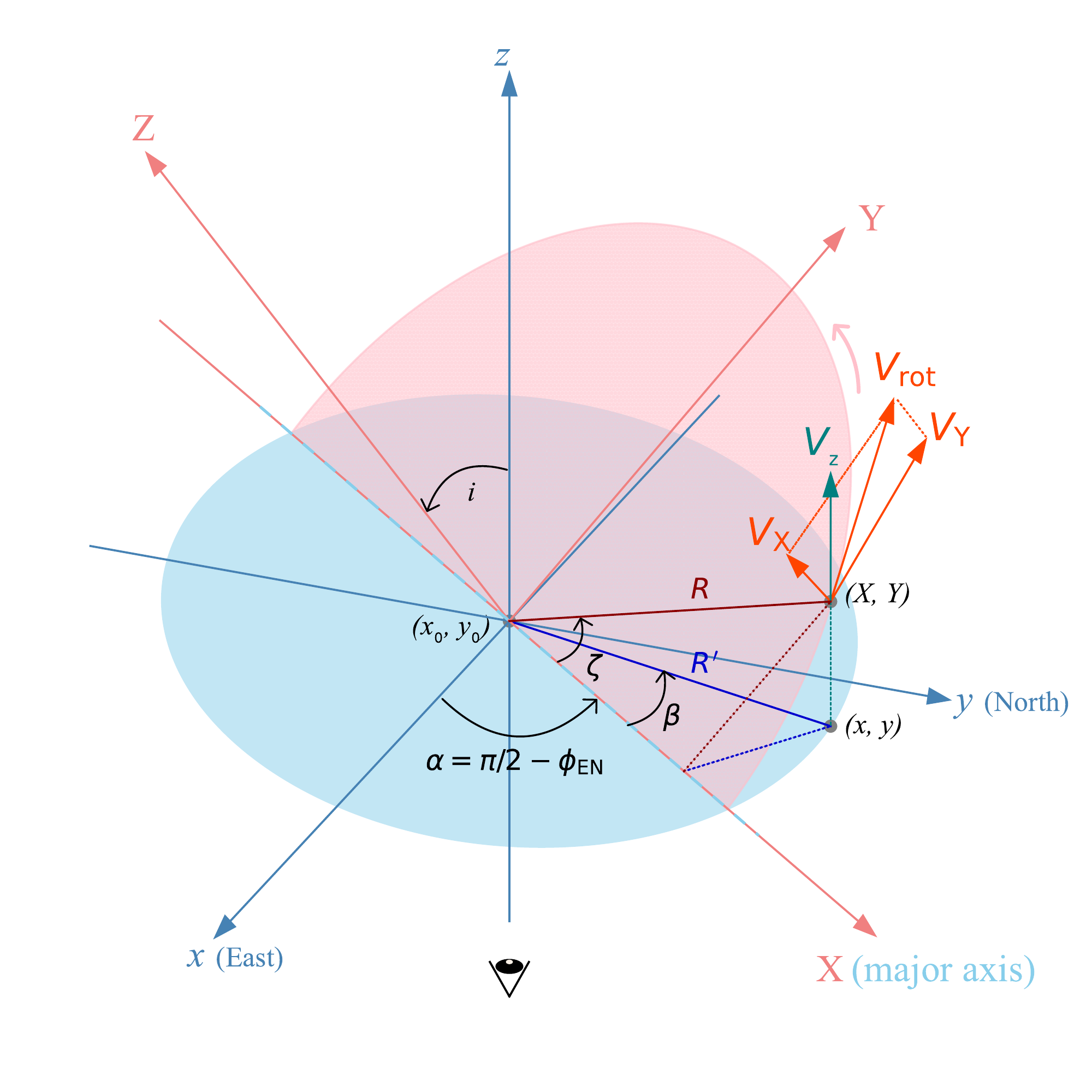}} 
\caption{Simple schematic diagram for the tilted ring model. The pink semiellipse represents the galaxy plane with a coordinate system of XYZ. The red-based color lines are on or perpendicular to the galaxy plane. The light-blue ellipse presents the sky plane with a coordinate system of xyz. The blue color lines are on or perpendicular to the sky plane. We takes the x direction as the east and the y direction as the north. The major axis on the sky plane coincides with the X axis on the galaxy plane. The line-of-sight direction is along the z direction from the bottom to the top. The galaxy plane is inclined with an angle of $i$ (the inclination angle, i.e., the angle between the line of sight and the perpendicular line to the galaxy plane XY). The data point ($X,\, Y$) with a distance of $R$ to the center rotates counterclockwise with a rotation velocity of $V_{\rm rot}$ on the galaxy plane. What we can see is the line-of-sight component $V_{\rm z}$ of the rotation velocity $V_{\rm rot}$. The angle between the major axis on the sky plane and the line connecting the galaxy center ($x_{0},\, y_{0}$) and the data point ($X,\, Y$) is the azimuth angle $\zeta$. The position angle $\phi$ is defined as the angle counting from the north (y axis) to the major axis of the receding half on the galaxy plane. After tilting, the data point ($X,\, Y$), the distance $R$ and the azimuth angle $\zeta$ on the galaxy plane project to ($x,\, y$), $R\arcmin$ and $\beta$ on the sky plane.}
\label{tilted}
\end{figure}

Here, we also put the $^{\rm 3D}$B{\scriptsize{AROLO}} modeling result of CO\,(9--8) line in Fig. \ref{co98datamodel}.

\begin{figure*}
\centering
\subfigure{\includegraphics[width=0.99\textwidth]{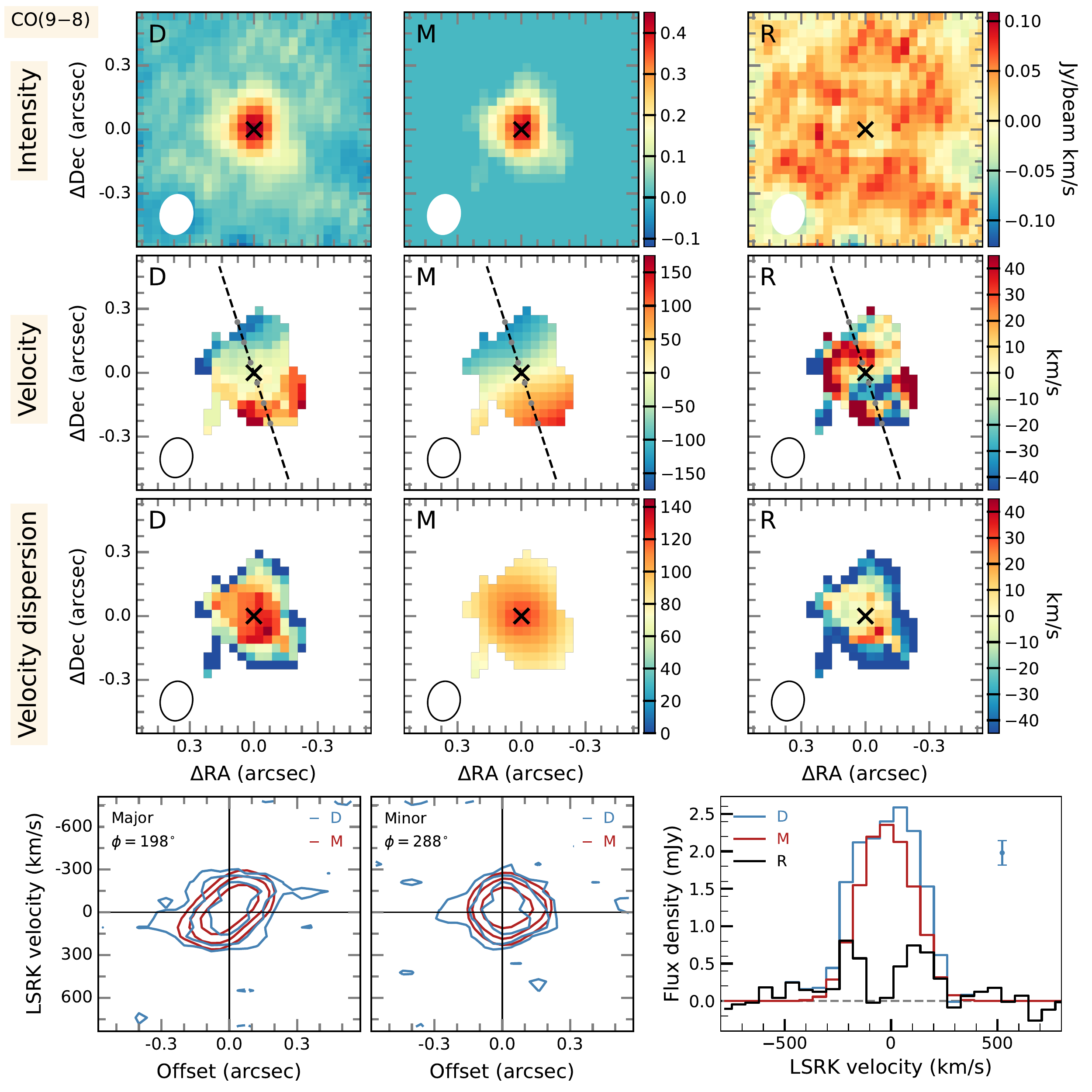}} 
\caption{Similar to that of  Fig. \ref{ciidatamodel}, but for the CO\,(9--8)  line kinematic modeling. {\bf Upper panels --} Shape of the synthesized beam with a FWHM size of $0\farcs187\times0\farcs153$ is plotted in the bottom-left corner of each panel. {\bf Lower panels --} Left and middle panels: Contour levels are [--2, 2, 4, 8, 16] $\times$ 0.093 mJy/beam for both the data and the model. Right panel: The spectral resolution is 31.25 MHz, corresponding to 64 km/s. }
\label{co98datamodel}
\end{figure*}

\end{appendix}

\end{document}